\documentclass[11pt,a4paper]{article}
\pdfoutput=1
\usepackage{jheppub}
\usepackage{amsthm,amsbsy,amsfonts,mathrsfs,enumerate,float,wrapfig,amsmath}
\usepackage{subfigure}

\newcommand{\be}{\begin{equation}}
\newcommand{\ee}{\end{equation}}
\newcommand{\bea}{\begin{eqnarray}}
\newcommand{\eea}{\end{eqnarray}}

\title{Discrete theta angle from an O5-plane}

\author[a]{Hirotaka Hayashi,}
\author[b]{Sung-Soo Kim,}
\author[c]{Kimyeong Lee,}
\author[d]{and Futoshi Yagi}

\affiliation[a]{Department of Physics, School of Science, Tokai University, 4-1-1 Kitakaname, Hiratsuka-shi, Kanagawa 259-1292, Japan}
\affiliation[b]{School of Physical Electronics, University of Electronic Science and Technology of China, North Jianshe Road, Chengdu 611731, China}
\affiliation[c]{School of Physics, Korea Institute for Advanced Study, 
85 Hoegi-ro Dongdaemun-gu, Seoul 02455, Korea}
\affiliation[d]{Department of Physics, Technion - Israel Institute of Technology, Haifa 32000, Israel}

\emailAdd{h.hayashi@tokai.ac.jp}
\emailAdd{sungsoo.kim@uestc.edu.cn}
\emailAdd{klee@kias.re.kr}
\emailAdd{fyagi@physics.technion.ac.il}

\abstract{
We consider 5d $\mathcal{N}=1$ $Sp(1)$ gauge theory based on a brane configuration with an O5-plane. At the UV fixed point, the theory with no matter enjoys enhanced global symmetry $SU(2)$ or $U(1)$ depending on the discrete theta angle $\theta=0, \pi$ (mod $2\pi$).  A naive brane configuration with an O5-plane, however, does not distinguish two different theories, as it describes the weak coupling region. We devise a technique for computing 5d Seiberg-Witten curve of the two theories from the brane web with an O5-plane. Their Seiberg-Witten curves show that their M5 configurations under the presence of OM5-planes are different. The decompactification limit of each Seiberg-Witten curve also shows distinct phase structures in their Coulomb branch leading to significantly different $(p,q)$ 5-brane configurations with an O5-plane in the strong coupling region. }

\begin{document}
\preprint{
\begin{flushright}
\tt 
KIAS-P17052\\
\end{flushright}
}

\maketitle


\section{Introduction}
\label{sec:intro}

A large class of five-dimensional (5d) supersymmetric quantum field theories with eight supercharges has been constructed
by 5-brane web diagrams, which were originally considered in \cite{Aharony:1997ju, Aharony:1997bh}.
The 5-brane web diagrams consist not only of D5-branes and NS5-branes but also of $(p, q)$ 5-branes. Furthermore we may include orientifolds such as an O7$^{\pm}$-plane, an O5$^{\pm}$-plane and also an ON$^0$-plane. The inclusion of O-planes can realize 5d gauge theories with gauge groups of $USp, SO$ for example. 

A simple example of 5d gauge theories is the pure $SU(2)$ gauge theory. 
In fact, we have two distinct pure $SU(2)$ gauge theories depending on the discrete theta angle $\theta$ \cite{Seiberg:1996bd, Morrison:1996xf, Douglas:1996xp}. The two theories exhibit a different flavor symmetry at the ultraviolet (UV) fixed point. One superconformal field theory (SCFT) has an $SU(2)$ flavor symmetry, called $E_1$ theory ($\theta = 0$), whereas the other SCFT has a $U(1)$ flavor symmetry, called $\widetilde{E}_1$ theory ($\theta = \pi$). The two theories are constructed by different 5-brane web diagrams \cite{Aharony:1997ju, Aharony:1997bh} and the flavor symmetry enhancement can be seen from 7-branes attached to the external 5-branes \cite{DeWolfe:1999hj}. The discrete theta angle is discussed in the context of the type I' string theory in \cite{Bergman:2013ala} as well as in terms of the decomposition of an O7$^-$-plane \cite{Bergman:2015dpa}.

The pure $SU(2)$ gauge theory has another 5-brane web construction using an O5-plane. A D5-brane on top of an O5$^+$-plane yields an $Sp(1)$ gauge group which is isomorphic to an $SU(2)$ gauge group. 
In fact, it has not been known how to encode the information of the discrete theta angle into the 5-brane web with an O5-plane. 
For maximally supersymmetric 5d $Sp(N)$ gauge theories, the difference between an O4$^+$-plane and an $\widetilde{\text{O4}}^+$-plane could explain the discrete theta angle \cite{Hori:1998iv, Tachikawa:2011ch}. Therefore, one might think that we may use an $\widetilde{\text{O5}}^+$-plane instead of an O5$^+$-plane to realize the pure $SU(2)$ gauge theory with $\theta = \pi$. However, it has been pointed out in \cite{Feng:2000eq, Zafrir:2015ftn} that such a construction rather yields the 5d $Sp(1)$ gauge theory with one massless flavor. Hence, it has not been clear how to distinguish the pure $Sp(1)$ gauge theories with a different discrete theta angle from the 5-brane web diagrams with an O5-plane.

In order to address this issue, we first move to the M-theory picture as follows:
we consider the compactification of the 5-brane web diagram with an O5-plane on a circle. Then, T-duality along the circle gives rise to D4/NS5-branes from $(p, q)$ 5-branes as well as two O4-planes from one O5-plane. An M-theory uplift of this type IIA picture yields the configuration of M5-branes under the existence of two OM5-planes \cite{Hori:1998iv, Ahn:1998qe, Gimon:1998be, Hanany:2000fq}, which
is nothing but the 5d Seiberg-Witten curve for the 5d theory realized on the 5-brane web diagram with the O5-plane \cite{Brandhuber:1997ua, Witten:1997sc}. Although the Seiberg-Witten curve for the 5d $Sp(N)$ theory with $N_f$ flavors has been proposed in \cite{Brandhuber:1997ua} with this method, the discrete theta angle has not been clear 
for the $N_f=0$ case.

We will see that the difference between the $E_1$ theory and the $\widetilde{E}_1$ theory is achieved by different boundary conditions for the two OM5-planes which arise by performing T-duality for the O5-plane and uplifting to M-theory, resulting in two different Seiberg-Witten curves.
This analysis indicates that 5-brane webs for the $E_1$ theory and the $\widetilde{E}_1$ theory are different at least when compactified on a circle.

Once we obtain the 5d Seiberg-Witten curve of the 5d theory on a circle, one can reproduce the original 5-brane web diagram by taking a decompactification limit of the circle \cite{Aharony:1997bh}. 
However, this procedure not just reproduces the originally expected 5-brane web diagram in the weak coupling region.
It is remarkable that the different shapes of web diagrams are obtained depending on the value of gauge coupling constant and the Coulomb moduli parameter.
Especially, it is also possible to obtain a 5-brane web diagram of the pure $Sp(1)$ gauge theory in the ``strong coupling region'', where the gauge coupling square is negative $1/g^2 < 0$, whose configuration have not been discussed before.\footnote{Rigorously, the region $1/g^2 < 0$ does not make sense as the original gauge theory. But it still makes sense as a relevant deformation of the SCFT realized on the UV fixed point of the original gauge theory. Throughout this paper, ``strong coupling'' denotes the case $1/g^2 < 0$.}
We will see that the strong coupling behavior of the 5-brane web diagram for the $E_1$ theory is different from that for the $\widetilde{E}_1$ theory. The analysis introduces various intriguing configurations for a 5-brane web diagram with an O5-plane. 

The difference of the web diagrams for the $E_1$ theory and the $\widetilde{E}_1$ theory are understood more uniformly by adding one flavor, which makes the $E_2$ theory.
We also study the Seiberg-Witten curve of this $E_2$ theory and its decompactification limit. We confirm that the web diagram for the $E_1$ theory and the $\widetilde{E}_1$ theory are obtained by the different limit where mass parameter goes to $-\infty$ and $+\infty$, respectively.
When shifting the mass parameter for the flavor, a generalized version of ``flop transition'' arises, which plays an important role in differentiating the flavor decoupling limit to the $E_1$ theory from the decoupling limit to the $\widetilde{E}_1$ theory by allowing different flop transitions in the strong coupling region.

The organization of this paper is as follows. In section \ref{sec:E1E1t}, we consider the  
way to distinguish the $E_1$ theory from the $\widetilde{E}_1$ theory from a circle compactification of the 5-brane web diagrams. We propose boundary conditions of 5d Seiberg-Witten curves at the two OM5-planes which originate from an O5-plane. We find two types of the boundary condition and argue that the different boundary conditions yield the 5d Seiberg-Witten curves for the pure $Sp(1)$ gauge theories with the two different discrete theta angles $\theta$. 
The method investigated in this section will be also useful to construct the 5d Seiberg-Witten curve from a generic 5-brane web diagram with an O5-plane.
In section \ref{sec:5dlimit}, we 
consider the decompactification limit of the 5d Seiberg-Witten curves of the $E_1$ and the $\widetilde{E}_1$ theory obtained in the previous section. We classify all the types of the 5-brane web diagrams with an O5-plane for the $Sp(1)$ gauge theories with $\theta = 0$ and $\theta = \pi$. In section \ref{sec:E2}, we also classify all the different types of the 5-brane web with an O5-plane for the $Sp(1)$ gauge theory with one flavor. We find that the phases of the $E_2$ theory consistently reduces to the phases of the $E_1$ and the $\widetilde{E}_1$ theory by decoupling one flavor. We then conclude in section \ref{sec:concl} with a summary of the results found in this paper. In appendix \ref{sec:E8}, we utilize the boundary conditions of two OM5-planes to compute the 5d Seiberg-Witten curves of the $E_{N}$ theory with $0 \leq N \leq 8$. In appendix \ref{sec:E2Web} we classify all the types of the 5-brane web diagram without orientifold of the $E_2$ theory and see the consistent results obtained in section \ref{sec:E2}. Appendix \ref{sec:E2Web} also summarizes the detailed structure of all the types of the 5-brane web diagram with an O5-plane for the $E_2$ theory.

\bigskip

\section{5d Seiberg-Witten curves of  the $E_1$ and $\widetilde{E}_1$ theories} 
\label{sec:E1E1t}
In this section, we discuss a method which enables us to compute 5d Seiberg-Witten curves based on a web diagram including an O5-plane.
 As a 5-brane web configuration with an O5-plane describes a 5d $\mathcal{N}=1$ $Sp(N)$ gauge theory, our method is about computing the 5d Seiberg-Witten curves of 5d $Sp(N)$ theories. Here, we focus on 5d $Sp(1)$ gauge theories as generalization to higher rank $Sp(N)$ as well as adding flavors is straightforward. 

\begin{figure}
\centering
\includegraphics[width=9cm]{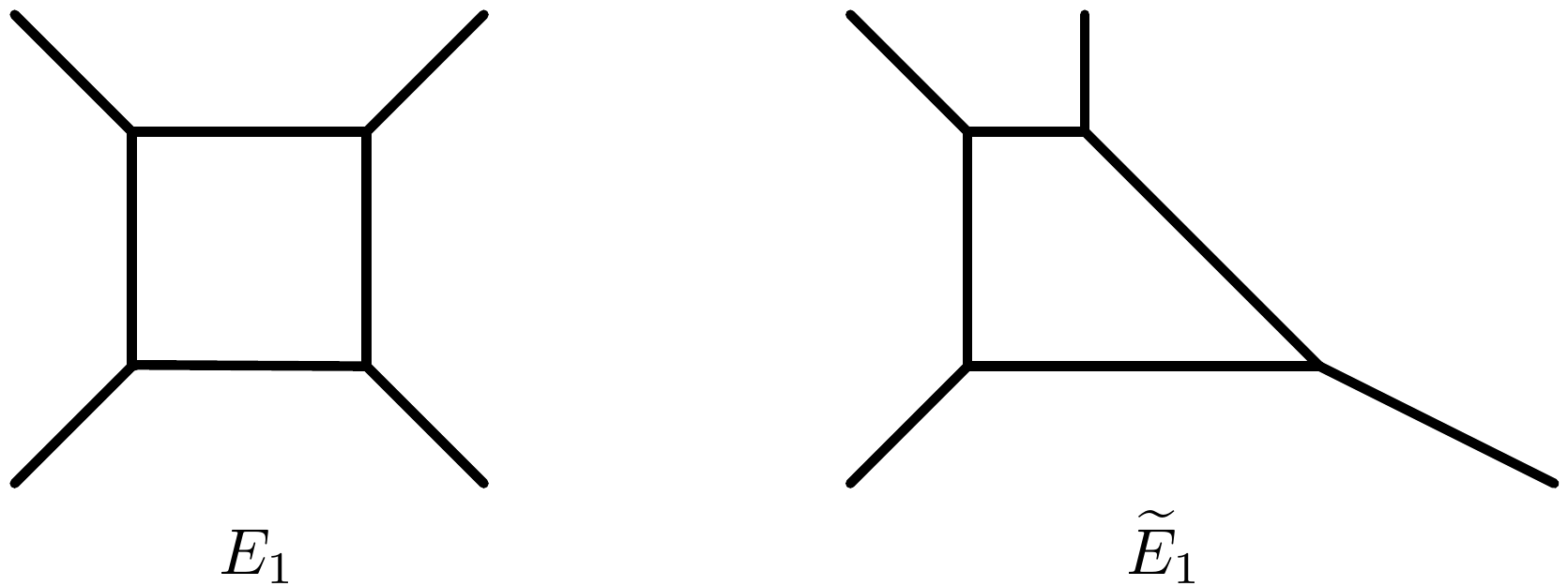}
\caption{Web diagram for 5d $SU(2)$ theory of enhanced $E_1=SU(2)$ and $\widetilde{E}_1=U(1)$ global symmetries}
\label{Fig:su2E1E1t}
\end{figure}

\begin{figure}
\centering
\includegraphics[width=11cm]{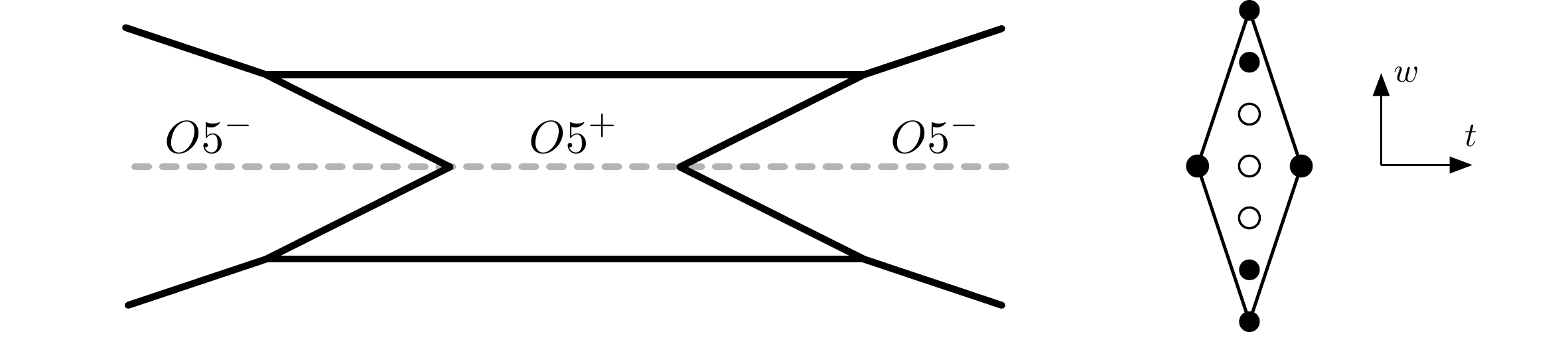}
\caption{Covering space web configuration for 5d $Sp(1)$ theory and corresponding toric-like diagram.}
\label{Fig:sp1toriclike}
\end{figure}

5d $\mathcal{N}=1$ $Sp(1)$ gauge theory with $N_f\le 7$ flavors enjoys global symmetry enhancement $SO(2N_f)\times U(1) \to E_{N_f+1}$ at UV fixed point\footnote{We use the following convention of $E_{N_f+1}$ : $E_8,\ E_7,\ E_6,\ E_5=SO(10),\ E_4=SU(5),\ E_3=SU(3)\times SU(2),\ E_2=SU(2)\times U(1),\ E_1=SU(2),\ \widetilde{E}_1=U(1)$.} \cite{Seiberg:1996bd, Morrison:1996xf, Douglas:1996xp, Ganor:1996pc, Kim:2012gu}. The theory has discrete theta angle $\theta=0, \pi$ (mod $2\pi$). When the theory has flavors, the effect of the non-trivial discrete theta angle is equivalent to flipping the sign of the mass of a flavor, and hence it is equivalent to the original theory. On the other hand, when the theory has no matter $(N_f=0)$, such discrete theta angle gives rise to two physically inequivalent theories. The enhanced global symmetries are different depending on discrete theta angle. It is $E_1=SU(2)$ for $\theta=0,$ while $\widetilde{E}_1=U(1)$ for $\theta=\pi$. The corresponding $SU(2)$ brane configurations show a clear difference as in Figure \ref{Fig:su2E1E1t}. The Seiberg-Witten curves for the $E_1$ theory and the $\widetilde{E}_1$ theory hence show clear differences as written in \cite{Minahan:1997ch, Kim:2014nqa}. Expressed in a brane web diagram with an O5-plane, on the other hand, the $E_1$ and $\widetilde{E}_1$ configurations seem indistinguishable. In other words, on $(p,q)$ 5-brane web shown in Figure \ref{Fig:sp1toriclike}, discrete theta angle seems not manifest, as the charges of ($p, q$) 5-branes toward O5-plane are fixed by charge conservation. We however claim that it is still possible to obtain $E_1$ and $\widetilde{E}_1$ Seiberg-Witten curves by giving different boundary conditions on an orientifold-plane.

\paragraph{Boundary condition for an O4-plane system:}
It is instructive to recall how Seiberg-Witten curves for 4d $\mathcal{N}=2$ pure $Sp(N)$ gauge theory can be obtained from brane configurations. We first uplift type IIA theory to M-theory, where a D4/NS5-brane becomes an M5-brane and an O4-plane becomes an OM5-plane.
As discussed in \cite{Landsteiner:1997vd, Brandhuber:1997cc}, the idea is to consider the covering space of the brane configuration with an OM5-plane which includes the mirror pair of the branes. The structure of Seiberg-Witten curves is then expressed in terms of an even function with respect to $v$ which parametrizes the transverse directions of an OM5-plane. The behavior of the M5-brane near the OM5-plane should be the M-theory uplift of the configuration of the two NS5-branes bending toward each other to be connected on the O4-plane. This leads to a form of the Seiberg-Witten curve which has a double root at $v=0$ \cite{Landsteiner:1997vd}
\begin{align}
t^2+ \big(v^2 B(v^2) -2 \big) t+ 1=0,
\end{align}
where $B(v^2)$ is an even function of $v$, respecting the invariance under $v\leftrightarrow -v$.

\paragraph{Boundary condition for an O5-plane system:}
In a similar fashion, we consider a web diagram with an O5-plane and its covering space which includes its mirror pair as shown in Figure \ref{Fig:sp1toriclike}.  Here, D5-branes are extended along the $012345$-directions and NS5-branes are extended along the $012356$-directions. 
Since the difference of the RR-charge of an O5$^+$-plane and that of an O5$^-$ plane is two, a $(2,1)$ 5-brane and a $(-2,1)$ 5-brane are attached to the O5-plane to satisfy the charge conservation.
By compactifying the $x_5$-direction on a circle and taking T-duality along this direction, one can consider a type IIA configuration where the O5-plane splits into two O4-planes on the circle. 
Uplifting this brane configuration to M-theory, we obtain an M5-brane with two OM5-planes. To describe this M-theory configuration which is 5d Seiberg-Witten curve, we use the coordinates
\begin{align}\label{eq:cplxcoord}
w= e^{-\frac{v}{R_A}} = e^{-\frac{1}{R_A}(x_6+ix_5)},\qquad t= e^{-\frac{1}{R_M} (x_4+ix_{11})}, 
\end{align}
where $R_A$ is the compactification circle radius along $x_5=x_5+2\pi R_A$, while $R_M$ is the M-theory circle radius $x_{11}=x_{11}+ 2\pi R_M$.
Parametrizing the circle by an angle, $\theta_A= {x_5}/{R_A}$, if one of the OM5-planes is located at $\theta_A=0$, the other OM5-plane is placed at $\theta_A=\pi$. 
In the following, we use the convention that two OM5-planes are located at $w=1$ and $w=-1$. It requires that the Seiberg-Witten curve should be invariant under $w \leftrightarrow w^{-1}$. As discussed above, a double root boundary condition is imposed at each position of an OM5-plane. Since we have two OM5-planes, we need to impose boundary conditions both at $w=1$ and $w=-1$.

As discussed in \cite{Aharony:1997bh}, the toric diagram, which is the dual graph of the original $(p,q)$ 5-brane web diagram, tells us which monomial should appear in the Seiberg-Witten curve. We assume that this method is available also for the toric-like diagram of the covering space for the 5d $Sp(1)$ theory with no flavors, given in Figure \ref{Fig:sp1toriclike}. 
Taking into account the invariance under  $w \leftrightarrow w^{-1}$, the corresponding toric-like diagram leads to a quadratic polynomial of $t$ as 
\begin{align}\label{eq:SWansatz}
t^2 + q^{-1}\Big( (w^3  +w^{-3}) +a (w^2+ w^{-2}) + b(w + w^{-1})+ c \Big) t + 1=0,
\end{align}
where $a,b,$ and $c$ are coefficients which will be determined by imposing the boundary conditions.
Here, we used possible rescalings of the curve. For example, overall rescaling is used to fix the constant to be 1, and a coordinate rescaling of $t$ is used to fix the coefficient of $t^2$ to be 1.
Note that asymptotic behaviors at large $w$ determine the coupling of the theory or equivalently the instanton factor $q$ \cite{Kim:2014nqa}.  
We extrapolate the configuration in the large $t \sim w^3$ region, which is originally the $(3,1)$ 5-brane, and that in the large $t^{-1} \sim w^3$ region, which is originally the $(3,-1)$ 5-brane, to the $w=1$ axis and identify the distance between them as $q^{-2}$ so that $q$ is the same instanton factor for the 5d $SU(2)$ theory \cite{Hayashi:2016abm}. 
This makes the coefficient of $w^3+w^{-3}$ to be given by $q^{-1}$.\\

\noindent\underline{$\widetilde{E}_1$ boundary condition:}
We now impose the boundary conditions on the OM5-planes for the pure $Sp(1)$ Seiberg-Witten curve as follows:
For $w=1$ and $w=-1$, the curve take the form of a complete square:
\begin{align}\label{eq:E1tbc}
	&{\rm For~} w=1:&& t^2+ q^{-1} (2+ 2a+ 2b+c)+1 = (t-1)^2,\crcr
	&{\rm For~} w=-1:&& t^2+ q^{-1} (-2+ 2a- 2b+c)+1 = (t-1)^2.
\end{align}
 The solution to the boundary conditions is $b=-1$ and $c=-2a-2q$ where $a$ is identified as the Coulomb branch moduli parameter $U$. 
 Hence this determines the Seiberg-Witten curve for the pure $Sp(1)$ theory
\begin{align}\label{eq:naivesp1curve}
	t^2 + q^{-1}\Big( (w^3  +w^{-3}) +U (w^2+ w^{-2}) -(w + w^{-1})-2(U+q) \Big) t + 1=0,
\end{align}

To compare it with the standard Weierstrass form, we introduce a coordinate which takes into account mirror image of the O5-plane so that it is manifestly invariant under $w\leftrightarrow w^{-1}$
\begin{align}
	x\equiv w+w^{-1}.
\end{align}
The Seiberg-Witten curve is then written as
\begin{align}
	t^2 + \Big(q^{-1} (x^2-4) (x+U)-2\Big) t + 1=0.
\end{align}
By making a complete square of the $t$-terms, we can rewrite it  to be a curve that describes a genus one curve written as a quartic curve 
\begin{align}\label{eq:o5curve1}
	y^2= \Big( (x^2-4) (x+U)-4q \Big)  (x+U),
\end{align}
where $y = 2q (x^2-4)^{-\frac12} \big[t+\frac12 \big(q^{-1} (x^2-4) (x+U)-2\big)\big]$.
Following the procedure converting a quartic curve into a Weierstrass form, given in  Appendix A in \cite{Huang:2013yta}, we have a 5d $Sp(1)$ Seiberg-Witten curve which is obtained from a web diagram with an O5-plane
\begin{align}
	Y^2 &= 4 X^3 - g_2^{O5} X - g_3^{O5},
\end{align}
where
\begin{align}
	g_2^{O5}&=\frac43\Big(U^4-8U^2-24\,q\,U+16 \Big),\crcr
	g_3^{O5}&=-\frac{8}{27} \Big( U^6 - 12 U^4- 36 \,q\,U^3+ 48 U^2+ 144 \,q\,U     + 216 q^2 -64\Big).
\end{align}
It is straightforward to check whether this curve is equivalent to the  5d SU(2) curve associated with  $\widetilde{E}_1=U(1)$ global symmetry whose Weierstrass form is given in \cite{Kim:2014nqa}  
\begin{align}\label{Huange1t}
g_2^{\widetilde{E}_1}=&~\frac{1}{12}U^4 - \frac{2}{3}U^2
- 2\chi_2^{\widetilde{E}_1}U
+\frac{4} {3} , \cr
g_3^{\widetilde{E}_1}=
    &~ \frac{1}{216} U^6  -\frac{1}{18}U^4
    - \frac{1}{6}\chi_2^{\widetilde{E}_1}U^3
    +\frac{2}{9}  U^2   + \frac{2}{3}\chi_2^{\widetilde{E}_1}U
   -\frac{8}{27}
 +\chi_2^{\widetilde{E}_1}\chi_2^{\widetilde{E}_1}.
\end{align}
The agreement can be seen by the following identification
\begin{align}
\chi_2^{\widetilde{E}_1} = q,\qquad 
g_2^{\widetilde{E}_1}=\big(-4\big)^{-2}	g_2^{O5}, \qquad
g_3^{\widetilde{E}_1}=\big(-4\big)^{-3}	g_3^{O5}.
\end{align}
We therefore find that the boundary condition \eqref{eq:E1tbc} gives rise to the Seiberg-Witten curve for the pure $Sp(1)$ theory with the $\widetilde{E}_1$ enhanced global symmetry.
\bigskip

\noindent\underline{${E}_1$ boundary condition:}
We now consider another boundary condition that leads to the theory of the $E_1$ global symmetry. Unlike \eqref{eq:E1tbc}, when we impose the boundary condition for $w=1$ and $w=-1$, \begin{align}\label{eq:E1bc}
	&{\rm For~} w=1:&& t^2+ q^{-1} (2+ 2a+ 2b+c)+1 = (t+1)^2,\crcr
	&{\rm For~} w=-1:&& t^2+ q^{-1} (-2+ 2a- 2b+c)+1 = (t-1)^2.
\end{align}
 The solution to the boundary conditions is $b=-1+q$ and $c=-2a=-2U$ where $U$ is the Coulomb moduli. This leads to the curve 
 \begin{align}\label{eq:E1curve}
	t^2 + q^{-1}\Big( w^3  +w^{-3} +U (w^2+ w^{-2}) -(1-q)(w + w^{-1})-2 U \Big) t + 1=0,
\end{align}
which is different from \eqref{eq:naivesp1curve}. As done for the previous case, it is easy to express it in a more familiar a genus one curve as
\begin{align}
y^2 = \Big((x-2) (x+U)+q\Big)\Big((x+2) (x+U)+q\Big) ,
\end{align}
where $x$ and $y$ are the same as defined before. 
As a Weierstrass form, the curve is given by
\begin{align}
	Y^2 &= 4 X^3 - g_2^{O5}{}_{(2)} X - g_3^{O5}{}_{(2)},
\end{align}
with
\begin{align}\label{eq:O5second}
	g_2^{O5}{}_{(2)} &=\frac{4}{3}\Big( U^4- 8 (1+q)U^2 +16(q^2 -  q +1)  \Big),\\
	g_2^{O5}{}_{(2)} &=-\frac{8}{27}\Big( U^6 - 12 (1+q) U^4 + 24(2q^2+q+2) U^2-32(2 q^3- 3 q^2- 3 q+ 2 ) \Big).\nonumber
\end{align}
It can be readily checked that it agrees with the Seiberg-Witten curve for $E_1$ \cite{Kim:2014nqa} given by
\begin{align}\label{eq:correctE1g2}
g_2^{E_1}=&~\frac{1}{12}U^4 - \frac{2}{3}\chi_1^{E_1}U^2
+\frac{4} {3}\chi_1^{E_1}\chi_1^{E_1} -4, \cr
g_3^{E_1}=
    &~ \frac{1}{216} U^6  -\frac{1}{18}\chi_1^{E_1}U^4
    + 
   \Big(\frac{2}{9}\chi_1^{E_1}\chi_1^{E_1}  -\frac13\Big)
       U^2   
- \frac{8}{27}\chi_1^{E_1}\chi_1^{E_1}\chi_1^{E_1} +\frac43\chi_1^{E_1},
\end{align}
where 
\begin{align}
	\chi_1^{E_1} = q^\frac12+q^{-\frac12}.
\end{align}
The agreement can be seen by the following identification
\begin{align}
	U^{E_1} = U^{O5} q^{-\frac14},\qquad
	g_2^{E_1} = (-4 q^\frac12)^{-2}g_2^{O5}{}_{(2)},\qquad
	g_3^{E_1}= (-4 q^\frac12)^{-3}g_3^{O5}{}_{(2)}.
\end{align}
We therefore find that the boundary condition \eqref{eq:E1bc} gives rise to the Seiberg-Witten curve for $Sp(1)$ theory of ${E}_1$ enhanced global symmetry.

\paragraph{Relation between 5d $\theta$-angle and $Sp(1)$ curve for $E_1$ and $\widetilde{E}_1$.}
We discuss other possible boundary conditions. 
The condition is that the Seiberg-Witten curve have double root at $w=1$ and at $w=-1$, 
where OM5-planes are located.
As discussed previously, there are two choices, $(t-1)^2$ or $(t+1)^2$, for each OM5-plane.
Here $\theta_A = {\rm Arg}(w)$ is the phase of Type IIA circle and $\theta_M= {\rm Arg}(t)$ is the phase of M-theory circle. Hence, all possible choices of the boundary condition are summarized as follows:
\begin{align}
\label{eq:possilbebc}
&{\rm (i)\, For~ the~ 1^{st~}~ b.c.,}\,\left\{
\begin{array}{cl}
	({\rm ia})\,& (\theta_A,\theta_M)=(0,0)~~{:~\rm at~} w=1,\, (t-1)^2=0,\\
	&~{\rm or }\\
	({\rm ib})\,& (\theta_A,\theta_M)=(0,\pi)~~{:~\rm at~} w=1,\, (t+1)^2=0;
\end{array}	
\right.\cr
\cr
&{\rm (ii) \,For~ the~ 2^{nd}~ b.c.,}\left\{
\begin{array}{cl}
	({\rm iia})\,& (\theta_A,\theta_M)=(\pi,0)~~{:~\rm at~} w=-1,\,(t-1)^2=0,\\
		&~{\rm or }\\
	({\rm iib)}\,& (\theta_A,\theta_M)=(\pi,\pi)~~{:~\rm at~} w=-1,\,(t+1)^2=0,\\
\end{array}	
\right.
\end{align}
where the 1$^{\rm st}$ boundary condition is for the first OM5-plane and the 2$^{\rm nd}$ boundary condition is for the second OM5-plane.
We have discussed above that (ia) and (iia) give the boundary conditions for $\widetilde{E}_1$ symmetry, while 
(ib) and (iia) give the boundary conditions for $E_1$ symmetry. 
We note that other two choices also yield the Seiberg-Witten curves with an Wilson line turned on  so that the instanton factor becomes $q\to -q$. More precisely, the choice (ib) and (iib) gives the $\widetilde{E}_1$ curve with $q \to - q$,
while the choice (ia) and (iib) gives the $E_1$ curve with $q \to - q$.
We observe hence that for the present case, there is an intriguing equality between 5d theta angle $\theta$ and the two angles $\theta_A, \theta_M$: 
\begin{align}
\theta &= \Big( (\theta_A)_{\rm (ii)} - (\theta_A)_{\rm (i)} \Big)
	+ \Big( (\theta_M)_{\rm (ii)} - (\theta_M)_{\rm (i)} \Big)
\qquad{\rm (mod}~2\pi),
\cr
&= \pi + \Big( (\theta_M)_{\rm (ii)} - (\theta_M)_{\rm (i)} \Big)
\qquad{\rm (mod}~2\pi),
\end{align}
where $\rm (i), (ii)$ refer to two boundary condition choices in \eqref{eq:possilbebc}.

We note that when adding flavors, the boundary conditions corresponding to theta angle $\theta=0$ or $\theta=\pi$ give rise to the same Seiberg-Witten curve up to sign flip of masses, as expected. See appendix \ref{sec:E8}.

\bigskip
\section{Decompactification limit of 5d Seiberg-Witten curves of $E_1$ and $\widetilde{E}_1$}
\label{sec:5dlimit}

In the previous section, we have seen how the discrete theta angle of the 5d pure $Sp(1)$ gauge theory is realized  
in terms of the Seiberg-Witten curves, which are the configurations of M5-branes under the existence of an OM5-plane.
In this section, we will see how the difference with respect to the discrete theta angle appears in the original $(p, q)$ 5-brane web with an O5-plane in type IIB string theory. 

\subsection{Strategy to recover 5-brane web from 5d Seiberg-Witten curve}

As discussed in \cite{Aharony:1997bh}, it is possible to reproduce $(p, q)$ 5-brane webs from the corresponding Seiberg-Witten curves. Indeed the two configurations are related by string dualities and a decompactification limit of the Seiberg-Witten curve reproduces a $(p, q)$ 5-brane web of the original 5d theory.
We expect that this property remains true even under the existence of an O5-plane and is applicable to the 5d Seiberg-Witten curves of the $E_1$ and $\widetilde{E}_1$ theories.

In order to take such a decompactification limit, we assume that the complex coordinates of the Seiberg-Witten curves of \eqref{eq:E1curve} and \eqref{eq:naivesp1curve} as well as their coefficients depend on the compactification radius $R$ in the following exponential form\footnote{
How to determine the radius $R$ dependence for the Coulomb moduli parameter $U$ is non-trivial. For example, we could have considered another choice \cite{Nekrasov:1996cz} as 
$$U = e^{ - R (u + i \phi)} + e^{ + R (u + i \phi)} \propto \cosh R (u + i \phi),$$
which makes invariant under the $Sp(1)$ Weyl transformation $u+i\phi \to -u-i\phi$. 
This choice may be even more natural because it naturally reproduce the classical Coulomb moduli parameter in the weak coupling limit $q \to 0$.
If we use this parametrization and take the decompactification limit $R \to \infty$,
$U$ is the same as that in \eqref{radiusdependence} in the region $u<0$ while the region $u>0$ is reproduced from the result in the region $u<0$. Since we will see later that the region $u>0$ is unphysical, the result is not sensitive to such difference of the parametrization.}
\begin{align}
& t = e^{ - R (x_4+ix_{11})}, \qquad w = e^{-  R (x_6+ix_5)},
\qquad
q = e^{ - R (m_0 + i \theta)}, \qquad U = e^{ - R (u + i \phi)}. \label{radiusdependence} 
\end{align}
Here $m_0 \propto 1/g_{\rm YM}^2$. The decompactification limit corresponds to taking $R \to \infty$ with the parameterization \eqref{radiusdependence}. Note that the complex coordinates $t, w$ in \eqref{radiusdependence} of the 5d Seiberg-Witten curves \eqref{eq:naivesp1curve}, \eqref{eq:E1curve} parameterize the 5-brane web diagram with an O5-plane. Therefore, such a limit should lead to the original 5-brane web with an O5-plane instead of the 5-brane web diagram without an O5-plane which also gives rise to the 5d $Sp(1)$ gauge theory\footnote{In fact, the decompactification is not unique. For example, the 5d Seiberg-Witten curve of the 5d pure $Sp(1)$ gauge theory with or without the discrete theta angle can have two types of the decompactification limit. One limit leads to a 5-brane web without an O5-plane. The other limit gives rise to a 5-brane web with an O5-plane. We consider the latter decompactification limit since we are interested in $(p, q)$ 5-branes with an O5-plane.}.

The periodicity condition is imposed on the imaginary part of the exponent and we can set their fundamental region to be $0 \leq x_{11}, x_5, \theta, \phi \leq 2 \pi / R$.
Supposing that we always use the values at their fundamental region, the values of the coordinates $x_{11}$, $x_5$, $\theta$ and $\phi$ are of the order $\mathcal{O} (R^{-1})$. Therefore, it is natural to rescale them by $R$ and redefine
 \begin{align}\label{eq:twqu}
& t = e^{ - R x_4 - i x_{11}' }, \qquad w = e^{ - R x_6 - ix_5' },
\qquad
q = e^{ - R m_0 - i \theta'}, \qquad U = e^{ - R u - i \phi'}. 
\end{align}
where  $x_{11}'$, $x_5'$, $\theta'$ and $\phi'$ are now of the order $\mathcal{O}(1)$.

When we use the parameterization \eqref{eq:twqu}, the 5d Seiberg-Witten curves \eqref{eq:E1curve} and \eqref{eq:naivesp1curve} of the $E_1$ and $\widetilde{E}_1$ theories can be written in the form
\begin{align}\label{eq:RAB}
\sum_{k, l, m} \exp( R A_{k, l}^{(m)} + i B_{k, l}^{(m)}) = 0,
\end{align}
where $A_{k, l}^{(m)} = -k x_4 - lx_6 + c_1^{(m)} m_0 + c_2^{(m)} u$ where $c_1^{(m)}, c_2^{(m)}$ are some integers, while $B_{k, l}^{(m)} = -kx_5' - lx_{11}' + c_1^{(m)} \theta' + c_2^{(m)} \phi'$. The label $m$ is relevant when there are multiple cases of $c_1^{(m)}, c_2^{(m)}$ for the same $k, l$. When there is a single case, we omit the label $m$ for simplicity of the notation. 

Then the $(p, q)$ 5-brane webs of the $E_1$ and $\widetilde{E}_1$ theories in the presence of an O5-plane should be reproduced by taking the limit $R \to \infty$ with $A_{k. l}^{(m)}, B_{k, l}^{(m)}$ fixed for \eqref{eq:RAB}. When we take the limit $R \to \infty$ while $A_{k, l}^{(m)}$, $B_{k, l}^{(m)}$ fixed, 
the terms in \eqref{eq:RAB} with the largest $A_{k,l}^{(m)}$ become dominant.
For example, if $A_{k_1, l_1}^{(m_1)} > A_{k_i, l_i}^{(m_i)}$ with $k_1 \neq k_i, l_1 \neq l_i$ and $m_1 \neq m_i$, we see that
\begin{align}
\exp( R A_{k_1, l_1}^{(m_1)})  \gg \sum_{k_i \neq k_1, l_i \neq l_1, m_i\neq m_1} \exp( R A_{k_i, l_i}^{(m_i)}),
\end{align}
in the limit $R \to \infty$. In this case, it is obvious that \eqref{eq:RAB} cannot be satisfied. In order words, there is no 5-brane inside the region where $A_{k, l}^{(m)}$ is the largest in the $(x_4, x_6)$-space. 
Note that such region depends on the values of the parameters $m_0$ and $u$. 
Let us then consider a case where $A_{k_1, l_1}^{(m_1)}$ is the largest in a region in the $(x_4, x_6)$-space and $A_{k_2, l_2}^{(m_2)}$ is the largest in an adjacent region. Since $A_{k, l}^{(m)}$'s are continuous parameters, along the boundary of the two regions we have 
\begin{align}
A_{k_1, l_1}^{(m_1)} = A_{k_2, l_2}^{(m_2)} \quad (> A_{k_i, l_i}^{(m_i)}) \qquad (k_i \neq i_1, i_2,\; l_i \neq l_1, l_2, \; m_i \neq m_1, m_2)
\label{eq:A12}
\end{align}
In this case, the decompactification limit of the Seiberg-Witten curve (\ref{eq:RAB}) reduces to the simple equation (\ref{eq:A12}) in the $R \to \infty$ limit if we drop the information of the phase $B_{k,l}^{(m)}$.
The equation \eqref{eq:A12} gives a line segment in the $(x_4,x_6)$ space, giving a part of a $(p, q)$ 5-brane web. 

Therefore we can reconstruct the 5-brane web diagram from the following strategy. First we choose some region in the $(m_0, u)$-space. Then we idetnfiy a region in the $(x_4, x_6)$-space where a particular $A_{k, l}^{(m)}$ is the largest. Depending on the regions in the $(x_4, x_6)$-space, which $A_{k, l}^{(m)}$ becomes the largest is different and we consider all the possible cases. Then along the boundaries of the regions, there is a linear condition between $A_{k, l}^{(m)}$'s and it yields a 5-brane web diagram. For example, the boundary of the two regions gives a line corresponding to a 5-brane whereas the boundary of the three regions gives a point corresponding to a vertex of a 5-brane web.

In the following, we explicitly calculate the decompactification limit of the 5d Seiberg-Witten curves of the $E_1$ and $\widetilde{E}_1$ theories for all the possible regions of 
the variables and coordinaters. We will first fix a parameter region of $m_0$, $u$, and then consider the regions in the $(x_4, x_6)$-space where each $A_{k, l}^{(m)}$ is the largest. 
 Since their boundaries yield a 5-brane web, it is enough to determine the regions where some $A_{k, l}^{(m)}$ becomes the largest. We will see that this calculation indeed reproduces the original $(p, q)$ 5-brane webs in the weak coupling region while generating a new diagram in the strong coupling region.

\subsection{Phase structure of the $E_1$ theory}
\label{sec:E1Phase}

We first start from the Seiberg-Witten curve \eqref{eq:E1curve} of the $E_1$ theory. 
Assuming the parametrization \eqref{eq:twqu}, the 5d Seiberg-Witten curve \eqref{eq:E1curve} is written in the form \eqref{eq:RAB} with
\begin{align}\label{eq:E1As}
&A_{2,0} = - 2 x_4, \quad
A_{1,3} = - (x_4 + 3 x_6 - m_0), \quad
A_{1,2} = - (x_4 + 2 x_6 - m_0 + u), \quad
\cr
&A_{1,1}^{(1)} = -(x_4 + x_6 - m_0), \quad
A_{1,1}^{(2)} = -(x_4 + x_6 ), \quad
A_{1,0} = -(x_4 - m_0 + u), \quad
\cr
& A_{1,-1}^{(1)} = -( x_4 - x_6 - m_0), \quad
A_{1,-1}^{(2)} = -( x_4 - x_6 ), 
\cr
&A_{1, -2} = -(x_4 - 2 x_6 - m_0 + u), \quad
A_{1, -3} = -( x_4 - 3 x_6 - m_0), \quad
A_{0,0} = 0
\end{align}
\begin{figure}
\centering
\includegraphics[width=9cm]{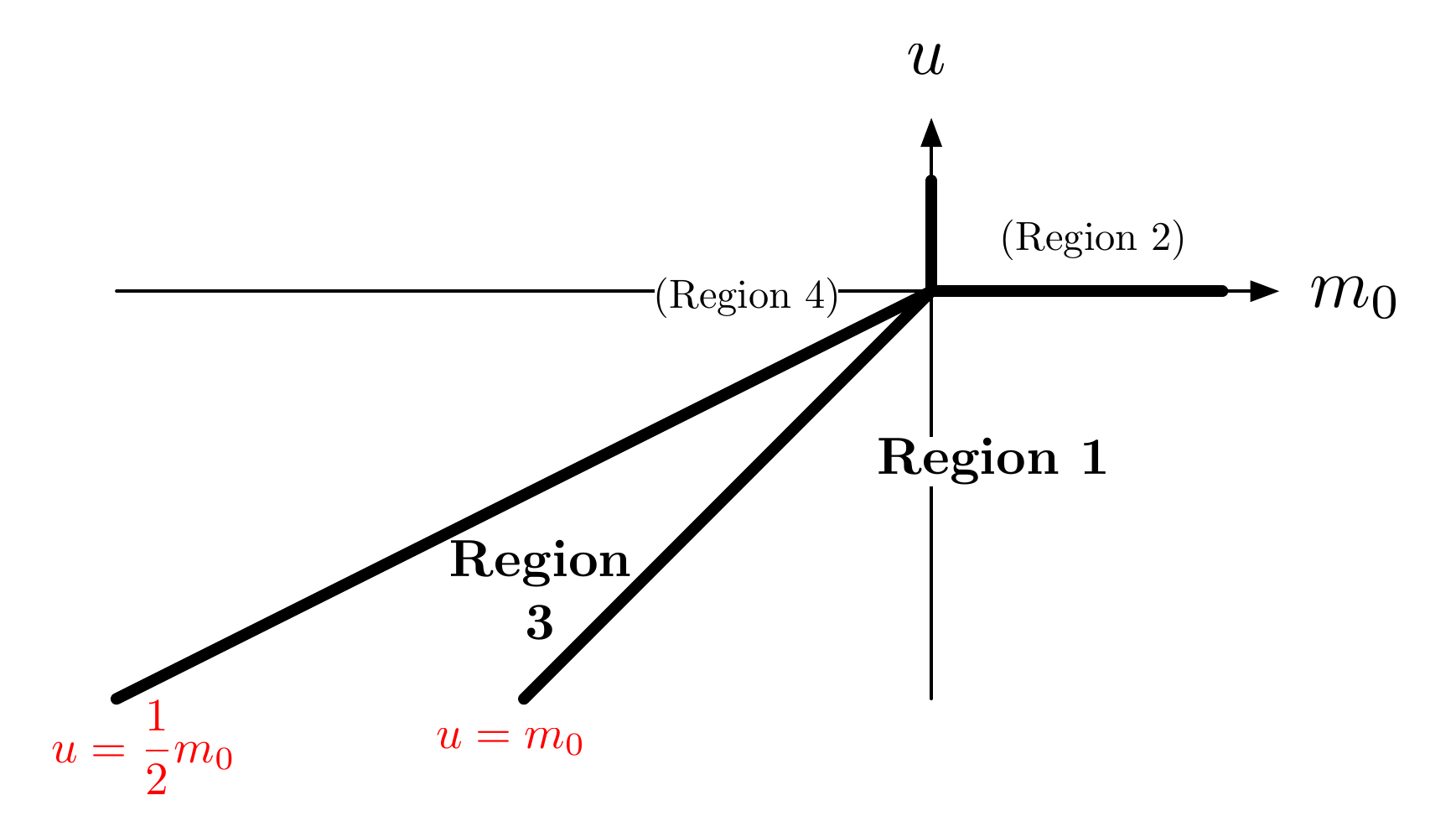}
\caption{The phase diagram of the $E_1$ theory. 
The line between the Region 1 and the Region 3 is $u=m_0$. The boundary of the physical parameter space is given by $u=\frac{1}{2}m_0$ for $m_0 \leq 0$ and $u=0$ for $m_0 \geq 0$.}
\label{Fig:E1Phase}
\end{figure}
We then consider all possible regions in the parameter space $(u, m_0)$ and determine the $(p, q)$ 5-brane web in the regions:
\begin{align}\label{eq:E1region1to4}
\text{Region 1: } &~\{\,u<0, \; m_0 > 0\,\} ~\cup ~ \{\,u < m_0 < 0\,\},
\cr
\text{Region 2: } &~  m_0 < u,\quad u > 0,
\cr
\text{Region 3: } &~ m_0 < u < \frac12 m_0 ~(< 0),
\cr
\text{Region 4: } &~ \frac12 m_0 < u , \quad m_0<0.
\end{align}
The corresponding region, which will turn out to give the ``phase diagram'' of the $E_1$ theory,
is depicted in Figure \ref{Fig:E1Phase}.

\paragraph{Region 1.}
The first region we consider is the following parameter region\footnote{Although the classical gauge coupling becomes negative in some region, the effective gauge coupling is always positive.}
\begin{align}\label{eq:E1region1}
\{\,u<0, \; m_0 > 0\,\} ~\cup ~ \{\,u < m_0 < 0\,\}.
\end{align}
In this parameter space, a particular $A_{k, l}^{(m)}$ in \eqref{eq:E1As} becomes the largest depending on a subspace in the $(x_4, x_6)$-space.
For example, the region where $A_{2,0}$ is the largest is obtained by solving $A_{2,0} > A_{i,j}^{(m)}$ for $\forall i,j,m$. However, we find that the independent conditions are only the following four, $A_{2,0} > A_{1,3}, A_{1,2}, A_{1,-2}, A_{1,-3}$ under the assumption \eqref{eq:E1region1}. All the others can be derived from these four conditions. Rewriting these four conditions explicitly in terms of $x_4$ and $x_6$, we obtain
\begin{align}\label{eq:E1region1-1}
&x_6 > \frac{x_4 + m_0}{3}, \quad
x_6 > \frac{x_4 + m_0 - u}{2}, \quad
x_6 < \frac{-x_4 - m_0 + u}{2}, \quad
x_6 < \frac{-x_4 - m_0}{3}.
\end{align}
In this way, we find the region where $A_{2,0}$ is the largest in the $(x_4, x_6)$-space.  We can compute for other $A_{k, l}^{(m)}$ in analogous way and the result is summarized in Table \ref{Table:E1Region1}. 
\begin{table}
\centering
\begin{tabular}{c|c|c}
$A_{k,l}^{(m)}$ & Independent conditions & Region for $A_{k,l}^{(m)}$ being the largest
\cr
\hline
$A_{2,0}$ & 
$\begin{array}{l}
A_{2,0} > A_{1,3}, A_{1,2}, \cr
\qquad \qquad A_{1,-2}, A_{1,-3}
\end{array}$ & 
$\begin{array}{l}
x_6 > \frac{x_4 + m_0}{3}, \,\,
x_6 > \frac{x_4 + m_0 - u}{2}, \cr
x_6 < \frac{-x_4 - m_0 + u}{2}, \,\, 
x_6 < \frac{-x_4 - m_0}{3}
\end{array}$
\cr
\hline
$A_{1,3}$ & $A_{1,3} > A_{2,0}, A_{1,2}, A_{0,0}$ & 
$x_6 < \frac{x_4 + m_0}{3}$, \,\,
$x_6 < u$, \,\,
$x_6 < \frac{-x_4 + m_0}{3}$
\cr
\hline
$A_{1,2}$ & 
$\begin{array}{l}
A_{1,2} > A_{2,0}, A_{1,3}, \cr
\qquad\qquad A_{1,-2}, A_{0,0}
\end{array}$ & 
$\begin{array}{l}
x_6 > \frac{x_4 + m_0 - u}{2}, \,\,
x_6 > u, \cr
x_6 < 0, \,\,
x_6 < \frac{-x_4 + m_0 + u}{2}
\end{array}$
\cr
\hline
$A_{1,-2}$ & 
$\begin{array}{l}
A_{1,-2} > A_{2,0}, A_{1,2}, \cr
\qquad \qquad A_{1,-3}, A_{0, 0}
\end{array}$ &
$\begin{array}{l}
x_6 > \frac{-x_4 - m_0 + u}{2}, \,\,
x_6 > 0, \cr
x_6 < -u, \,\, 
x_6 < \frac{x_4 - m_0 + u}{2}
\end{array}$
\cr
\hline
$A_{1,-3}$ & $A_{1,-3} > A_{2,0}, A_{1,-2}, A_{0,0}$ &
$x_6 < \frac{-x_4 - m_0}{3}$, \,\,
$x_6 < -u$, \,\,
$x_6 < \frac{x_4 - m_0}{3}$
\cr
\hline
$A_{0,0}$ & 
$\begin{array}{l}
A_{0,0} > A_{1,3}, A_{1,2}, \cr
\qquad\qquad A_{1,-2}, A_{1,-3}
\end{array}$ &
$\begin{array}{l}
x_6 > \frac{-x_4 + m_0}{3}, \,\,
x_6 > \frac{-x_4 + m_0 - u}{2}, \cr
x_6 < \frac{x_4 -m_0 + u}{2}, \,\, 
x_6 < \frac{x_4 - m_0}{3}
\end{array}$
\cr
\hline
$\begin{array}{c}
A_{1,1}^{(m)}, A_{1,0}, \cr
A_{1,-1}^{(m)}
\end{array}$ &
n/a & No region 
\cr \hline
\end{tabular}
\caption{The regions where some $A_{k,l}^{(m)}$ becomes larger than any other $A_{k',l'}^{(m')}$'s for the Region 1 \eqref{eq:E1region1} of the $E_1$ theory. The second column gives independent relations for ensuring that $A_{k, l}^{(m)}$ in the first column becomes the largest. The last column indicates a region in the $(x_4, x_6)$-space where the $A_{k,l}^{(m)}$ in the first column becomes the largest. $m$ in this table is either $1$ or $2$. }
\label{Table:E1Region1}
\end{table}

\begin{figure}
\centering
\subfigure[]{
\includegraphics[width=10cm]{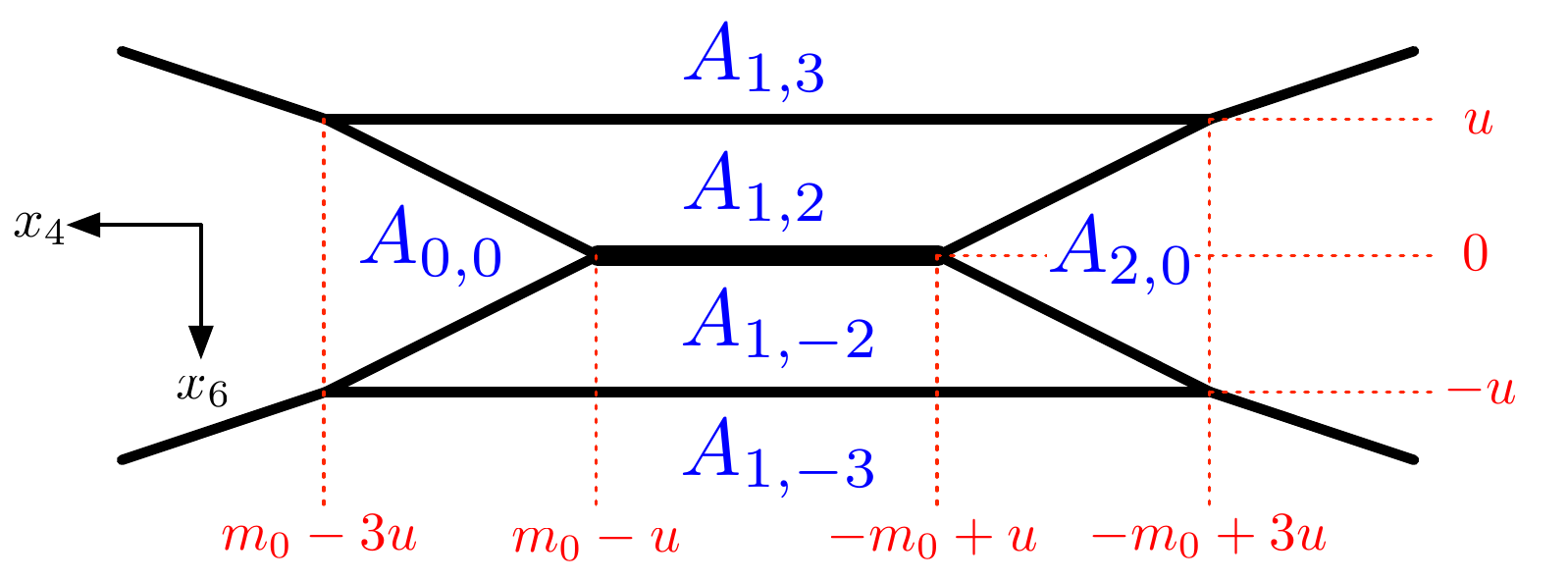}}
\subfigure[]{
\includegraphics[width=2.4cm]{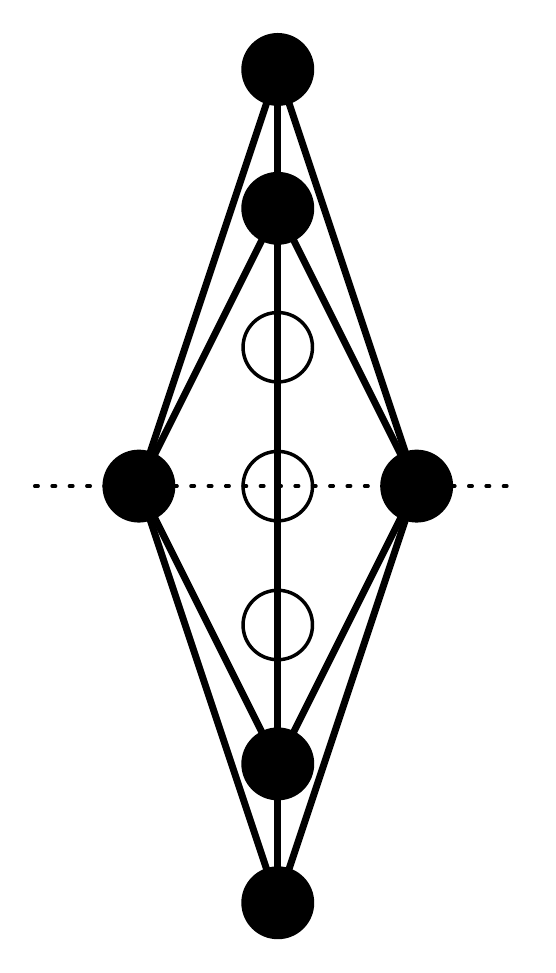}}
\caption{(a): The blue letter $A_{k, l}^{(m)}$ indicates the largest $A_{k, l}^{(m)}$ in the region. The black solid lines are the boundaries of those regions and give the 5-brane web in the decompactification limit of the $E_1$ curve in Region 1 \eqref{eq:E1region1}. The thick line denotes the coincident 5-branes. (b): The toric-like diagram of the $E_1$ theory with a triangulation given by the web in Figure (a). }
\label{Fig:E1Region1}
\end{figure}

From Table \ref{Table:E1Region1}, one needs to look at boundaries between two regions to recover a 5-brane web. For example, the boundary between the region where $A_{2,0}$ is the largest and the region where $A_{1,3}$ is the largest is characterized by $A_{2,0} = A_{1,3}$, which is rewritten as
\begin{align}\label{eq:E1region1p}
&x_6 = \frac{x_4 + m_0}{3}.
\end{align}
This means that the 5d Seiberg-Witten curve reduces to this line in some subspace of $(x_4, x_6)$-space,
which corresponds to a partial 5-brane configuration. One can repeat the same analysis for other possible largest $A^{(m)}_{k,l}$ cases and 
combine all the corresponding partial configurations together to make a complete 5-brane web digram that is consistent with  
all the boundaries of the regions listed in Table \ref{Table:E1Region1}. The result is depicted in Figure \ref{Fig:E1Region1} (a). It is then easy to see that the region labeled $A_{2,0}$ on the right side of the Figure \ref{Fig:E1Region1} (a) corresponds \eqref{eq:E1region1-1} and the line segment \eqref{eq:E1region1p} is the boundary between the regions labeled $A_{2,0}$ and $A_{1,3}$.
Unlike the case for the usual toric diagram, some of the $A_{k, l}^{(m)}$ does not become largest in any region in $(x_4,x_6)$ space.
In fact, it turns out that $A_{1,1}^{(1)}, A_{1,1}^{(2)}, A_{1,0}, A_{1,-1}^{(1)}$, and $A_{1,-1}^{(2)}$ never become the largest in the case of \eqref{eq:E1region1}.

It is also instructive to write down the corresponding toric-like diagram (also known as dot diagram) \cite{Benini:2009gi, Kim:2014nqa} as in Figure \ref{Fig:E1Region1} (b). 
From the construction of the 5d Seiberg-Witten curve from the toric-like diagram, 
each monomial in the Seiberg-Witten curve corresponds to either a black dot or a white dot.
In other word, $A_{k, l}^{(m)}$ given in \eqref{eq:E1As} with different $k,l$ corresponds to different dot in the toric-like diagram.
We assign a black dot to $A_{k, l}^{(m)}$ which becomes largest at some region in $(x_4,x_6)$ space while white dot corresponds to $A_{k, l}^{(m)}$ which does not become largest in any region in $(x_4,x_6)$ space.
Therefore, each face of the 5-brane web corresponds to a black dot.
For example, the monomial associated to the top black dot gives a term related to $A_{1,3}$, hence when the $A_{1,3}$ is the largest compared to the other $A_{k, l}^{(m)}$, we have a face at the top in the 5-brane web. 
On the other hand, the white dots correspond to ``shrunken face'', which indicates coincident 5-branes.
For example, there is no region where $A_{1,1}^{(m)}, A_{1,0}$, or $A_{1,-1}^{(m)}$ become the largest and therefore, they correspond to white dots.
The existence of the three ``shrunken face'' indicates that there are four coincident branes at $x_6=0$. 
In Figure \ref{Fig:E1Region1} (a) and also in the following figures, thick lines denote such coincident 5-branes.
Here, note that $x_6=0$ is the location of the O5-plane and we also include the mirror image in the web diagram. 
The four coincident D5-branes including the mirror images together with an O5$^-$-plane should be interpreted as an O5$^+$-plane, giving consistent picture with Figure \ref{Fig:sp1toriclike}.
Also, if the region where $A_{k,l}^{(m)}$ is the largest 
and the region where $A_{k',l'}^{(m')}$ is the largest share their boundaries,   
we connect the corresponding two black dots in the toric-like diagram.
For example, the dot corresponding to $A_{1,2}$ and $A_{1,-2}$ should be connected by the line in this case.
This rule determines the ``triangulation''\footnote{For generic toric-like diagram, the smallest unit is not necessarily triangle.} of the toric-like diagram.

\begin{figure}
\centering
\includegraphics[width=10cm]{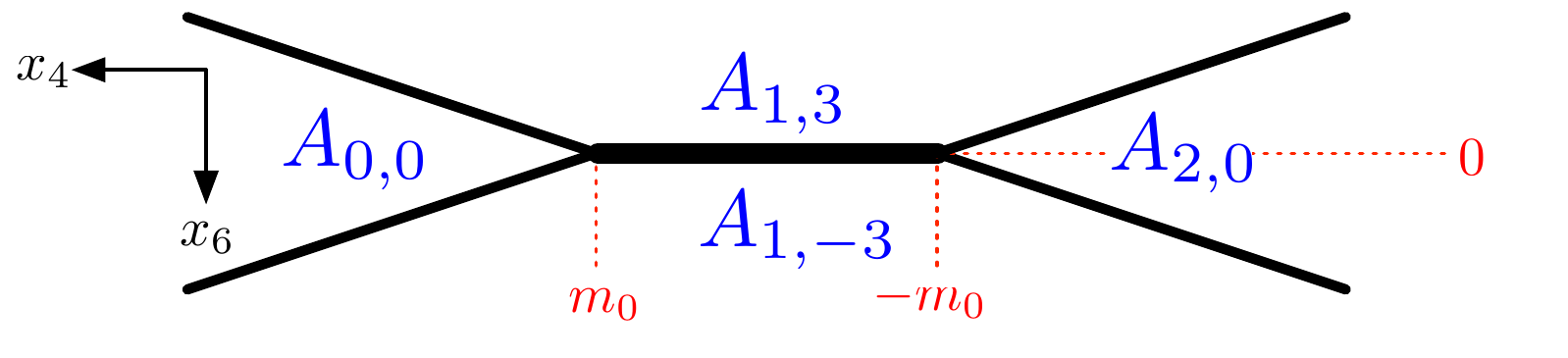}
\caption{The blue letter $A_{k, l}^{(m)}$ indicates the largest $A_{k, l}^{(m)}$ in the region. The black solid lines are the boundaries of those regions and give the 5-brane web in the decompactification limit of the $E_1$ curve in Region 2 \eqref{eq:E1region2}.}
\label{Fig:E1Region2}
\end{figure}

\paragraph{Region 2.}
Next, we consider the region
\begin{align}\label{eq:E1region2}
u>0, \quad m_0 > 0.
\end{align}
In this parameter space, we determine regions in $(x_4, x_6)$ in which a particular $A_{k, l}^{(m)}$ becomes the largest. The result is given in Table \ref{Table:E1Region2}. 
\begin{table}
\centering
\begin{tabular}{c|c|c}
$A_{k,l}^{(m)}$ & Independent conditions & Region for $A_{k,l}^{(m)}$ being the largest
\cr
\hline
$A_{2,0}$ & $A_{2,0} > A_{1,3}, A_{1,-3}$ & 
$x_6 > \frac{x_4 + m_0}{3}$, \,\,
$x_6 < \frac{-x_4 - m_0}{3}$
\cr \hline
$A_{1,3}$ & $A_{1,3} > A_{2,0}, A_{1,-3}, A_{0, 0}$ & 
$x_6 < \frac{x_4 + m_0}{3}$, \,\,
$x_6 < 0$, \,\,
$x_6 < \frac{-x_4 + m_0}{3}$
\cr \hline
$A_{1,-3}$ & $A_{1,-3} > A_{2,0}, A_{1,3}, A_{0,0}$ &
$x_6 > \frac{-x_4 - m_0}{3}$, \,\,
$x_6 > 0$, \,\,
$x_6 > \frac{x_4 - m_0}{3}$
\cr \hline
$A_{0,0}$ & $A_{0,0} > A_{1,3}, A_{1,-3}$ &
$x_6 > \frac{-x_4 + m_0}{3}$, \,\,
$x_6 < \frac{x_4 - m_0}{3}$
\cr \hline
$
\begin{array}{c}
A_{1,2}, A_{1,1}^{(m)}, A_{1,0}
\cr
A_{1,-1}^{(m)}, A_{1,-2}
\end{array}
$ 
& n/a & No region
\cr \hline
\end{tabular}
\caption{The regions where some $A_{k,l}^{(m)}$ becomes larger than any other $A_{k',l'}^{(m')}$'s for Region 2 \eqref{eq:E1region2} of the $E_1$ theory. The second column gives independent relations for ensuring that $A_{k, l}^{(m)}$ in the first column becomes the largest. The last column indicates a region in the $(x_4, x_6)$-space where the $A_{k,l}^{(m)}$ in the first column becomes the largest. $m$ in this table is either $1$ or $2$. }
\label{Table:E1Region2}
\end{table}
The 5-brane segments in this region \eqref{eq:E1region2} can be realized from the boundaries between the regoins in Table \ref{Table:E1Region2}. The web diagram is given in Figure \ref{Fig:E1Region2}.

Remarkably, the result does not depend on the Coulomb branch moduli parameter $u$.
Even if we move the value of $u$ in the region $u>0$, the web diagram does not change at all.
Furthermore, this result in Table \ref{Table:E1Region2} can be reproduced from Table \ref{Table:E1Region1} by substituting $u=0$, where some of the regions in Table \ref{Table:E1Region1} disappear. 
Correspondingly, the web diagram in Figure \ref{Fig:E1Region2} 
can be also reproduced from Figure \ref{Fig:E1Region1} by considering $u=0$.

The fact that the web diagram does not depend on $u$ in the Region 2 \eqref{eq:E1region2} and the result can be reproduced by setting $u=0$ for the regoins in Table \ref{Table:E1Region1} implies that this region is merely the boundary $u=0$ of the Region 1 \eqref{eq:E1region1} and the region with $u > 0, m_0 > 0$ does not exist. Indeed there is no phase transition from the web diagram in Figure \ref{Fig:E1Region1} to the web diagram in Figure \ref{Fig:E1Region2} but the web in Figure \ref{Fig:E1Region2} is simply a special case of the web in Figure \ref{Fig:E1Region1}. As argued in \cite{Jefferson:2017ahm}, we can interpret that the Region 2 \eqref{eq:E1region2} corresponds to an ``unphysical region''. Along the boundary, $a_D = \partial F / \partial a$ vanishes and hence monopole strings become tensionless, implying that the effective description breaks down. Hence we cannot go over this boundary.

\begin{figure}
\centering
\subfigure[]{\includegraphics[width=10.1cm]{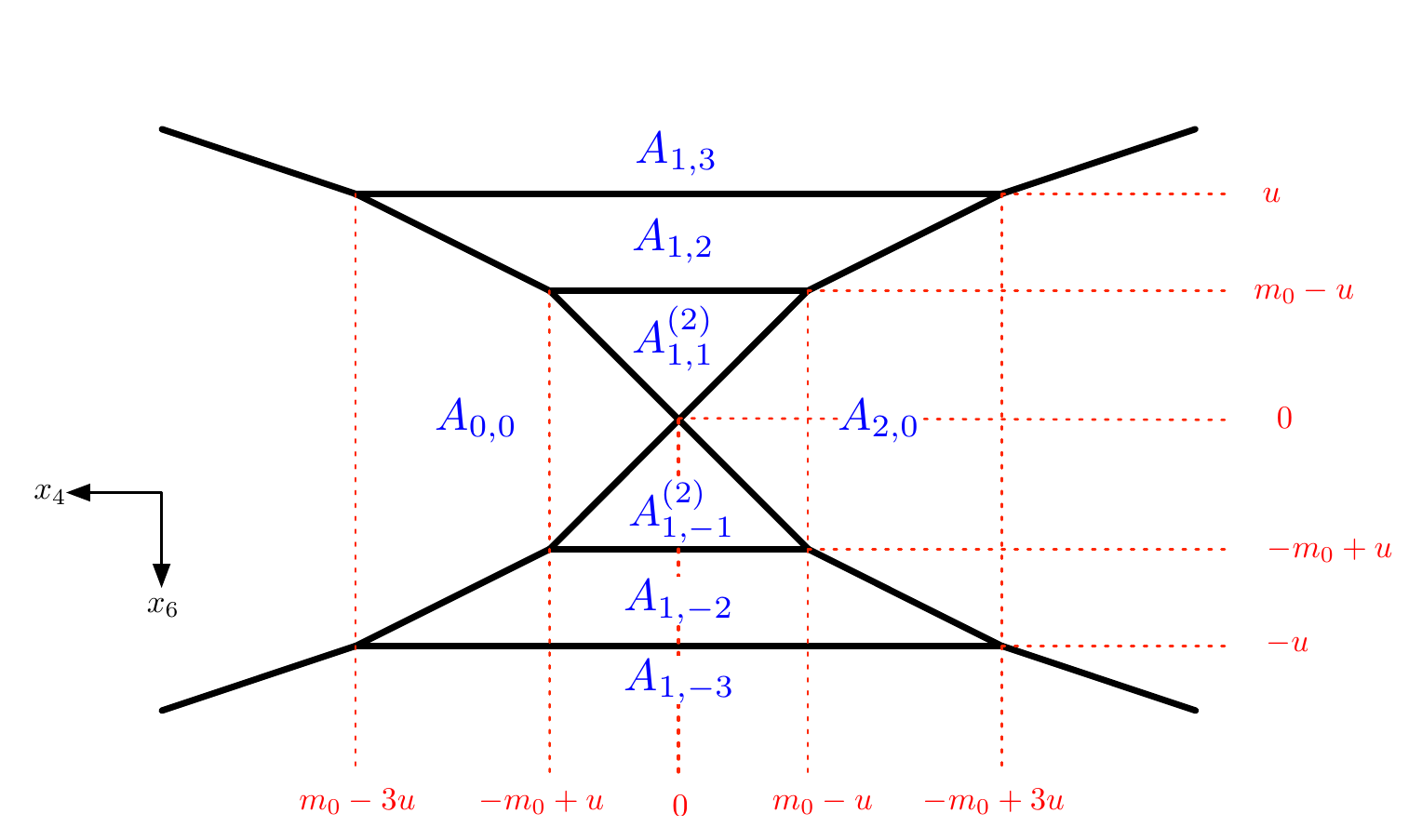}}
\subfigure[]{\includegraphics[width=3cm]{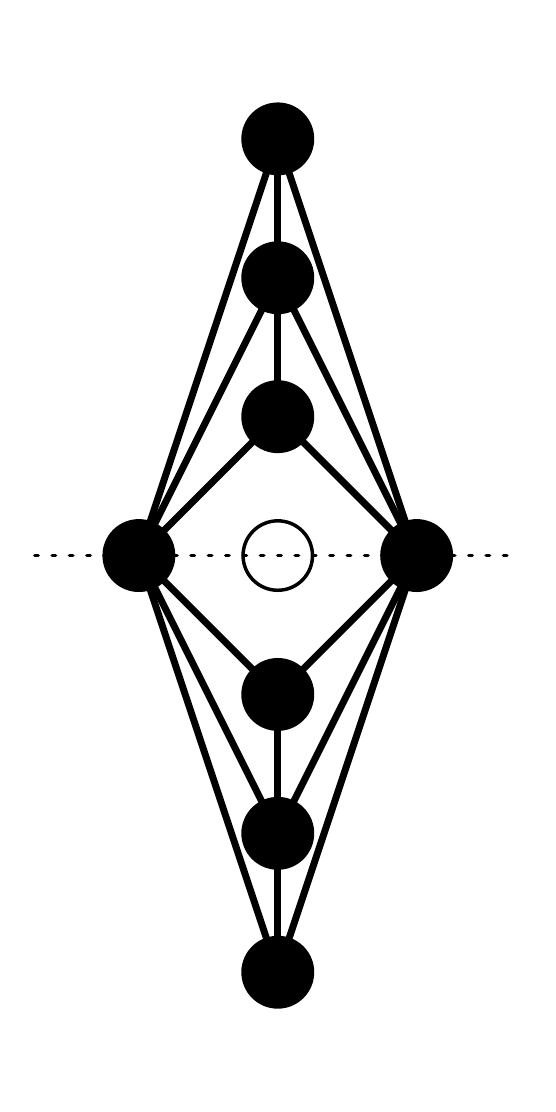}}
\caption{(a): The blue letter $A_{k, l}^{(m)}$ indicates the largest $A_{k, l}^{(m)}$ in the region. The black solid lines are the boundaries of those regions and give the 5-brane web in the decompactification limit of the $E_1$ curve in Region 3 \eqref{eq:E1region3}.  (b): The toric-like diagram of the $E_1$ theory with a triangulation given by the web in Figure (a). }
\label{Fig:E1Region3}
\end{figure}

\paragraph{Region 3.}
We then move on to the third region given by
\begin{align}\label{eq:E1region3}
m_0 < u < \frac{1}{2}m_0 \,\, ( <0 ).
\end{align}
In this parameter region, the regions in the $(x_4, x_6)$-space where some $A_{k, l}^{(m)}$ is the largest are summarized in Table \ref{Table:E1Region3}.  
\begin{table}
\centering
\begin{tabular}{c|c|c}
$A_{k, l}^{(m)}$ & Independent conditions & Region for $A_{k, l}^{(m)}$ being the largest
\cr
\hline
$A_{2,0}$ & 
$\begin{array}{l}
A_{2,0} > A_{1,3}, A_{1,2}, A_{1,1}^{(2)} \cr
\qquad \qquad A_{1,-1}^{(2)}, A_{1,-2}, A_{1,3}
\end{array}$ 
& 
$\begin{array}{l}
x_6 > \frac{x_4 + m_0}{3}, \,\,
x_6 > \frac{x_4 + m_0 - u}{2}, \,\, 
x_6 > x_4, 
\cr
x_6 < -x_4, \,\,
x_6 < \frac{-x_4 - m_0 +u}{2}, \,\,
x_6 < \frac{-x_4 - m_0}{3}
\end{array}$
\cr
\hline
$A_{1,3}$ & $A_{1,3} > A_{2,0}, A_{1,2}, A_{0,0}$ & 
$x_6 > \frac{x_4 + m_0}{3}$, \,\,
$x_6 > u $, \,\,
$x_6 > \frac{-x_4 + m_0}{3}$
\cr
\hline
$A_{1,2}$ & 
$\begin{array}{l}
A_{1,2} > A_{2,0}, A_{1,3}, \cr
\qquad \qquad A_{1,1}^{(2)}, A_{0,0}
\end{array}$ 
& 
$\begin{array}{l}
x_6 < \frac{x_4 + m_0 - u}{2}, \,\,
x_6 > u, 
\cr
x_6 < m_0 - u, \,\,
x_6 < \frac{-x_4 + m_0  - u}{2}
\end{array}$
\cr \hline
$A_{1,1}^{(2)}$ & $A_{1,1}^{(2)}>A_{2,0}, A_{1,2}, A_{0,0}$ & 
$x_6 < x_4$, \,\,
$x_6 > m_0 -u$, \,\,
$x_6 < -x_4$
\cr \hline
$A_{1,-1}^{(2)}$ & $A_{1,-1}^{(2)}>A_{2,0}, A_{1,-2}, A_{0,0} $ & 
$x_6 > -x_4$, \,\,
$x_6 < -m_0 + u$, \,\,
$x_6 > x_4$
\cr \hline
$A_{1,-2}$ & 
$\begin{array}{l}
A_{1,-2} > A_{2,0}, A_{1,-1}^{(2)}, \cr
\qquad \qquad A_{1,-3}, A_{0,0}
\end{array}$ & 
$\begin{array}{l}
x_6 > \frac{-x_4 - m_0 + u}{2}, \,\,
x_6 > -m_0 + u, \cr
x_6 < -u, \,\,
x_6 < \frac{x_4 - m_0  + u}{2}
\end{array}$
\cr \hline
$A_{1,-3}$ & $A_{1,-3} > A_{2,0}, A_{1,-2}, A_{0,0}$ & 
$x_6 > \frac{-x_4 - m_0}{3}$, \,\,
$x_6 > -u $, \,\,
$x_6 > \frac{x_4 - m_0}{3}$
\cr \hline
$A_{0,0}$ & 
$\begin{array}{l}
A_{0,0} > A_{1,3}, A_{1,2}, A_{1,1}^{(2)} \cr
\qquad \qquad A_{1,-1}^{(2)}, A_{1,-2}, A_{1,-3}
\end{array}$
& 
$\begin{array}{l}
x_6 > \frac{-x_4 + m_0}{3}, \,\,
x_6 > \frac{-x_4 + m_0 - u}{2}, \,\,
x_6 > -x_4, \cr
x_6 < x_4, \,\,
x_6 < \frac{x_4 - m_0 +u}{2}, \,\,
x_6 < \frac{x_4 - m_0}{3}
\end{array}$
\cr \hline
$\begin{array}{c}
A_{1,1}^{(1)}, A_{1,0}, \cr
A_{1,-1}^{(1)}
\end{array}$ & 
n/a & No region 
\cr \hline
\end{tabular}
\caption{The regions where each $A_{k,l}^{(m)}$ becomes larger than any other $A_{k',l'}^{(m')}$'s for Region 3 \eqref{eq:E1region3} of the $E_1$ theory. The second column gives independent relations for ensuring that $A_{k, l}^{(m)}$ in the first column becomes the largest. The last column indicates a region in the $(x_4, x_6)$-space where the $A_{k,l}^{(m)}$ in the first column becomes the largest.}
\label{Table:E1Region3}
\end{table}
The corresponding 5-brane web and the toric-like diagram are depicted in Figure \ref{Fig:E1Region3} (a).

When we go from the Region 1 \eqref{eq:E1region1} to the Region 3 \eqref{eq:E1region3}, the web diagram undergoes a ``phase transition''. From the web in Figure \ref{Fig:E1Region1} (a), two 5-branes intersecting with an O5-plane are separated by a finite distance $d = 2m_0 - 2u > 0$. As we approach the Region 3 \eqref{eq:E1region3}, the distance becomes smaller and smaller and it becomes zero along the boundary $m_ 0 = u$. Then in the Region 3 \eqref{eq:E1region3}, the naive distance becomes negative but the analysis in Table \ref{Table:E1Region3} indicates that a 5-brane web diagram exists in this region and it is given by one in Figure \ref{Fig:E1Region3}. The diagram in Figure \ref{Fig:E1Region3} (a) shows a particular pattern for a strong coupling behavior of the web diagram for the $E_1$ theory. In fact, we will see that the strong coupling behavior of the web diagram is quite different between the web diagram of the $E_1$ theory and the web diagram of the $\widetilde{E}_1$ theory. In terms of the toric-like diagrams in Figure \ref{Fig:E1Region1} (b) and in Figure \ref{Fig:E1Region3} (b), the different strong coupling behavior is associated to different triangulations of the toric-like diagram\footnote{Note that some of the white dots in Figure \ref{Fig:E1Region1} (b) turn to black dots in Figure \ref{Fig:E1Region3} (b). In general, whether the dots in the toric-like diagram are black or white depend on its triangulation as mentioned in \cite{Kim:2014nqa}. Such black dots do not increase the dimension of Coulomb moduli space, in this case.}. 
This may be interpreted as a generalization of a flop transition in the usual toric diagram which contains only black dots. 

\begin{figure}
\centering
\includegraphics[width=8cm]{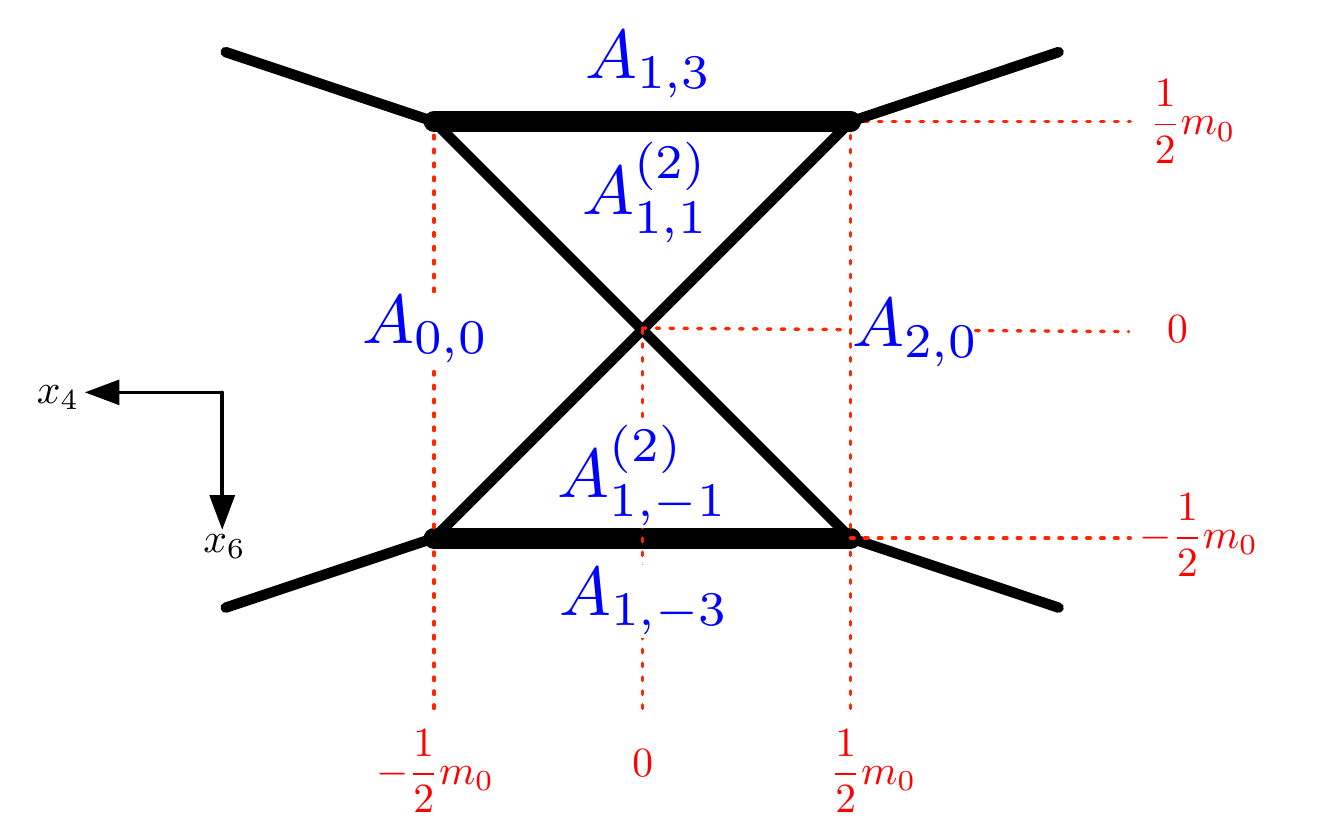}
\caption{The blue letter $A_{k, l}^{(m)}$ indicates the largest $A_{k, l}^{(m)}$ in the region. The black solid lines are the boundaries of those regions and give the 5-brane web in the decompactification limit of the $E_1$ curve in Region 4 \eqref{eq:E1region4}. }
\label{Fig:E1Region4}
\end{figure}

\paragraph{Region 4.}
The final region for the parameter space of the $E_1$ curve is given by
\begin{align}\label{eq:E1region4}
u > \frac{1}{2}m_0, \quad m_0 < 0.
\end{align}
The relations between the regoins in $(x_4, x_6)$-space and the largest $A_{k,l}^{(m)}$ are shown in Table \ref{Table:E1Region4}. 
\begin{table}
\centering
\begin{tabular}{c|c|c}
$A_i$ & Independent conditions & Region for $A_i$ being the largest
\cr
\hline
$A_{2,0}$ & 
$\begin{array}{l}
A_{2,0} > A_{1,3}, A_{1,1}^{(2)}, \cr
\qquad \qquad A_{1,-1}^{(2)}, A_{1,-3}
\end{array}$ & 
$\begin{array}{l}
x_6 > \frac{x_4 + m_0 }{3}, \,\,
x_6 > x_4, \cr
x_6 < -x_4, \,\,
x_6 < \frac{-x_4 - m_0}{3}
\end{array}$
\cr \hline
$A_{1,3}$ & $A_{1,3} > A_{2,0}, A_{1,1}^{(2)}, A_{0,0}$ & 
$x_6 < \frac{x_4 + m_0 }{3}$, \,\,
$x_6 < \frac{m_0}{2}$, \,\,
$x_6 < \frac{-x_4 + m_0}{3}$
\cr \hline
$A_{1,1}^{(2)}$ & $A_{1,1}^{(2)}> A_{2,0}, A_{1,3}, A_{0,0}$ & 
$x_6 < x_4$, \,\,
$x_6 > \frac{m_0}{2}$, \,\,
$x_6 < -x_4$
\cr \hline
$A_{1,-1}^{(2)}$ & $A_{1,-1}^{(2)}>A_{2,0}, A_{1,-3}, A_{0,0} $ & 
$x_6 > -x_4$, \,\,
$x_6 < \frac{-m_0}{2}$, \,\,
$x_6 > x_4$
\cr \hline
$A_{1,-3}$ & $A_{1,-3} > A_{2,0}, A_{1,-1}^{(2)}, A_{0,0}$ & 
$x_6 > \frac{-x_4 -m_0}{3}$, \,\,
$x_6 > \frac{-m_0}{2}$, \,\,
$x_6 > \frac{x_4 - m_0}{3}$
\cr \hline
$A_{0,0}$ & 
$\begin{array}{l}
A_{0,0} > A_{1,3}, A_{1,1}^{(2)}, \cr
\qquad \qquad A_{1,-1}^{(2)}, A_{1,-3}
\end{array}$ & 
$\begin{array}{l}
x_6 > \frac{-x_4 + m_0 }{3}, \,\,
x_6 > -x_4, \cr
x_6 < x_4, \,\,
x_6 < \frac{x_4 - m_0}{3}
\end{array}$
\cr \hline
$\begin{array}{c}
A_{1,-2}, A_{1,-1}^{(1)}, \cr
A_{1,0}, A_{1,2}, A_{1,1}^{(1)}
\end{array}$
& n/a & No region 
\cr \hline
\end{tabular}
\caption{The regions where each $A_{k,l}^{(m)}$ becomes larger than any other $A_{k',l'}^{(m')}$'s for the Region 4 \eqref{eq:E1region4} of the $E_1$ theory. The second column gives independent relations for ensuring that $A_{k, l}^{(m)}$ in the first column becomes the largest. The last column indicates a region in the $(x_4, x_6)$-space where the $A_{k,l}^{(m)}$ in the first column becomes the largest.}
\label{Table:E1Region4}
\end{table}
The boundaries of the regions in Table \ref{Table:E1Region4} give rise to a 5-brane web given in Figure \ref{Fig:E1Region4}

Note that in this case also like the region \eqref{eq:E1region3}, the shape of the 5-brane web does not depend on the Coulomb branch parameter $u$. Furthermore the results in Table \ref{Table:E1Region4} can be reproduced by simply setting $u = \frac{1}{2}m_0$ for the results in Table \ref{Table:E1Region3}. Therefore the web diagram in Figure \ref{Fig:E1Region4} is merely a special case $u = \frac{1}{2}m_0$ of the diagram in Figure \ref{Fig:E1Region3} (1) and the shape does not change further when we move $u$. This implies that the Region 4 \eqref{eq:E1region4} is an unphysical region. 

\paragraph{Phase diagram.} By combining the results of the Region 1 \eqref{eq:E1region1}, Region 2 \eqref{eq:E1region2}, Region 3 \eqref{eq:E1region3} and Region 4 \eqref{eq:E1region4}, the phase structure of the $E_1$ theory in the $(m_0, u)$-space is given in Figure \ref{Fig:E1Phase}. 
Note that the Region 2 \eqref{eq:E1region2} and Region 4 \eqref{eq:E1region4} are unphysical regions.

\paragraph{Effective coupling.}
After identifying the 5-branes web diagrams of the $E_1$ theory in each parameter region, it is also possible to compute the effective coupling of the $E_1$ theory. In order to determine the effective coupling, we first consider the tension of monopole strings. The monopole string can be realized by a D3-brane stretched over a face of a 5-brane web. Therefore, the tension is related to the area $a_D$ of the face. Then the effective coupling can be given by taking a derivative of $a_D$ with respect to a Couomb branch modulus. 

In the case of the $E_1$ theory, the $a_D$ is given by the area of the face where $A_{1, 2}$ becomes the largest. For the Region 1, the area is given by
\be
a_{D, 1} =  -u(2m_0 - 4u).
\ee 
Since $u < 0$, we consider taking a derivative with respect to $-u$ in order to obtain a positive effective coupling and the result becomes
\be\label{eq:effE1Region1}
\tau_{\text{eff},1} = \frac{\partial a_D}{\partial (-u)} = 2m_0 - 8u.
\ee
This expression looks different from the standard expression of the $E_1$ theory for example in \cite{Seiberg:1996bd, Morrison:1996xf}, but the difference can be absorbed by the redefinition of $m_0, u$.

Let us then move on to the Region 3. The area where $A_{1, 2}$ becomes the largest is 
\be
a_{D, 3} = -2u(m_0-2u).
\ee
Hence the effective coupling in this region is 
$
\tau_{\text{eff},3} = \frac{\partial a_D}{\partial (-u)} = 2m_0 - 8u.
$
Namely the effective coupling in the Region 3 is the same as that \eqref{eq:effE1Region1} in the Region 1. 
In summary, the effective coupling constant for the $E_1$ theory is given by 
\be\label{eq:E1taueff}
\tau_{\text{eff}} = 2m_0 - 8u,
\ee
everywhere inside the physical Coulomb moduli parameter region.

\subsection{Phase structure of the $\widetilde{E}_1$ theory}

\begin{figure}
\centering
\includegraphics[width=10cm]{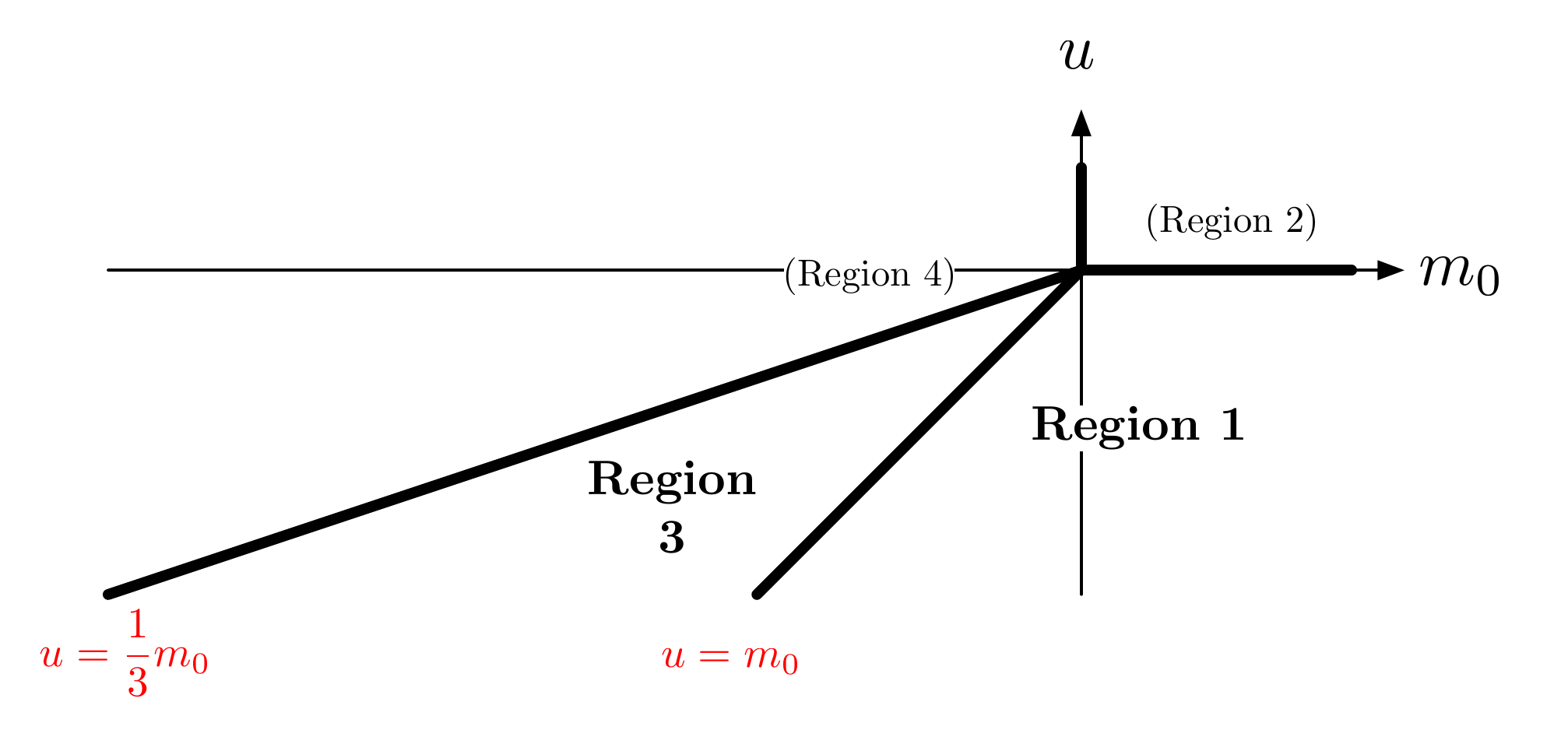}
\caption{The phase diagram of the $\widetilde{E}_1$ theory. 
The line between Region 1 and Region 3 is $u=m_0$. The boundary of the physical parameter region is given by $u=\frac{1}{3}m_0$ for $m_0 \leq 0$ and $u=0$ for $m_0 \geq 0$.}
\label{Fig:E1tildePhase}
\end{figure}

We next consider the 5d Seiberg-Witten curve \eqref{eq:naivesp1curve} of the $\widetilde{E}_1$ theory. 
When we use the parametrization \eqref{eq:twqu}, the 5d Seiberg-Witten curve \eqref{eq:naivesp1curve} can be written in the form \eqref{eq:RAB} with
\begin{align}\label{eq:E1tildeAs}
&A_{2,0} = - 2 x_4, \quad
A_{1,3} = - (x_4 + 3 x_6 - m_0), \quad
A_{1,2} = - (x_4 + 2 x_6 - m_0 + u), \quad
\cr
&A_{1,1} = -(x_4 + x_6-m_0), \quad
A_{1,0}^{(1)} = -(x_4 -m_0 + u), \quad
A_{1,0}^{(2)} = -x_4, 
\cr
&A_{1,-1} = -(x_4 - x_6 - m_0 ), \quad
A_{1,-2} = -(x_4 - 2 x_6  -m_0 + u), 
\cr
&A_{1,-3} = -(x_4  - 3 x_6 - m_0), \quad
A_{0,0} = 0.
\end{align}
We determine the $(p, q)$ 5-brane web in each region by taking the decompactification limit in the all the regions in the $(m_0, u)$-space:
\begin{align}\label{eq:E1tilderegion1to4}
\text{Region 1: } &~ \{\,u < m_0 < 0\,\}~ \cup~  \{\,u < 0, \;m_0>0\,\}\quad
\cr
\text{Region 2: } &~  m_0 > 0,\quad u > 0,
\cr
\text{Region 3: } &~ m_0 < u <\frac13 m_0 ~~(<0),
\cr
\text{Region 4: } &~ m_0 < 0, \quad u > \frac13 m_0.
\end{align}
The corresponding phase diagram is depicted in Figure  \ref{Fig:E1tildePhase}.

\paragraph{Region 1.}
The first region we consider is the following,
\begin{align}\label{eq:E1tilderegion1}
\{\,u<0, \; m_0 > 0\,\}~ \cup~ \{\,u < m_0 < 0\,\}.
\end{align}
The strategy to obtain the 5-brane web is again the same as we did in section \ref{sec:E1Phase}. We determined the regions where a particular $A_{k,l}^{(m)}$ becomes the largest in the $(x_4, x_6)$-space under the condition \eqref{eq:E1tilderegion1} for the Region 1. In fact, in the case of \eqref{eq:E1tilderegion1} it turns out that the regions where a $A_{k,l}^{(m)}$ becomes the largest are exactly the same ones in Table \ref{Table:E1Region1} and we obtain the same web diagram in Figure \ref{Fig:E1Region1} (a). Note that the decompactification limit yields the same web diagram in Figure \ref{Fig:E1Region1} although the original Seiberg-Witten curve \eqref{eq:naivesp1curve} of the $\widetilde{E}_1$ theory is different from the Seiberg-Witten curve \eqref{eq:E1curve} of the $E_1$ theory in this case.

\paragraph{Region 2.}
The second region is 
\begin{align}\label{eq:E1tilderegion2}
u>0, \quad m_0 > 0.
\end{align}
In this case also, the decompactification limit in fact leads to the same web diagram as the one in Figure \ref{Fig:E1Region2}. Again the shape does not depend on the parameter $u$ and the web diagram can be thought of as a special case of the Region 1 \eqref{eq:E1tilderegion1} characterinzed by $u=0, m_0 > 0$. Hence this region \eqref{eq:E1tilderegion2} corresponds to an unphysical region. 

\paragraph{Region 3.}
The third region we consider is 
\begin{align}\label{eq:E1tilderegion3}
m_0 < u < \frac{1}{3}m_0 \,\, ( <0 ).
\end{align}
In this parameter space \eqref{eq:E1tilderegion3}, a particular $A_{k,l}^{(m)}$ in \eqref{eq:E1tildeAs} becomes the largest in some region in the $(x_4, x_6)$-space. The results are summarized in Table \ref{Table:E1tildeRegion3}.
\begin{table}
\centering
\begin{tabular}{c|c|c}
$A_{k, l}^{(m)}$ & Independent conditions & Region for $A_{k, l}^{(m)}$ being the largest
\cr
\hline
$A_{2,0}$ & 
$\begin{array}{l}
A_{2,0} > A_{1,3}, A_{1,2}, A_{1,-2},  
\cr
\qquad \qquad A_{1,-3}, A_{0,0}
\end{array}$
& 
$\begin{array}{c}
x_6 > \frac{x_4 + m_0 }{3}, \,\,
x_6 > \frac{x_4 + m_0 - u}{2}, \,\,
x_6 > \frac{-x_4 - m_0 + u}{2}, \cr
x_6 < \frac{-x_4 - m_0}{3}, \,\,
x_4 < 0 
\end{array}$
\cr
\hline
$A_{1,3}$ & $A_{1,3} > A_{2,0}, A_{1,2}, A_{0,0}$ & 
$x_6 < \frac{x_4 + m_0 }{3}$, \,\,
$x_6 < u$, \,\,
$x_6 < \frac{-x_4 + m_0}{3}$
\cr
\hline
$A_{1,2}$ & $A_{1,2} > A_{2,0}, A_{1,3}, A_{0,0}$ & 
$x_6 < \frac{x_4 + m_0 - u }{2}$, \,\,
$x_6 >  u$, \,\,
$x_6 < \frac{-x_4 + m_0 - u}{2}$
\cr \hline
$A_{1,-2}$ & $A_{1,-2} > A_{2,0}, A_{1,-3}, A_{0,0}$ & 
$x_6 > \frac{-x_4 - m_0 + u }{2}$, \,\,
$x_6 <  -u$, \,\,
$x_6 > \frac{x_4 - m_0 + u}{2}$
\cr \hline
$A_{1,-3}$ & $A_{1,-3} > A_{2,0}, A_{1,-2}, A_{0,0}$ & 
$x_6 > \frac{-x_4 - m_0 }{3}$, \,\,
$x_6 > -u$, \,\,
$x_6 > \frac{x_4 - m_0}{3}$
\cr \hline
$A_{0, 0}$ & 
$\begin{array}{l}
A_{0, 0} > A_{2,0}, A_{1,3},A_{1,2}, \cr
\qquad \qquad A_{1,-2}, A_{1,-3}
\end{array}$
& 
$\begin{array}{c}
x_4 > 0, \,\,
x_6 > \frac{-x_4 + m_0}{3}, \,\,
x_6 > \frac{-x_4 + m_0 - u}{2}, \cr
x_6 < \frac{x_4 - m_0 + u}{2}, \,\,
x_6 < \frac{x_4 - m_0}{3}
\end{array}$
\cr \hline
$\begin{array}{l}
A_{1,1}, A_{1,0}^{(1)},  \cr
A_{1,0}^{(2)}, A_{1,-1}
\end{array}$
& n/a & No region 
\cr\hline
\end{tabular}
\caption{The regions where each $A_{k,l}^{(m)}$ becomes larger than any other $A_{k',l'}^{(m')}$'s for Region 3 \eqref{eq:E1tilderegion3} of the $\widetilde{E}_1$ theory. The second column gives independent relations for ensuring that $A_{k, l}^{(m)}$ in the first column becomes the largest. The last column indicates a region in the $(x_4, x_6)$-space where the $A_{k,l}^{(m)}$ in the first column becomes the largest.}
\label{Table:E1tildeRegion3}
\end{table} 
\begin{figure}
\centering
\subfigure[]{
\includegraphics[width=10cm]{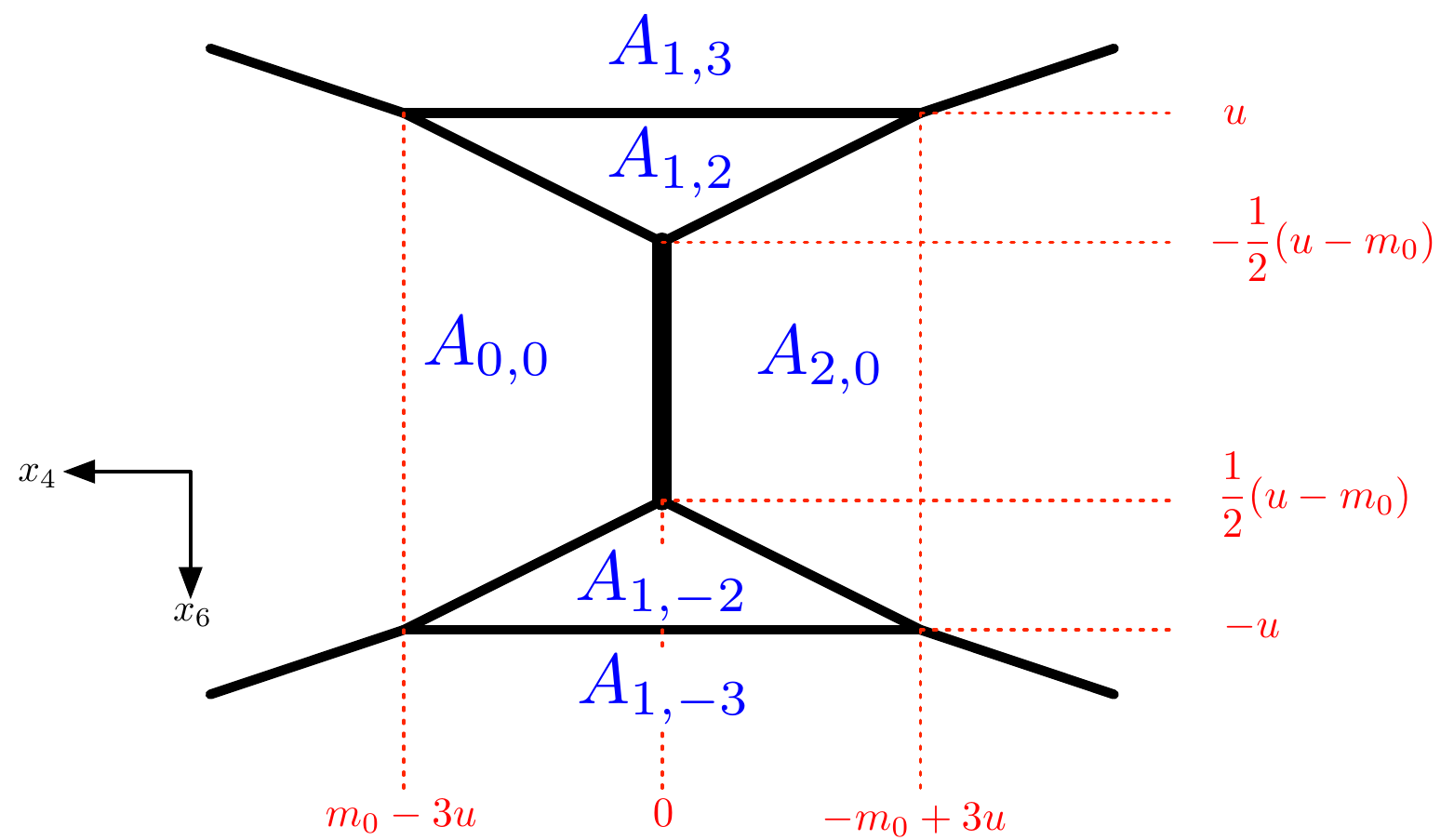}}
\subfigure[]{\includegraphics[width=3.2cm]{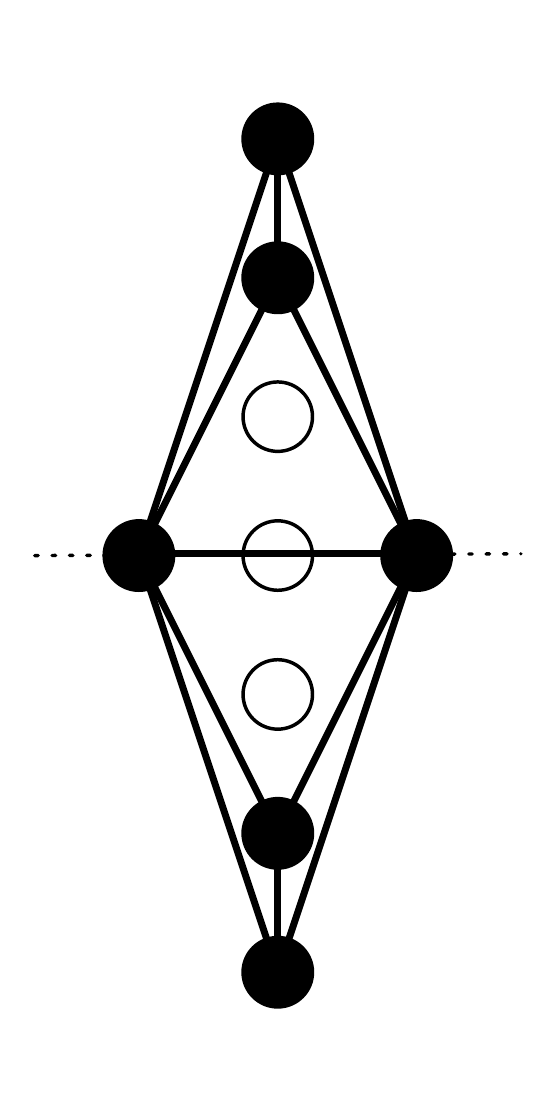}}
\caption{(a): The blue letter $A_{k, l}^{(m)}$ indicates the largest $A_{k, l}^{(m)}$ in the region. The black solid lines are the boundaries of those regions and give the 5-brane web in the decompactification limit of the $\widetilde{E}_1$ curve in Region 3 \eqref{eq:E1tilderegion3}.  (b): The toric-like diagram of the $\widetilde{E}_1$ theory with a triangulation given by the web in Figure (a). }
\label{Fig:E1tildeRegion3}
\end{figure}
The boundaries of the regions in Table \ref{Table:E1tildeRegion3} give rise to the 5-brane web in the Region 3 \eqref{eq:E1tilderegion3} and it is depicted in Figure \ref{Fig:E1tildeRegion3}. 

From the web diagram in Figure \ref{Fig:E1Region1} to the diagram in Figure \ref{Fig:E1tildeRegion3}, the $\widetilde{E}_1$ theory undergoes a phase transition as the distance $d = 2m_0 - 2u$ becomes negative in the Region 3 \eqref{eq:E1tilderegion3}. However the 5-brane web diagram in Figure \ref{Fig:E1tildeRegion3} of the $\widetilde{E}_1$ theory shows a different characteristic compared to the 5-brane web diagram in Figure \ref{Fig:E1Region3} of the $E_1$ theory after the phase transition. The two NS5-branes come out of the O5-plane in Figure \ref{Fig:E1tildeRegion3} whereas a $(1, -1)$ 5-brane and a $(1, 1)$ 5-brane come out of the O5-plane in Figure \ref{Fig:E1Region3}. Hence the 5-brane diagrams of the $E_1$ theory and the $\widetilde{E}_1$ theory exhibit different structure in the strong coupling region although they appear to be the same in the weak coupling region as in Figure \ref{Fig:E1Region1}.

\begin{figure}
\centering
\includegraphics[width=8cm]{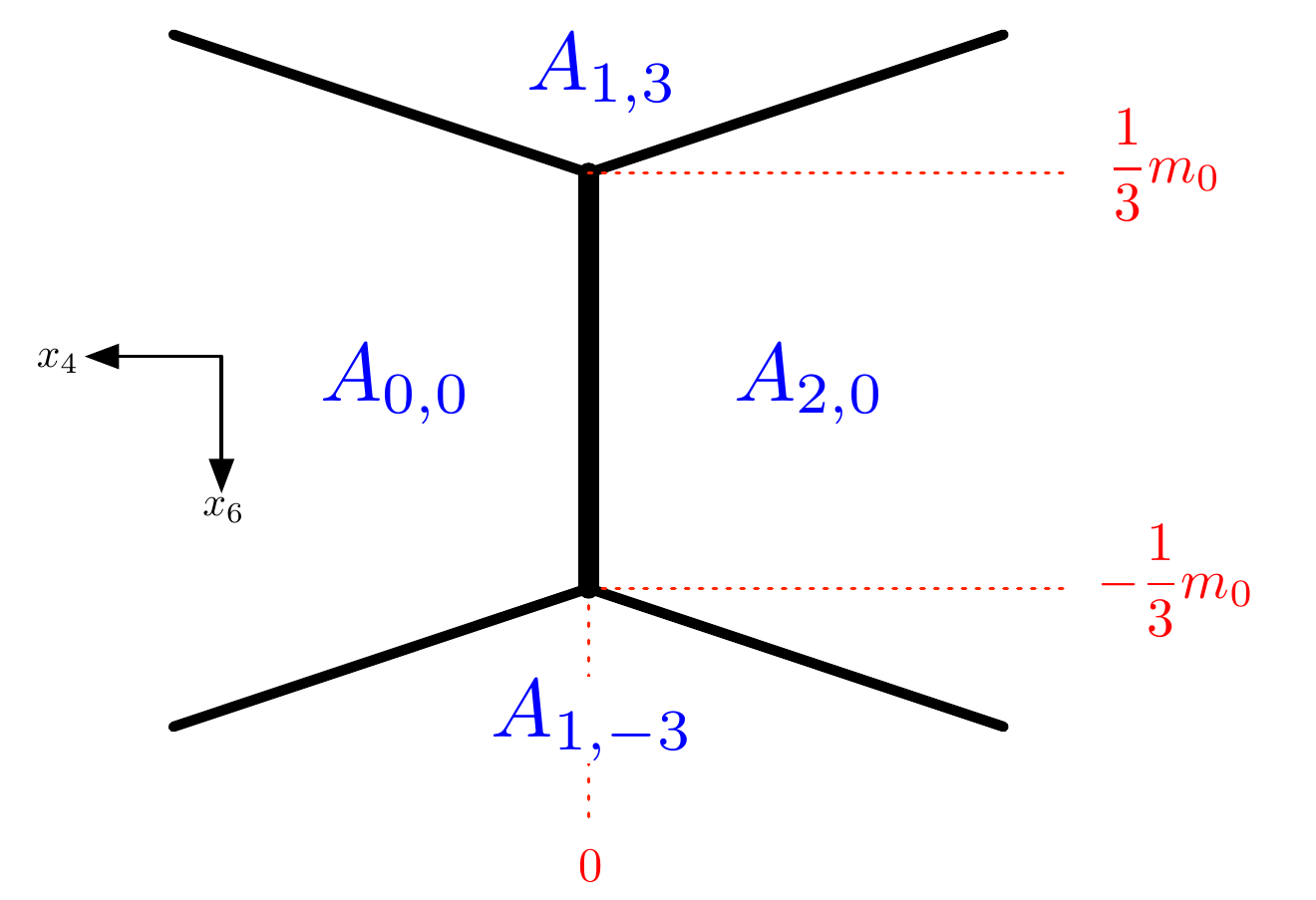}
\caption{The blue letter $A_{k, l}^{(m)}$ indicates the largest $A_{k, l}^{(m)}$ in the region. The black solid lines are the boundaries of those regions and give the 5-brane web in the decompactification limit of the $\widetilde{E}_1$ curve in Region 4 \eqref{eq:E1tilderegion4}. }
\label{Fig:E1tildeRegion4}
\end{figure}

\paragraph{Region 4.}
The final region we consider is 
\begin{align}\label{eq:E1tilderegion4}
m_0 < 0, \,\, u > \frac{1}{3}m_0. 
\end{align}
Each $A_{k,l}^{(m)}$ in \eqref{eq:E1tilderegion4} becomes the largest in the subregion in the $(x_4, x_6)$-space as in Table \ref{Table:E1tildeRegion4}.
\begin{table}
\centering
\begin{tabular}{c|c|c}
$A_{k, l}^{(m)}$ & Independent conditions & Region for $A_{k, l}^{(m)}$ being the largest
\cr
\hline
$A_{2,0}$ & $A_{2,0} > A_{1,3}, A_{1,-3}, A_{0,0}$ & 
$x_6 > \frac{x_4 + m_0 }{3}$, \,\,
$x_4 < 0$, \,\,
$x_6 < \frac{-x_4 - m_0 }{3}$
\cr
\hline
$A_{1,3}$ & $A_{1,3} > A_{2,0}, A_{0,0}$ & 
$x_6 < \frac{x_4 + m_0 }{3}$, \,\,
$x_6 < \frac{-x_4 + m_0}{3}$
\cr
\hline
$A_{1,-3}$ & $A_{1,-3} > A_{2,0}, A_{0,0}$ & 
$x_6 > \frac{-x_4 - m_0 }{3}$, \,\,
$x_6 > \frac{x_4 - m_0}{3}$
\cr \hline
$A_{0, 0}$ & $A_{0, 0} > A_{2,0}, A_{1,3}, A_{1,-3}$ & 
$x_4 > 0$, \,\,
$x_6 > \frac{-x_4 + m_0 }{3}$, \,\,
$x_6 < \frac{x_4 - m_0 }{3}$
\cr \hline
$\begin{array}{c}
A_{1,2}, A_{1,1}, A_{1,0}^{(1)} \cr
A_{1,0}^{(2)}, A_{1,-1}, A_{1,-2}
\end{array}$ & 
n/a & No region
\cr \hline
\end{tabular}
\caption{The regions where each $A_{k,l}^{(m)}$ becomes larger than any other $A_{k',l'}^{(m')}$'s for Region 4 \eqref{eq:E1tilderegion4} of the $\widetilde{E}_1$ theory. The second column gives independent relations for ensuring that $A_{k, l}^{(m)}$ in the first column becomes the largest. The last column indicates a region in the $(x_4, x_6)$-space where the $A_{k,l}^{(m)}$ in the first column becomes the largest.}
\label{Table:E1tildeRegion4}
\end{table} 
The boundaries of the regions in Table \ref{Table:E1tildeRegion4} again yields the 5-brane web given in Figure \ref{Fig:E1tildeRegion4}.
In this region, the 5-brane web diagram does not depend on the parameter $u$ and the result in Table \ref{Table:E1tildeRegion4} can be obtained by substituting $u=\frac{m_0}{3}$ to the result of Region 3 in Table \ref{Table:E1tildeRegion3}. Hence the Region 4 \eqref{eq:E1tilderegion4} corresponds to an unphysical region. 

\paragraph{Phase diagram.}
By summarizing the results in the Region 1 \eqref{eq:E1tilderegion1}, Region 2 \eqref{eq:E1tilderegion2}, Region 3 \eqref{eq:E1tilderegion3} and Region 4 \eqref{eq:E1tilderegion4}, we can construct a phase diagram for the $\widetilde{E}_1$ theory. A phase transition occurs when we go from the Region 1 \eqref{eq:E1tilderegion1} to the Region 3 \eqref{eq:E1tilderegion3}. The other two regions are unphysical regions. The corresponding phase diagram is depicted in Figure \ref{Fig:E1tildePhase}.

\paragraph{Effective coupling.}

As we obtain the 5-brane web diagram in each region of the parameter space of the $\widetilde{E}_1$
 theory, it is possible to compute the effective coupling in the each region. 
 
 For the Region 1, the 5-brane web diagram is the same as the one for the Region 1 of the $E_1$ theory and hence the effective coupling is $2m_0 - 8u$ from \eqref{eq:effE1Region1}. On the other hand, it turns out that the effective coupling in the Region 3 of the $\widetilde{E}_1$ theory is different from $2m_0 - 8u$. The area of the face where $A_{1, 2}$ is the largest is 
 \be
 a_{D, 3} = \frac{1}{2}(m_0 - 3u)^2.
 \ee
Therefore, the effective coupling is 
\be\label{eq:effE1tildeRegion3}
\tau_{\text{eff},3} = \frac{\partial a_{D}}{\partial (-u)} = 3m_0 - 9u,
\ee
which is different from \eqref{eq:effE1Region1}. 
Combining the result in \eqref{eq:effE1Region1} and \eqref{eq:effE1tildeRegion3}, we can rewrite the effective coupling constant in a unified manner as%
\footnote{We would like to interpret the last term as the contribution from the massless particle with instanton charge. We thank Kazuya Yonekura for related discussion.}
\be\label{eq:E1tildetaueff}
\tau_{\text{eff}} = \frac{5}{2} m_0 - \frac{17}{2} u - \frac{1}{2} \left| u - m_0 \right|.
\ee
Now we see a sharp phase transition from Region 1 to Region 3 for the $\widetilde{E}_1$ theory due to the change of the effective coupling. 
This is consistent with the fact that the Calabi--Yau threefold for the $\widetilde{E}_1$ admit a flop transition, while the Calabi--Yau threefold for the $E_1$ theory does not admit.

\bigskip
\section{Decompactification limit of 5d Seiberg-Witten curve of $E_2$}
\label{sec:E2}

In section \ref{sec:5dlimit}, we reproduced 5-brane web diagram with an O5-plane for the $E_1$ and the $\widetilde{E}_1$ theories from the decompactification limit of the 5d Seiberg-Witten curves \eqref{eq:E1curve} and \eqref{eq:naivesp1curve}, respectively. We then determined the phase diagrams in the parameter space $m_0, u$ of the $E_1$ and the $\widetilde{E}_1$ theories which show a phase transition between the weak coupling region and the strong coupling region. In this section, we discuss the $Sp(1)$ theory with one flavor whose UV fixed point has an enhanced symmetry $E_2=SU(2)\times U(1)$. Depending on how one decouples the flavor, one can obtain either $E_1$ or $\widetilde{E}_1$ theory. We repeat the same analysis to study the phase structure of the $E_2$ theory. We then take two different flavor decoupling limits on the phase diagram of the $E_2$ theory and we show that the decoupling precisely reproduce the phase diagrams of the $E_1$ and $\widetilde{E}_1$ theories discussed in section \ref{sec:5dlimit}. We also discuss a realization of the $E_0$ theory from the point of view of the brane configuration with an O5-plane.

\subsection{Phase structure of the $E_2$ theory}
\label{sec:E2Phase}
We first obtain the 5-brane web diagram of the 
$Sp(1)$ gauge theory with one flavor from the 5d Seiberg-Witten curve and then determine its phase structure in the parameter space. The 5d Seiberg-Witten curve of the $E_2$ theory is given by \eqref{eq:sp1nf1} in appendix \ref{sec:E8}, which is obtained a successive flavor decoupling from the curve for $N_f=7$ flavors. After rescaling $t$, the Seiberg-Witten curve for $Sp(1)$ curve with $N_f=1$ flavor takes the following form
\begin{align}\label{eq:sp1nf1curve}
t^2 +  
\Big[ q^{-1} (w^3 + w^{-3}) &+ q^{-1} U (w^2 + w^{-2}) - (q^{-1} - 
M_1{}^{\frac{1}{2}}) (w+w^{-1})
\cr &- (2q^{-1} U+2M_1{}^{-\frac{1}{2}}) \Big] t 
+ (- w + M_1{}^{-1} + M_1 -  w^{-1}) = 0.
\end{align}
As the form of \eqref{eq:RAB} of the Seiberg-Witten curve, $A_{k, l}^{(m)}$'s read
\begin{align}\label{eq:E2As}
&A_{2,0} = - 2 x_4, &&
\\
& A_{1,3} = - ( x_4 + 3 x_6 - m_0 ),  &&
A_{1,2} = - ( x_4 + 2 x_6 - m_0 + u ), \,\,
\cr
&A_{1,1}^{(1)} = - ( x_4 + x_6 - m_0 ), &&
A_{1,1}^{(2)} = - ( x_4 + x_6 + \frac{1}{2} m_1 ),  \,\,
\cr
& A_{1,0}^{(1)} = - ( x_4 - m_0 + u ), &&
A_{1,0}^{(2)} = - ( x_4 - \frac{1}{2} m_1  ), 
\cr
& A_{1,-1}^{(1)} = - ( x_4 - x_6 - m_0 ), &&A_{1,-1}^{(2)}  = - ( x_4 - x_6 + \frac{1}{2} m_1 ),  \,\,
\cr
& A_{1,-2} = - ( x_4 - 2 x_6 - m_0 + u ), &&
A_{1,-3} = - ( x_4 - 3 x_6 - m_0 ),  \,\, 
\cr
& A_{0,1} = - x_6, \qquad
A_{0,0}^{(1)} = - m_1,  &&
A_{0,0}^{(2)} = - (-m_1), \qquad 
A_{0,-1} = - (-x_6),\nonumber
\end{align}
where we used the parameterization \eqref{eq:twqu} and $M_1 = e^{-Rm_1 -i\psi'}$. 
The $E_2$ theory and its vacua are parametrized by the threee parameters $m_0, m_1$ and $u$. 
Following the same procedure as done in section \ref{sec:5dlimit}, determining a 5-brane diagram for the $E_2$ theory is straightforward.
Namely, we divide 
the parameter region into subregions and then determine which $A_{k, l}^{(m)}$ in \eqref{eq:E2As} becomes the largest in each region, as the largest $A_{k, l}^{(m)}$ varies depending on the region in the $(x_4, x_6)$-space.  The boundaries of the regions then yield the corresponding 5-brane web in the presence of an O5-plane. 

\begin{figure}
\centering
\subfigure[$m_1 \leq 0$]{
\includegraphics[width=12cm]{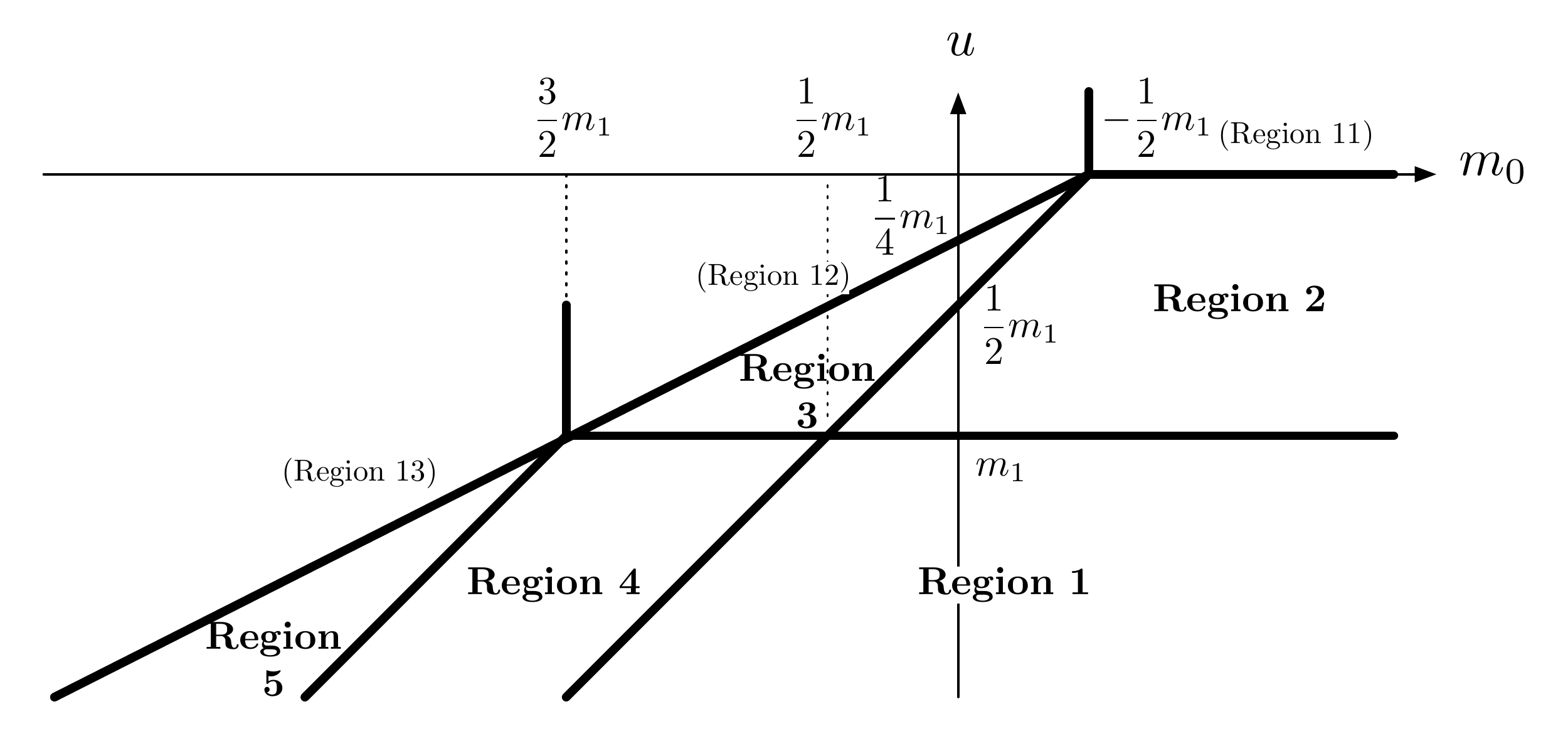}}
\subfigure[$m_1 \geq 0$]{
\includegraphics[width=14cm]{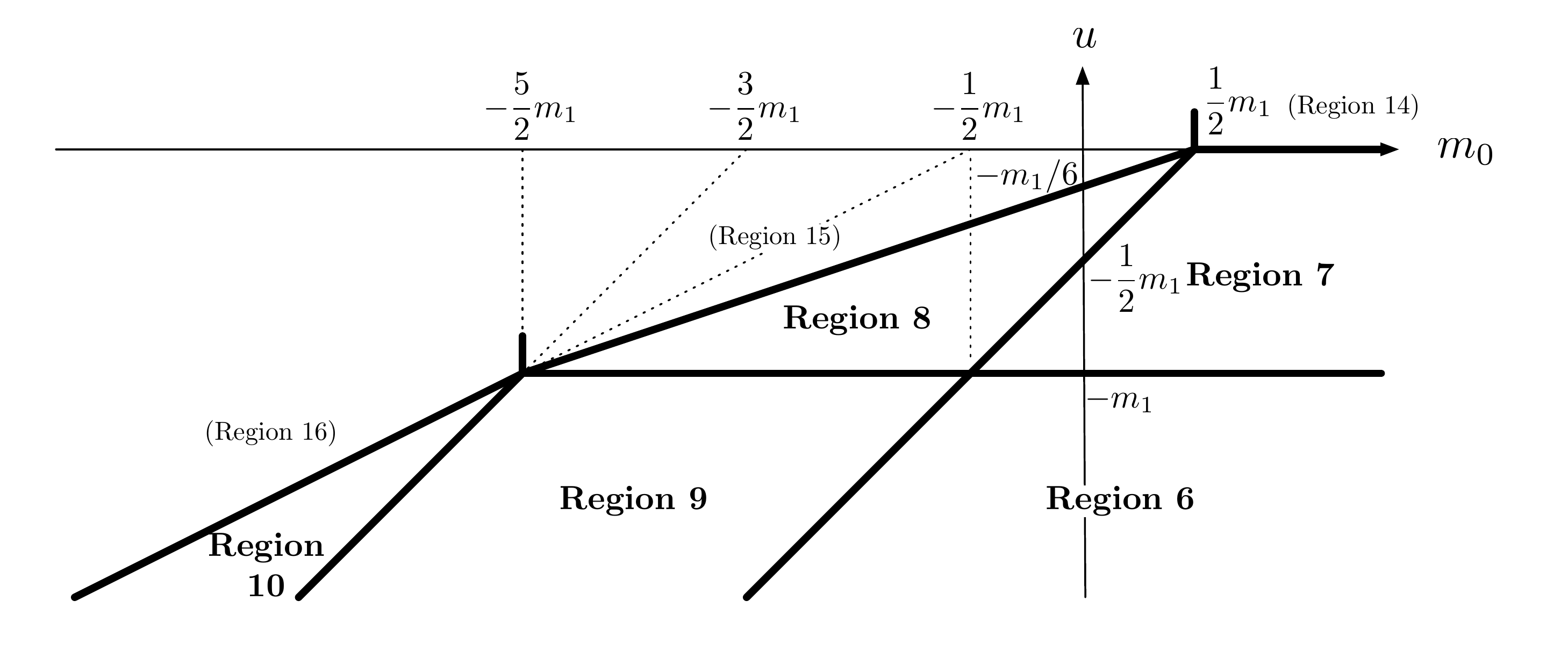}}
\caption{The phase diagram for the $E_2$ theory with a fixed $m_1$.}
\label{Fig:E2Phase3}
\end{figure}

We divide the $(m_0, m_1, u)$-space into the following sixteen regions
 and the corresponding phase diagram is depicted in Figure \ref{Fig:E2Phase3}:
\begin{align}\label{eq:E2region1to16}
\begin{array}{llll}
\text{Region 1: } & m_1<0, \,\, u < m_1, \,\,  u < m_0 + \frac{1}{2} m_1,
\cr
\text{Region 2: } & m_1<0, \,\, u > m_1, \,\, u < 0, \,\,  u < m_0 + \frac{1}{2} m_1,
\cr
\text{Region 3: } & m_1<0, \,\, u > m_1, \,\,  m_0 + \frac{1}{2} m_1 < u < \frac{1}{2} m_0 + \frac{1}{4} m_1,
\cr
\text{Region 4: } & m_1<0, \,\, u < m_1, \,\,  m_0 + \frac{1}{2} m_1 < u <  m_0 - \frac{1}{2} m_1 ,
\cr
\text{Region 5: } & m_1 < 0, \,\,  m_0 - \frac{1}{2} m_1  < u < \frac{1}{2} m_0 + \frac{1}{4} m_1,
\cr
\text{Region 6: } & m_1>0, \,\, u < - m_1, \,\,  u < m_0 - \frac{1}{2} m_1,
\cr
\text{Region 7: } & m_1>0, \,\, u > - m_1, \,\,  u < 0, \,\, u < m_0 - \frac{1}{2} m_1,
\cr
\text{Region 8: } & m_1>0, \,\, u > - m_1, \,\,  m_0 - \frac{1}{2} m_1 < u < \frac{1}{3} m_0 - \frac{1}{6} m_1,
\cr
\text{Region 9: } & m_1>0, \,\, u < - m_1, \,\,  m_0 -  \frac{1}{2} m_1 < u <  m_0 + \frac{3}{2} m_1,
\cr
\text{Region 10: } & m_1 > 0, \,\,  m_0 + \frac{3}{2} m_1  < u < \frac{1}{2} m_0 + \frac{1}{4} m_1,
\cr
\text{Region 11: } & m_1 < 0, \,\,  u > 0, \,\,  m_0 > -\frac{1}{2} m_1,
\cr
\text{Region 12: } & m_1 < 0, \,\,  u > \frac{1}{2} m_0 + \frac{1}{4} m_1,  \,\,  \frac{3}{2} m_1 < m_0 < -\frac{1}{2} m_1,
\cr
\text{Region 13: } & m_1 < 0, \,\,  u > \frac{1}{2} m_0 + \frac{1}{4} m_1,  \,\,  m_0 < \frac{3}{2} m_1,
\cr
\text{Region 14: } & m_1 > 0, \,\,  u > 0, \,\,  m_0 >  \frac{1}{2} m_1,
\cr
\text{Region 15: } & m_1 > 0, \,\,  u > \frac{1}{3} m_0 - \frac{1}{6} m_1,  \,\,  -  \frac{5}{2} m_1 < m_0 < \frac{1}{2} m_1,
\cr
\text{Region 16: } & m_1 > 0, \,\,  u > \frac{1}{2} m_0 + \frac{1}{4} m_1,  \,\,  m_0 < -  \frac{5}{2} m_1.
\end{array}
\end{align}
We note that the Region 11 -- Region 16, in fact, correspond unphysical Coulomb branch moduli space. In other words, the 5-brane web diagram reproduced in these regions do not depend on the Coulomb modulus $u$. Furthermore, the web diagram in Region 11 is merely a special case $u=0$ of the web in Region 2. This means that Region 11 is a boundary of Region 2. 
Similarly, the 5-brane web diagrams in Region 12, 13, 14, 15, and  16 are realized as a special case (or a boundary) of Region 3, 5, 7, 8, and 10, respectively. Therefore, the relevant regions are from the Region 1 to Region 10. 
\begin{figure}
\centering
\subfigure[Phase diagram with fixed $u<0$]{
\includegraphics[width=8cm]{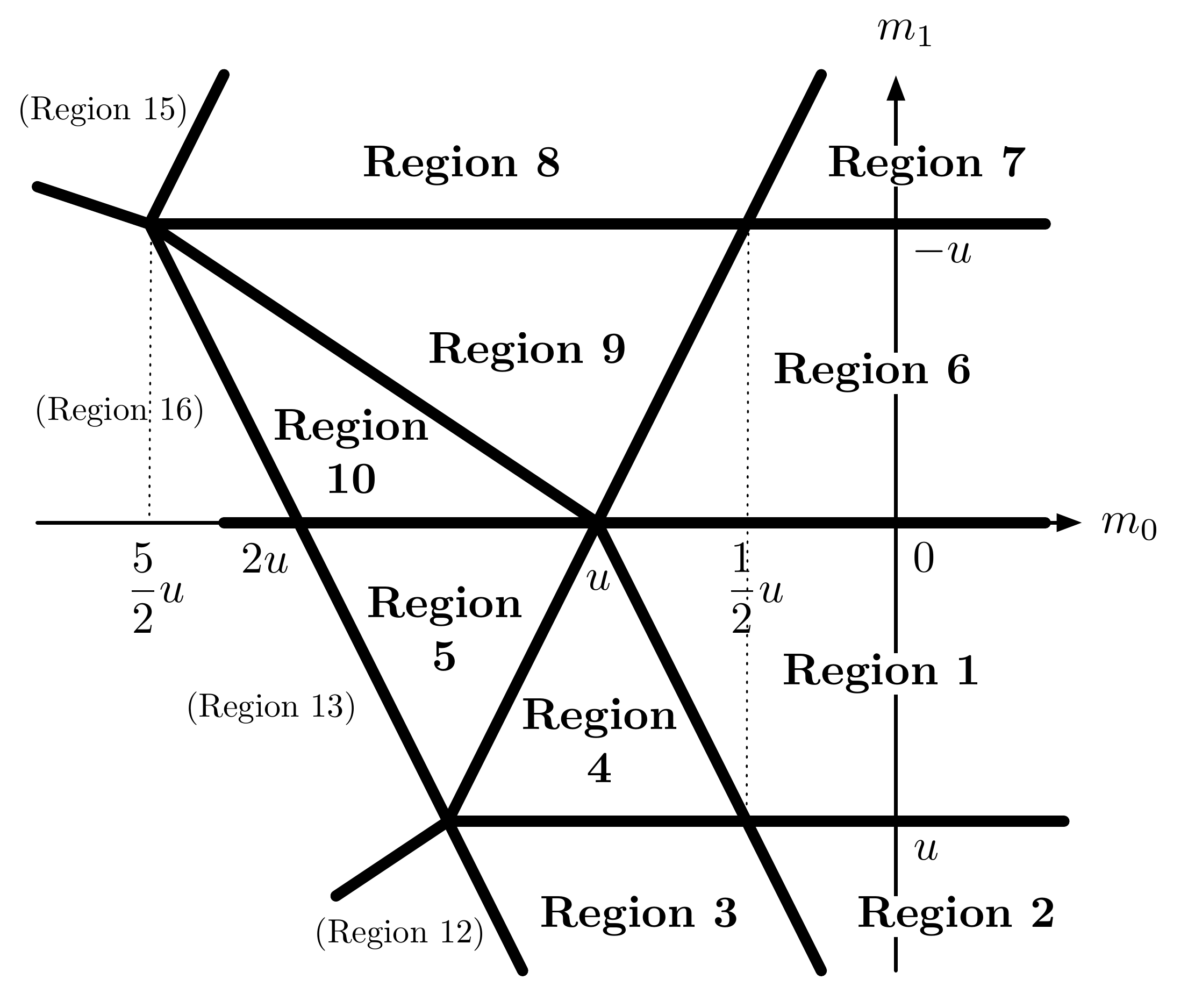}}
\\
\subfigure[Region 1]{\includegraphics[width=2.9cm]{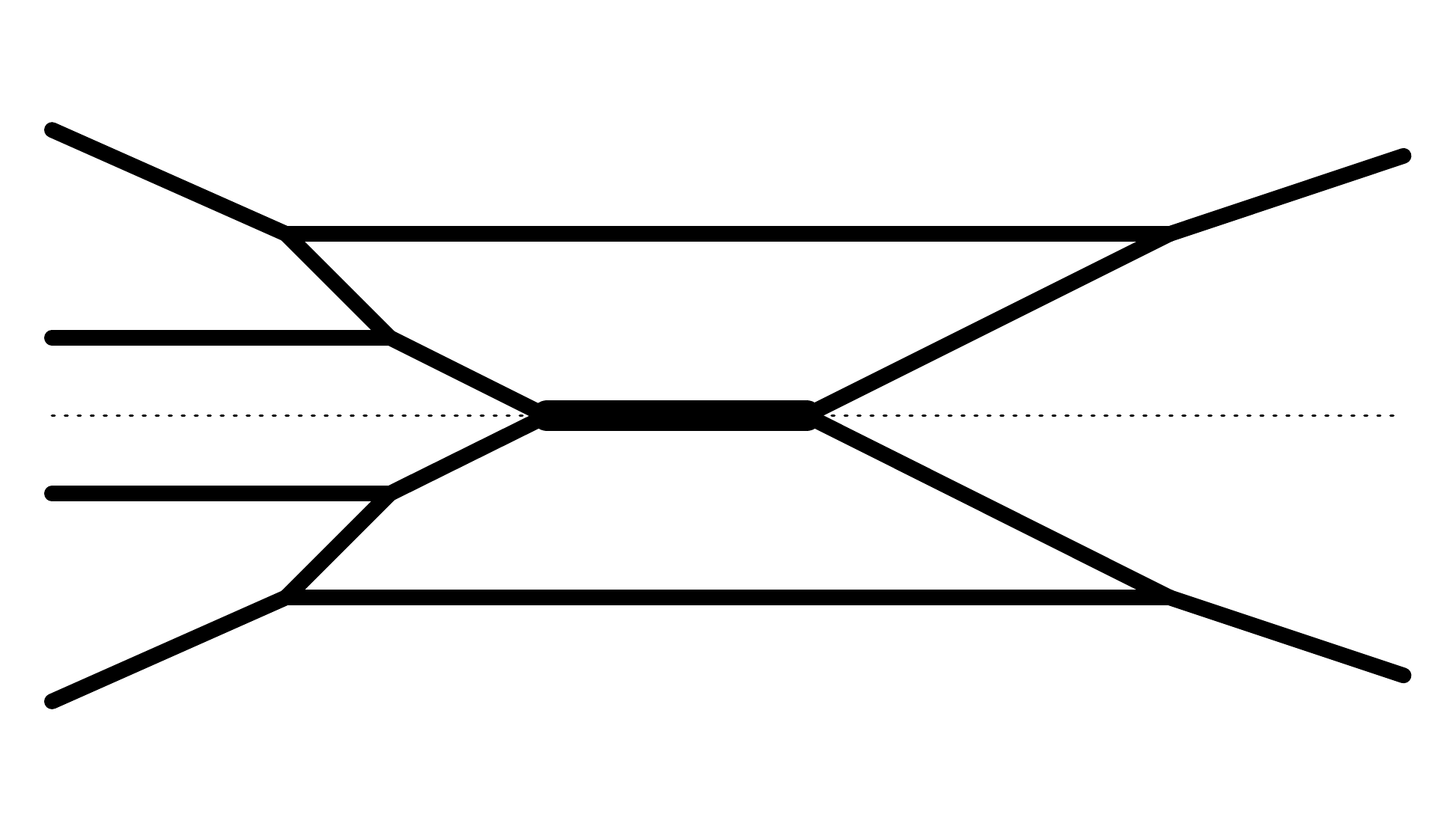}}
\subfigure[Region 2]{\includegraphics[width=2.9cm]{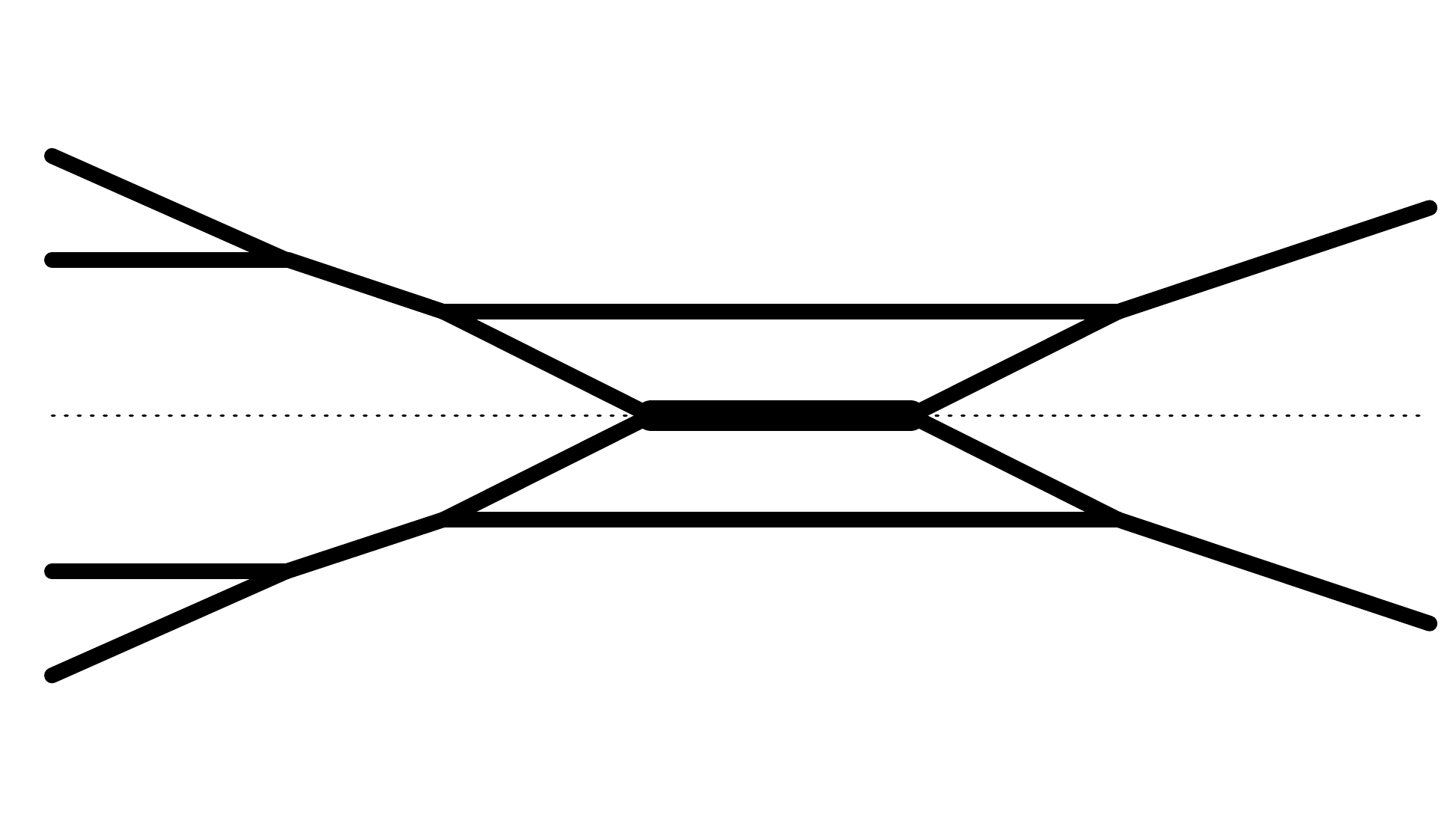}}
\subfigure[Region 3]{\includegraphics[width=2.9cm]{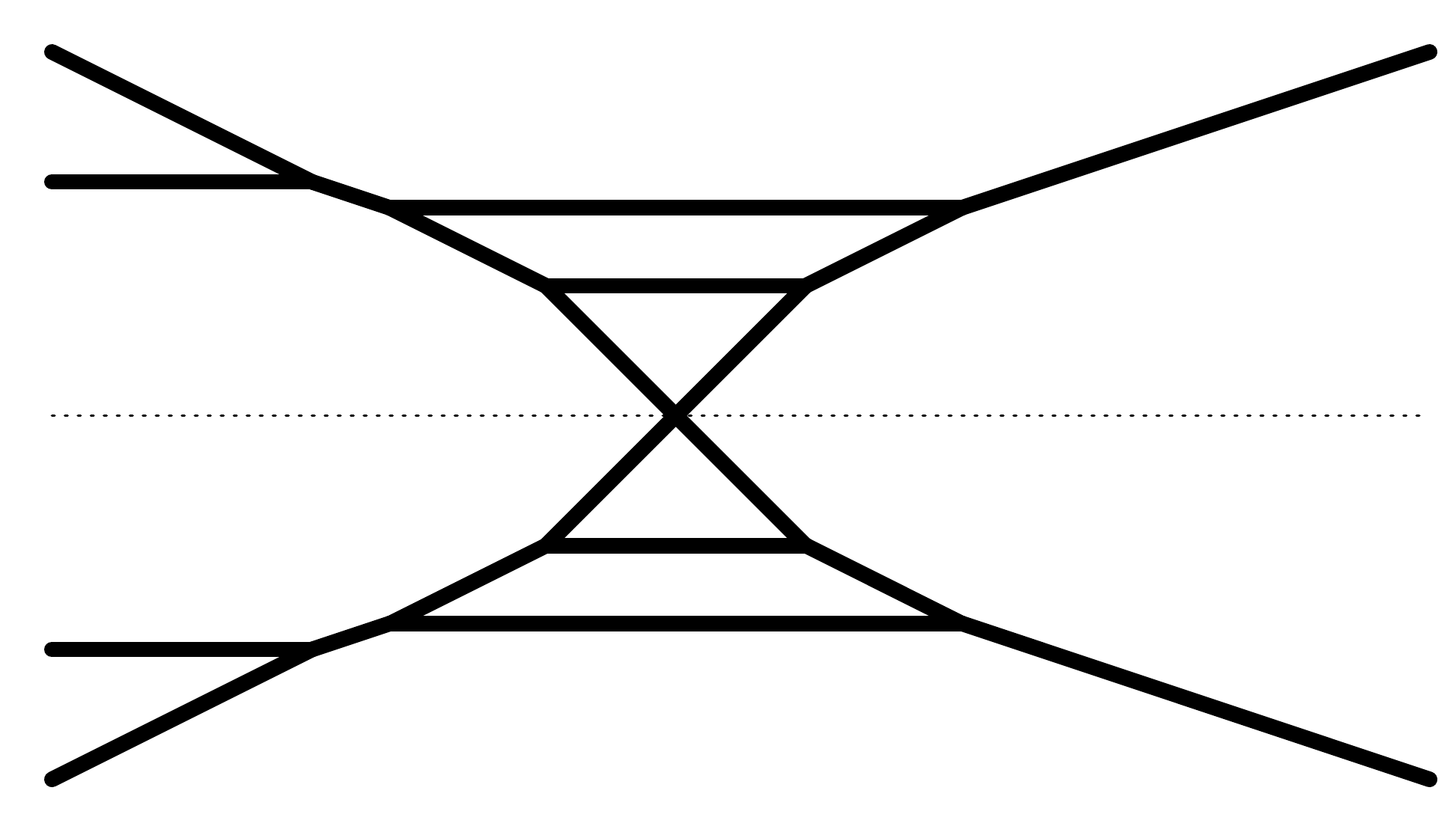}}
\subfigure[Region 4]{\includegraphics[width=2.9cm]{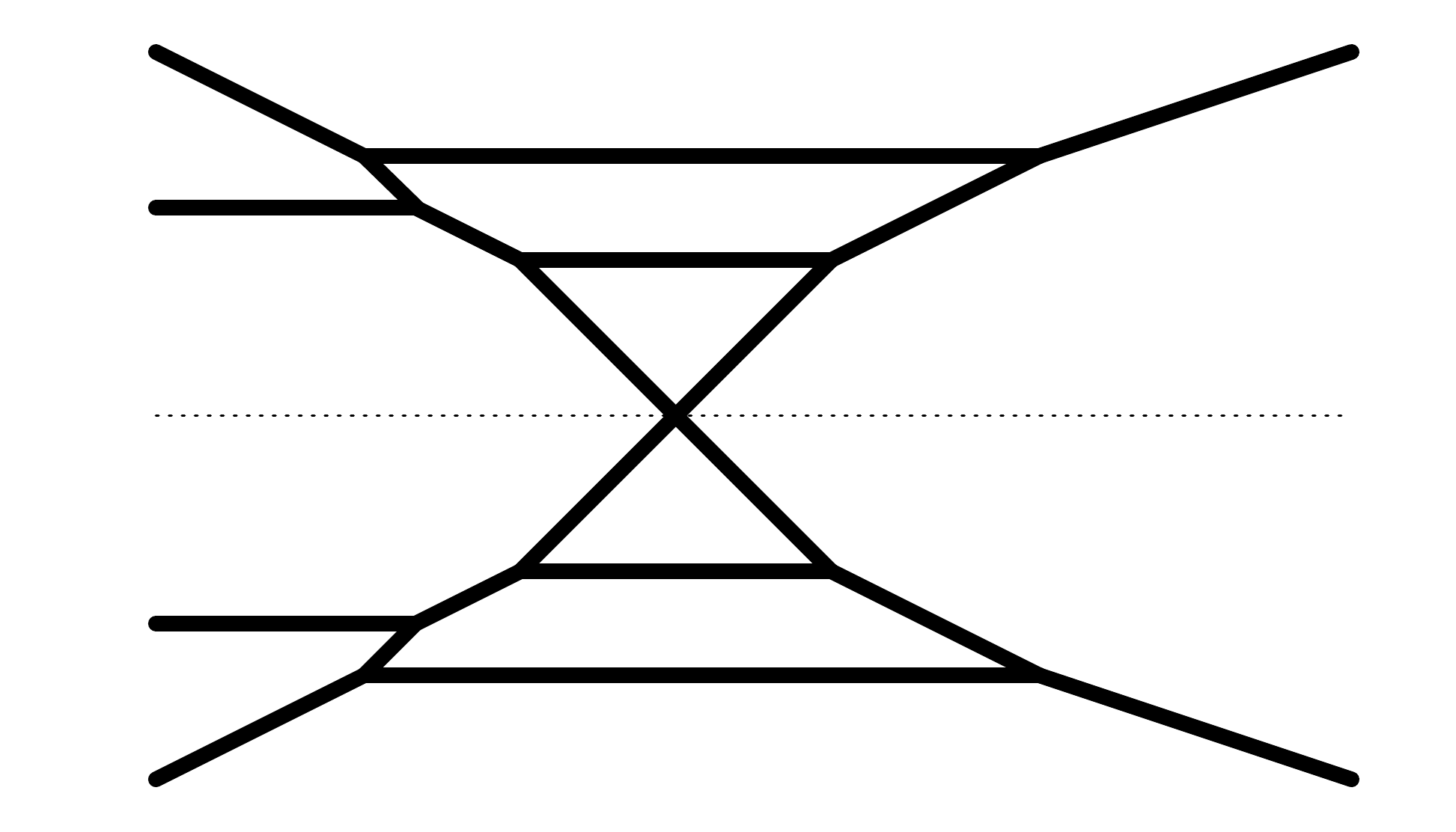}}
\subfigure[Region 5]{\includegraphics[width=2.9cm]{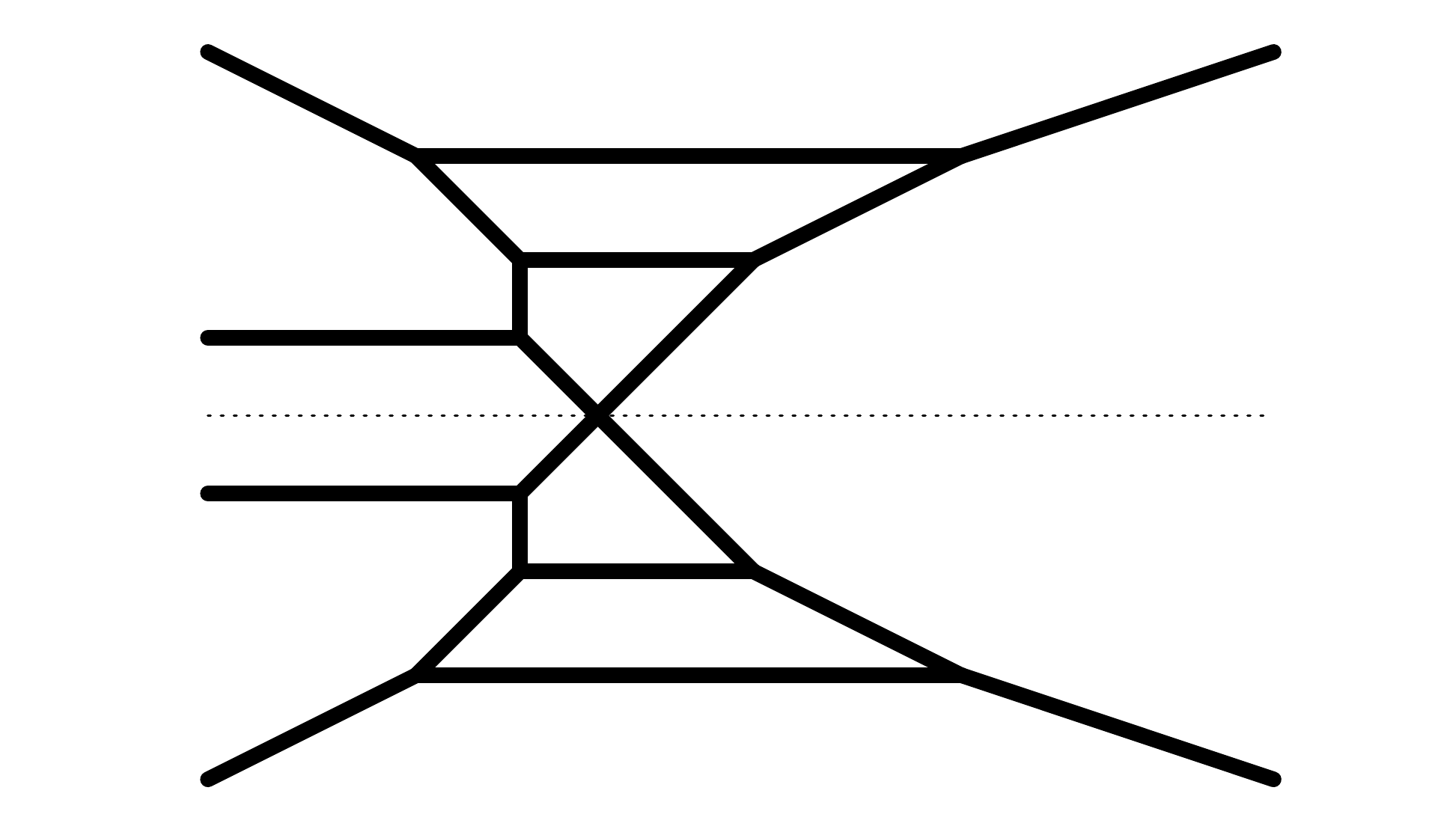}}
\\
\subfigure[Region 6]{\includegraphics[width=2.9cm]{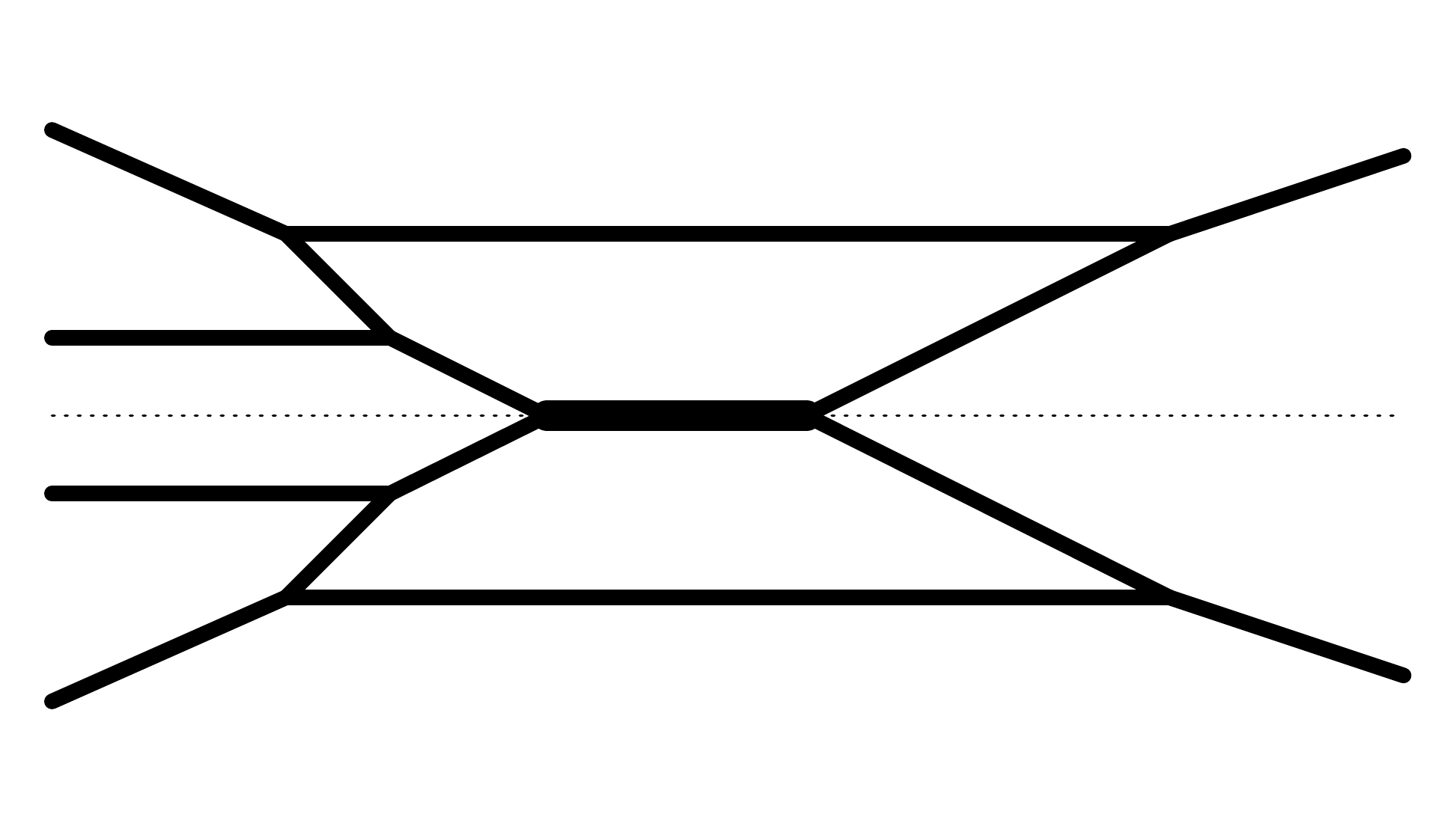}}
\subfigure[Region 7]{\includegraphics[width=2.9cm]{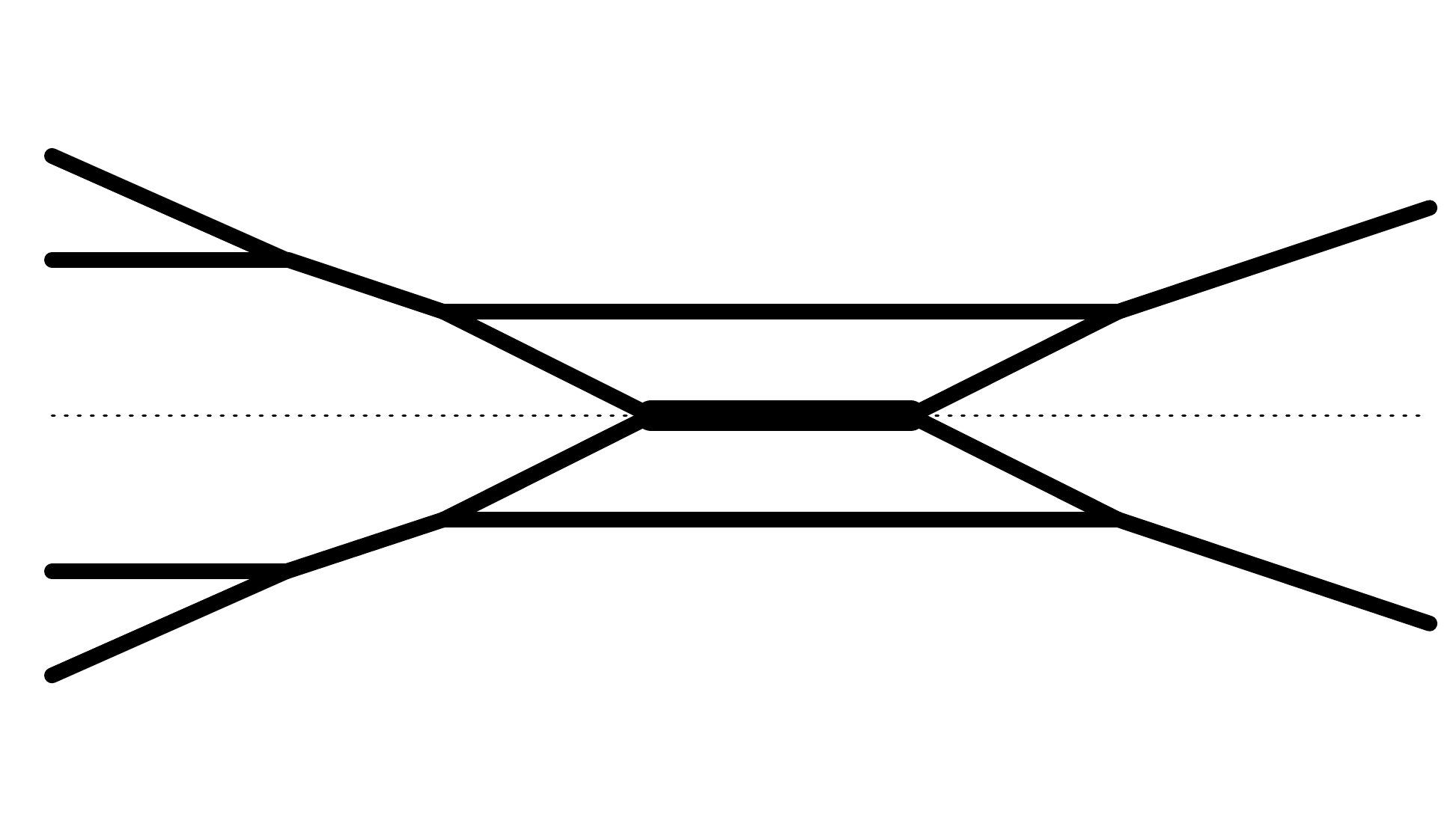}}
\subfigure[Region 8]{\includegraphics[width=2.9cm]{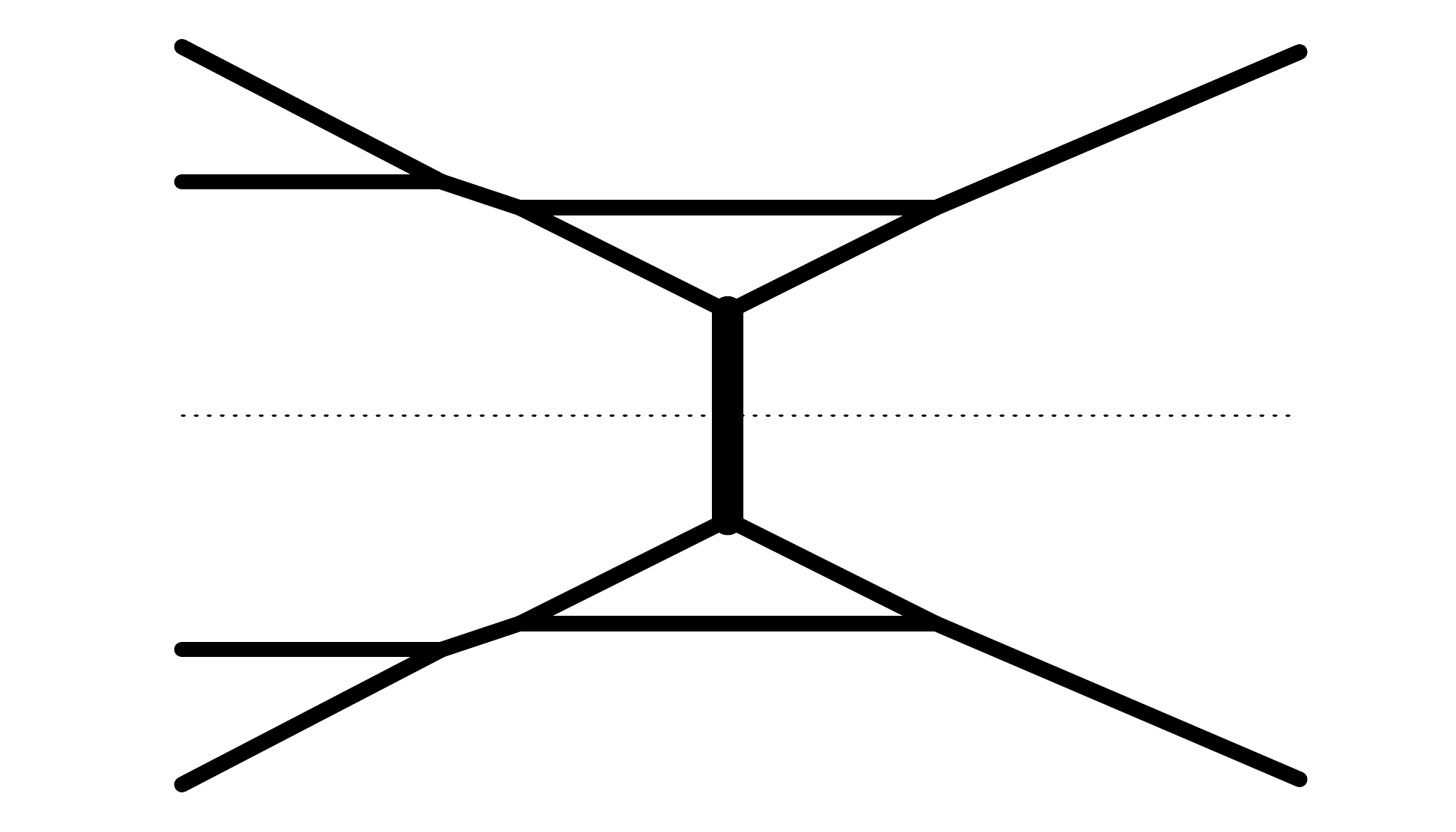}}
\subfigure[Region 9]{\includegraphics[width=2.9cm]{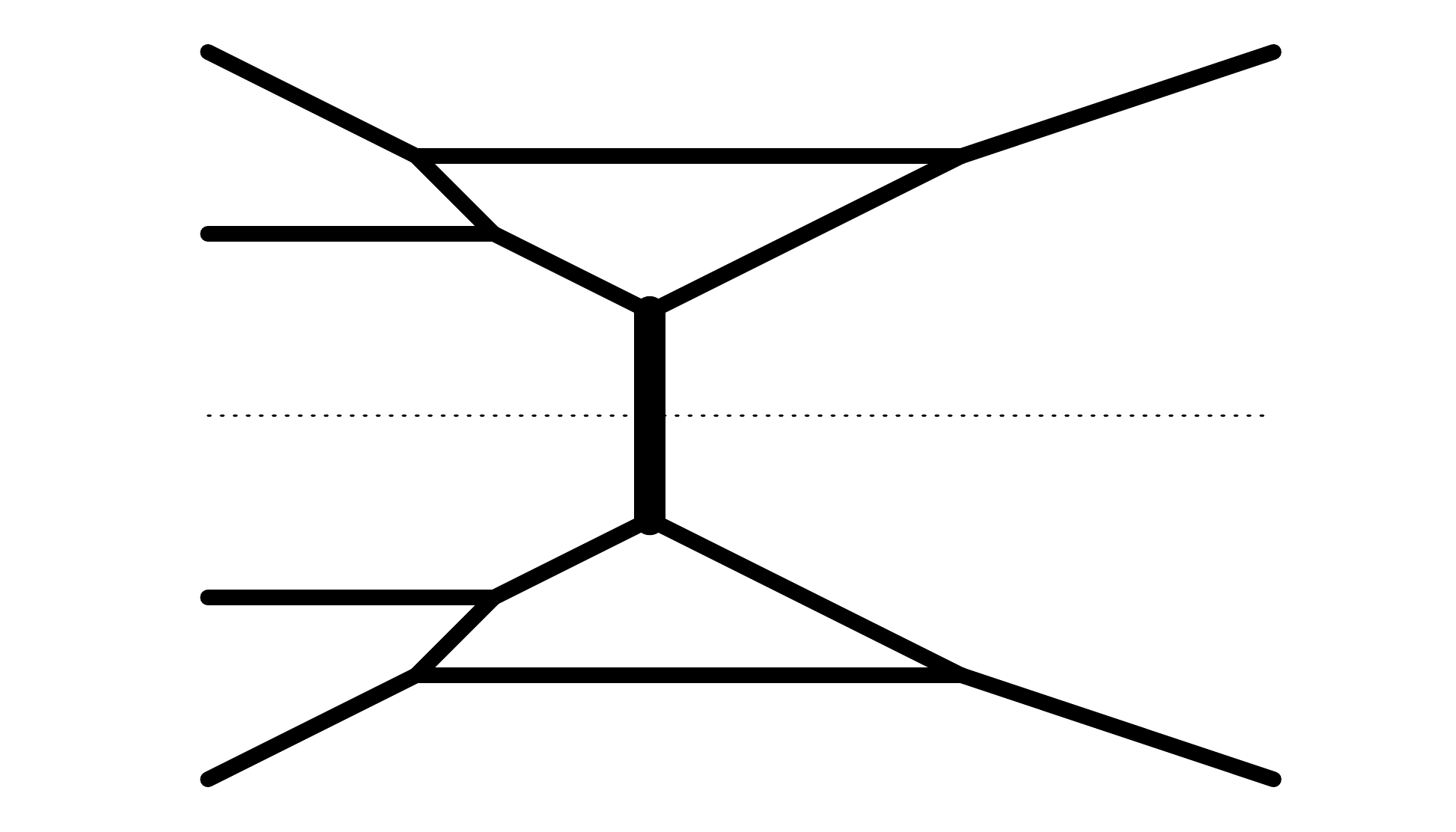}}
\subfigure[Region 10]{\includegraphics[width=2.9cm]{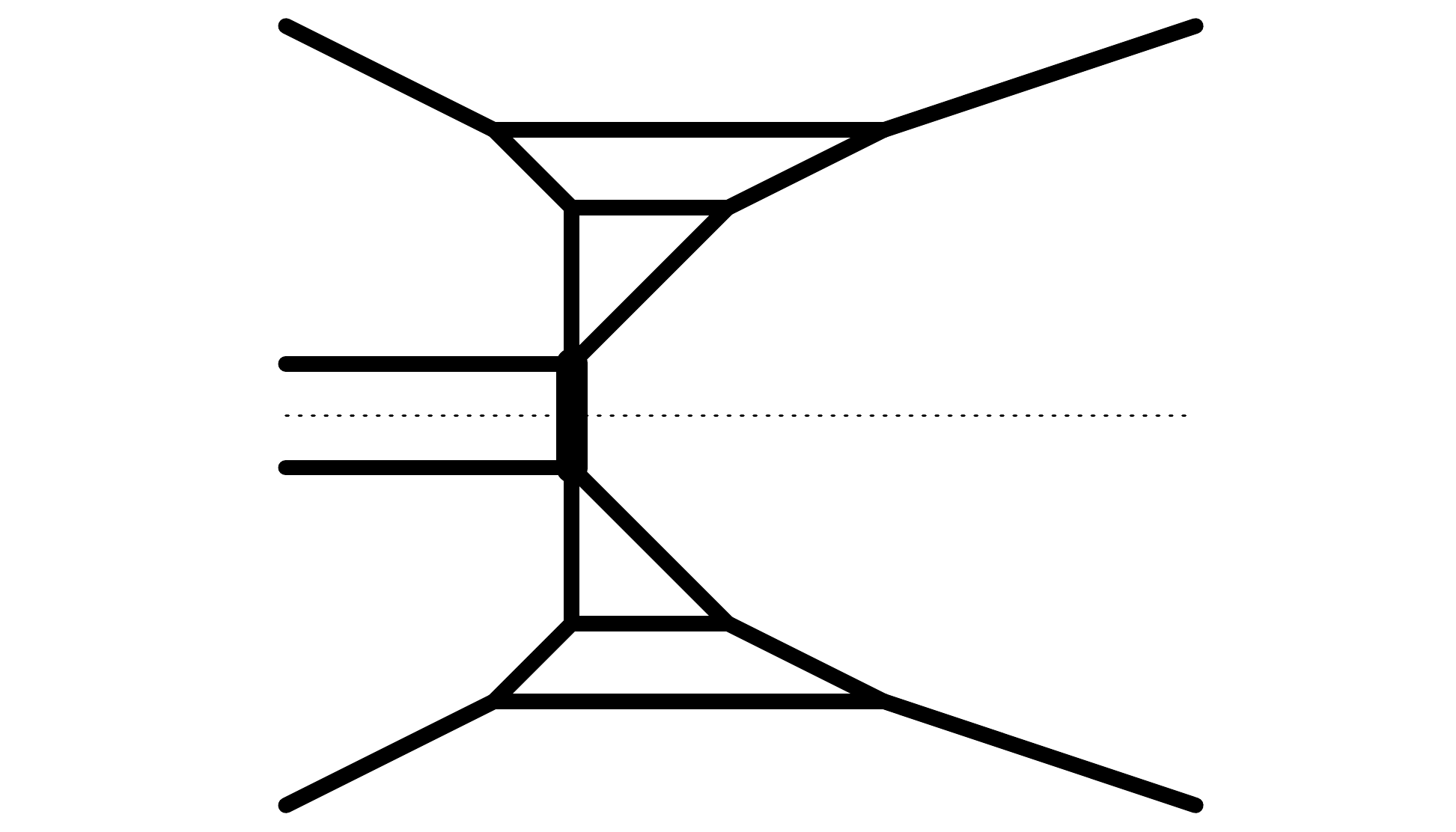}}
\caption{The phase diagram of the $E_2$ theory. The 5-brane web diagram in each region is depicted in Figure (b) -- (k). }
\label{Fig:E2Phase1}
\end{figure}

Since Region 1--10 are all inside the region $u<0$, it is enough to consider only this region.
The resultant phase diagrams in the $(u, m_0, m_1)$-space with fixed Coulomb moduli parameter $u<0$
are summarized in Figure \ref{Fig:E2Phase1}. 
The corresponding 5-brane web diagrams are depicted in Figure \ref{Fig:E2Phase1} (b) -- (k).
We remark that here we omit the information of the largest $A_{k, l}^{(m)}$'s and the explicit locations of the boundaries in the figures for simplicity, and such detailed information is relegated to appendix \ref{sec:E2Web}.

\subsection{Interpretation of the phase diagram}\label{sec:flop}

\begin{figure}
\centering
\subfigure[Region 2 $\leftrightarrow$ Region 3]{\includegraphics[width=10cm]{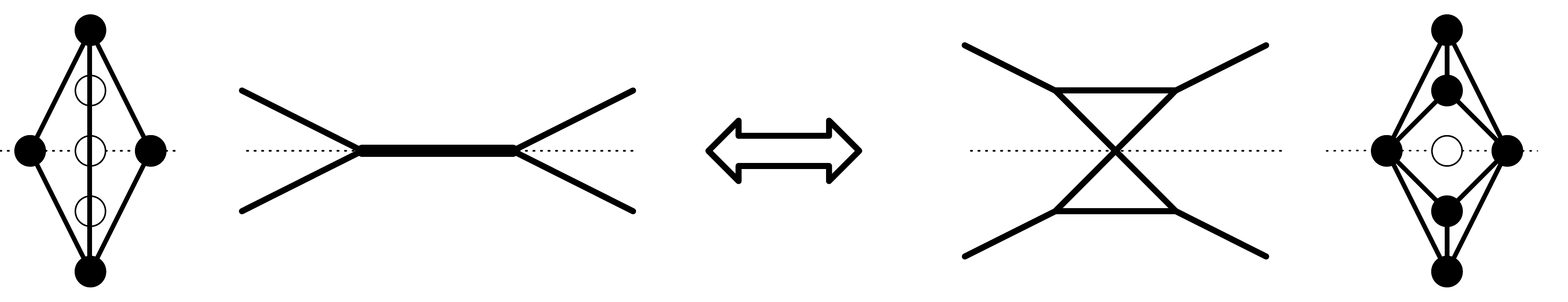}} \\
\subfigure[Region 5 $\leftrightarrow$ Region 10]{\includegraphics[width=10cm]{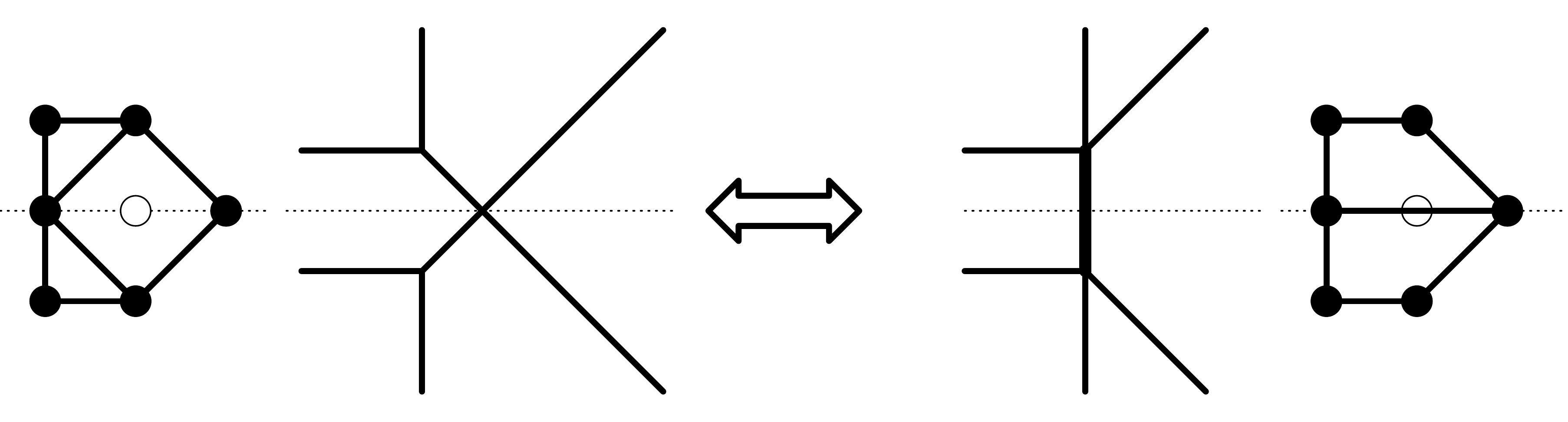}} \\
\subfigure[Region 7 $\leftrightarrow$ Region 8]{\includegraphics[width=10cm]{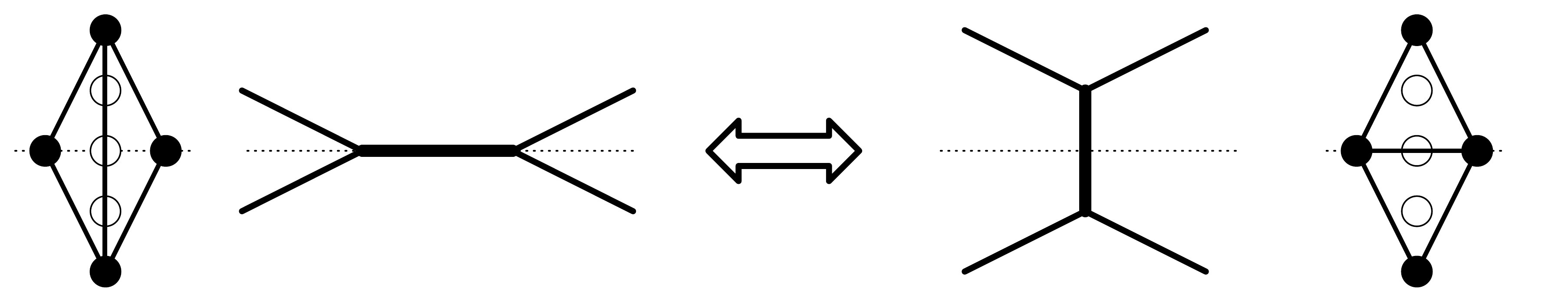}} \\
\subfigure[Region 9 $\leftrightarrow$ Region 10]{\includegraphics[width=10cm]{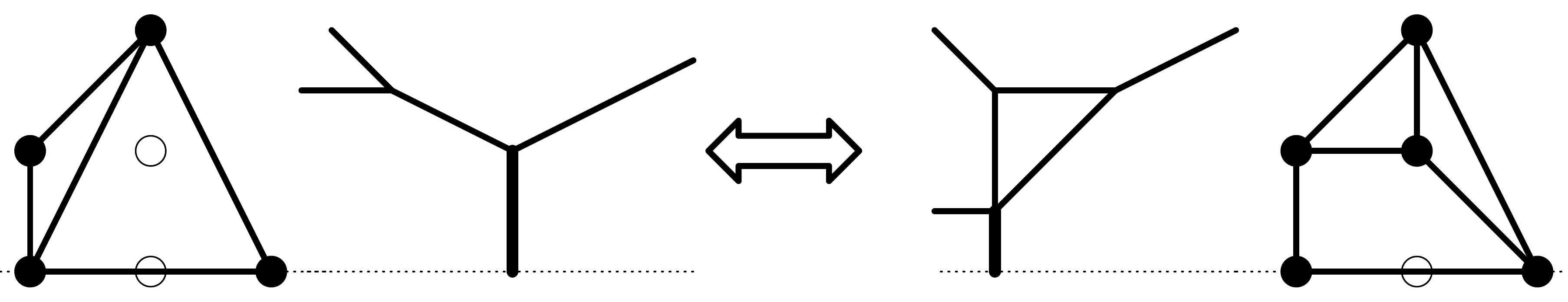}} \\
\caption{Four new transitions occurring in the phase diagram of the $E_2$ theory which can be interpreted as a generalized flop transition on a 5-brane web diagrams with an O5-plane.}
\label{Fig:newflops}
\end{figure}

We found that different regions for the masses $m_0,m_1$ and for the Coulomb moduli parameter $u$ yield various different phases for the $E_2$ SCFT as given in Figure \ref{Fig:E2Phase1}.
We see that some of them are related by a standard flop transition on a usual web diagram. A familiar example is the transition between Region 1 and Region 2.
On the other hand, others are related by new kinds of transitions,
which we would like to interpret as generalized version of the flop transitions for a web diagram including an O5-plane. 
We find four types of such transitions.
We depict only the relevant part of the web diagram in Figure \ref{Fig:newflops} (a) -- (d). 
We propose that these transitions are generally valid for any web diagrams.

From the argument of the Seiberg-Witten curves those transitions in Figure \ref{Fig:newflops} indeed exist and one can perform generalized flop transitions if a 5-brane web diagram contains these local structures. In particular the transition in Figure \ref{Fig:newflops} (d) does not involve an O5-plane and hence one can use the transition even if we an O5-plane is replaced by a $[0,1]$ 7-brane to which two $NS5$-branes attach. We note that in section 2.2 (also Figure 4) of \cite{Hayashi:2015vhy} we introduced a transition in similar to those in Figure \ref{Fig:newflops}, which is in fact nothing but a successive transition of those in Figure \ref{Fig:newflops} (c) and (d). Therefore, the analysis in section \ref{sec:E2Phase} shows that the transition considered in \cite{Hayashi:2015vhy} is indeed possible.

\begin{figure}
\centering
\includegraphics[width=10cm]{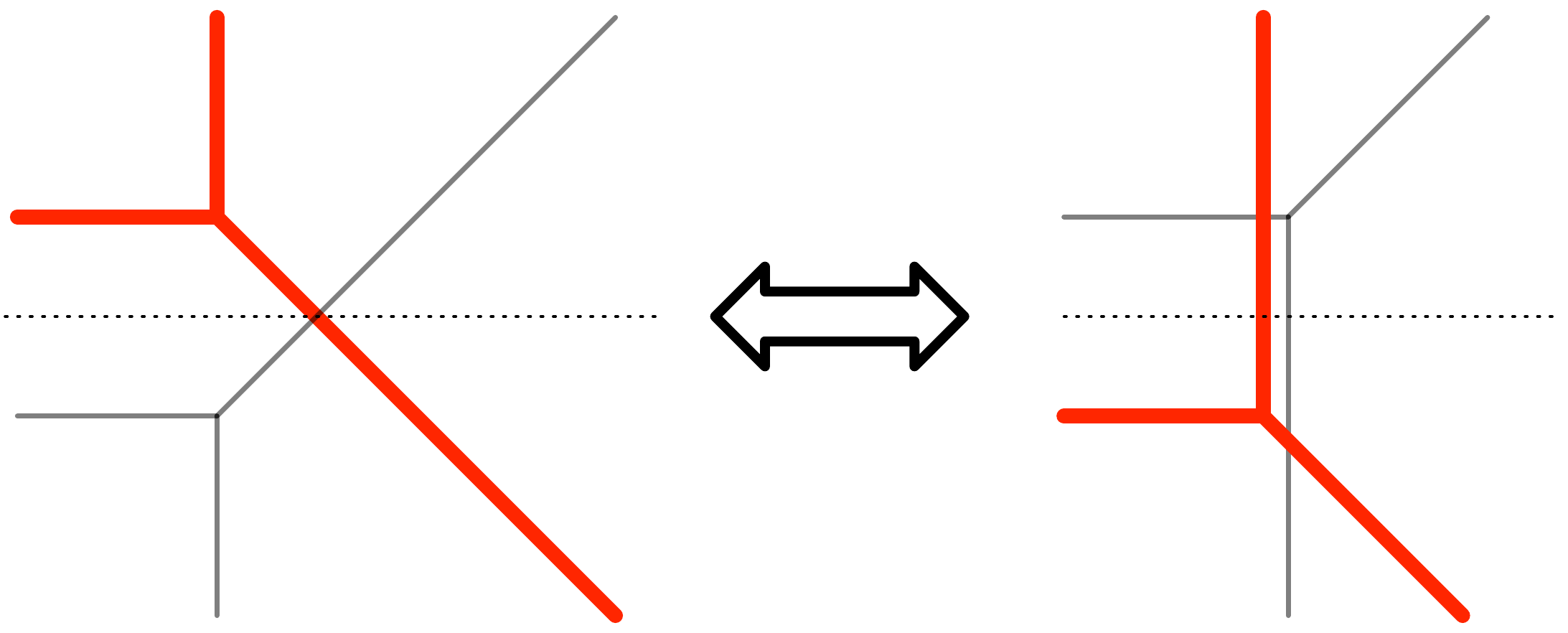}
\caption{Interpretation of the flop transition in Figure \ref{Fig:newflops} (b).}
\label{Fig:flop2}
\end{figure}

Among the four transitions in Figure \ref{Fig:newflops}, the transition in Figure \ref{Fig:newflops} (b) may have an intriguing interpretation as a continuous movement of 5-branes across the O5-plane. 
This flop transition can be naturally understood by interpreting that the $(1, -1)$ 5-brane is connected to 
the mirror image of (1, 1) 5-brane as depicted in the thick red line in Figure \ref{Fig:flop2}. The thin black lines are just another copy of this red ones with reflected along O5-plane. When we move this red lines down until D5 brane goes below the O5-plane, while moving the black part in a way to be consistent with the reflection, we obtain the right of Figure \ref{Fig:flop2}. In this picture in the right, we see that there are two coincident NS5-branes stuck to the O5-plane and that one of the NS5-brane is connected to the mirror image of the other NS5-brane. This configuration is consistent with the S-dual of the configuration 
with two coincident D5-branes stuck to ON$^-$ plane discussed in section 2 of \cite{Hayashi:2015vhy}.
This observation tells us what happens when we move flavor D5-brane below the O5-plane, which corresponds to move from positive mass to negative mass (or vice versa).

\subsection{Flow to the $E_1$, $\widetilde{E}_1$ and $E_0$ theories}
\label{sec:FlowfromE2}
Recalling that for $SU(2)$ theory with $N_f$ flavors, a flavor decoupling is to take $|m_i| \to \infty$ which decouples the hypermultiplet in the fundamental representation associated with mass $m_i$, and hence make the $SU(2)$ theory of one less $N_f-1$ flavors. It is, however, for $SU(2)$ theory with $N_f=1$ flavor, there are two possible flavor decouplings, giving rise to two different SCFTs with different global symmetries, known as $E_1$ and $\widetilde{E}_1$. The decoupling limits are to take $m_1 \to -\infty$ and $m_1 \to \infty$ which result in two different pure $SU(2)$ theory with discrete theta angle $0$ and $\pi$, respectively \cite{Morrison:1996xf}. See also Figure \ref{Fig:su2E1E1t} for usual 5-brane diagrams.

For configuration of 5-brane web with an O5-plane, it is natural to expect to reproduce two different phase diagrams for $E_1$ and $\widetilde{E}_1$ theories by taking the decoupling limits on the phase diagram of the $E_2$ theory. As we will show, it is indeed the case that the limits $m_1 \to -\infty$ and $m_1 \to \infty$ of the $E_2$ theory reproduce the phase diagram for the $E_1$ theory and the $\widetilde{E}_1$ theory respectively. In particular, the two different strong coupling behaviors in the $E_1$ and $\widetilde{E}_1$ theories will be also precisely reproduced from the decouplings of the $E_2$ theory. This further supports for the 5-brane web picture for $Sp(1)$ theory in section \ref{sec:5dlimit}.

\paragraph{Flow to $E_1$ theory.} 
First consider the flavor decoupling limit $m_1 \to -\infty$ on the phase diagram of the $E_2$ theory with $m_1\le 0$, given in Figure \ref{Fig:E2Phase3} (a), together with the redefinition of $m_0$ for the $E_2$ theory,
\be\label{eq:m0E2toE1}
m_1 + 2m_0^{(E_2)} = 2m_0^{(E_1)}.
\ee
The relation can be understood from the identification of the Seiberg-Witten curves of the $E_2$ theory into the $E_1$ theory in the decoupling limit as shown in appendix \ref{sec:E8}. It follows that in the phase diagram for the $E_2$ theory, $(m_0^{(E_2)}, u) = \left(-\frac{1}{2}m_1, 0\right)$ becomes the origin in the $(m_0^{(E_1)}, u)$-space. With the relation \eqref{eq:m0E2toE1}, it is possible to see that applying the limit $m_1 \to -\infty $ to the phase diagram of the $E_2$ theory in Figure \ref{Fig:E2Phase3} (a) gives rise to the phase diagram of the $E_1$ theory in Figure \ref{Fig:E1Phase}. After taking the limit, only Region 2 and Region 3 of the $E_2$ theory remain. Region 2 of the $E_2$ theory becomes the Region 1 of the $E_1$ theory, while Region 3 of the $E_2$ theory becomes Region 3 of the $E_1$ theory. 
As for the 5-brane webs in these regions,
we can take the limit $m_1 \to -\infty$ 
together with proper shift in $x_4$ direction
for the 5-brane webs for Region 2 and Region 3 depicted in 
Figure \ref{Fig:E2Region2} and Figure \ref{Fig:E2Region3} in Appendix \ref{sec:detailE2}.
The resulting 5-brane web diagrams 
of Region 2 in Figure \ref{Fig:E2Phase1} (c) and Region 3 in Figure \ref{Fig:E2Phase1} (d) 
in the limit $m_1 \to -\infty$ exactly yield the 5-brane webs for Region 1 in Figure \ref{Fig:E1Region1} and Region 3 in Figure \ref{Fig:E1Region3} of the $E_1$ theory respectively. In particular the strong coupling behavior of the $E_1$ theory in Figure \ref{Fig:E1Region3} is reproduced from the limit $m_1 \to -\infty$ of the 5-brane web in Figure \ref{Fig:E2Phase1} (d).

\paragraph{Flow to $\widetilde{E}_1$ theory.} Next consider the decoupling limit $m_1 \to \infty$ on the phase diagram for the $E_2$ theory with $m_1\ge0$, given in Figure \ref{Fig:E2Phase3} (b), together with the redefinition of $m_0$ for the $E_2$ theory,
\be\label{eq:m0E2toE1tilde}
- m_1 + 2m_0^{(E_2)} = 2m_0^{(\widetilde{E}_1)}.
\ee
This is again the identification of the decoupling of the $E_2$ theory to the $\widetilde{E}_1$ theory shown in appendix \ref{sec:E8}. 
It follows form \eqref{eq:m0E2toE1tilde} that $(m_0^{(E_2)}, u) = (\frac{1}{2}m_1, 0)$ becomes the origin in the $(m_0^{(\widetilde{E}_1)}, u)$-space. The relevant regions of the phase diagram of the $E_2$ theory in this decoupling limit are Region 7 and Region 8 in Figure \ref{Fig:E2Phase3} (b). It is then readily to see that the resulting phase diagram is exactly the same diagram as that of the $\widetilde{E}_1$ theory in Figure \ref{Fig:E1tildePhase}. Namely, Region 7 of the $E_2$ theory becomes Region 1 of the $\widetilde{E}_1$ theory;  Region 8 of the $E_2$ theory becomes Region 3 of the $\widetilde{E}_1$ theory. 
Similarly, after taking the limit $m_1 \to \infty$ the 5-brane web diagrams for Region 7 in Figure \ref{Fig:E2Region7} and for Region 8 in Figure \ref{Fig:E2Region8}
of the $E_2$ theory become the 5-brane web diagrams for Region 1 in Figure \ref{Fig:E1Region1} for Region 3 in Figure \ref{Fig:E1tildeRegion3} of the $\widetilde{E}_1$ theory, respectively. Again the strong coupling behavior of the $\widetilde{E}_1$ theory in Figure \ref{Fig:E1tildeRegion3} is reproduced from the limit $m_1 \to \infty$ of the 5-brane web in Figure \ref{Fig:E2Phase1} (i). 

\paragraph{From $E_1$ to $\widetilde{E}_1$ via $E_2$.}
From the discussion above, we find that, by changing the mass parameter $m_1$ from $-\infty$ to $+\infty$, we can continuously change the web diagram of the $E_1$ theory, which is identified as the $E_2$ theory with $m_1=-\infty$, to that of the $\tilde{E}_1$ theory, which is identified as the $E_2$ theory with $m_1=+\infty$. In this process, a sequence of standard or generalized flop transitions arises, as can be seen from the discussion in sections \ref{sec:E2Phase} and \ref{sec:flop}. 
Especially, the web diagrams in the strong coupling region, Figure \ref{Fig:E1Region3} and Figure \ref{Fig:E1tildeRegion3}, are connected by the transitions as Region $3 \to 4 \to 5 \to 10 \to 9 \to 8$ in Figure \ref{Fig:E2Phase1}. Most of the transitions are standard flop transitions while the transitions in Region $5 \to 10$ and in Region $10 \to 9$ are the new ones, which are depicted in Figure \ref{Fig:newflops} (b) and (d). 
Thus, these new transitions 
help us to understand the difference of the discrete theta angle at the level of web diagram with an O5-plane in the strong coupling region.

\paragraph{Flow to $E_0$ theory.}

\begin{figure}
\begin{minipage}[t]{5cm}
\centering
\includegraphics[width=4.5cm]{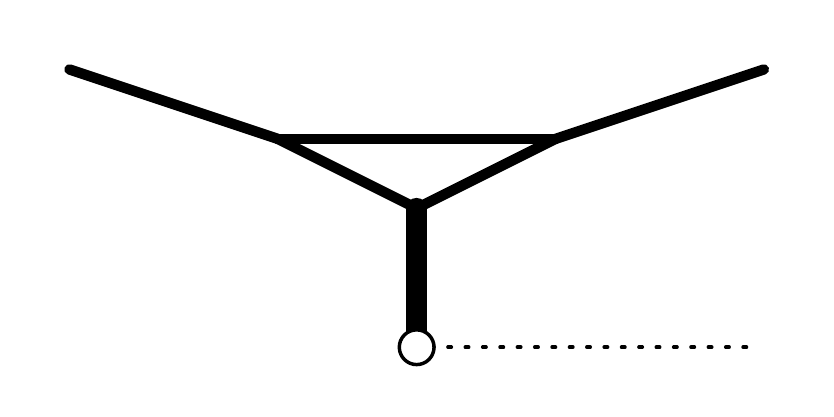}
\caption{The 5-brane web diagram given by taking the limit $m_0 \to \infty$ for the 5-brane web of the $\widetilde{E}_1$ in Figure \ref{Fig:E1tildeRegion3}. A $(0, 1)$ 7-brane is attached to the end of the two NS5-branes.}
\label{Fig:E0no1}
\end{minipage}
\hfill
\begin{minipage}[t]{5cm}
\centering
\includegraphics[width=4.5cm]{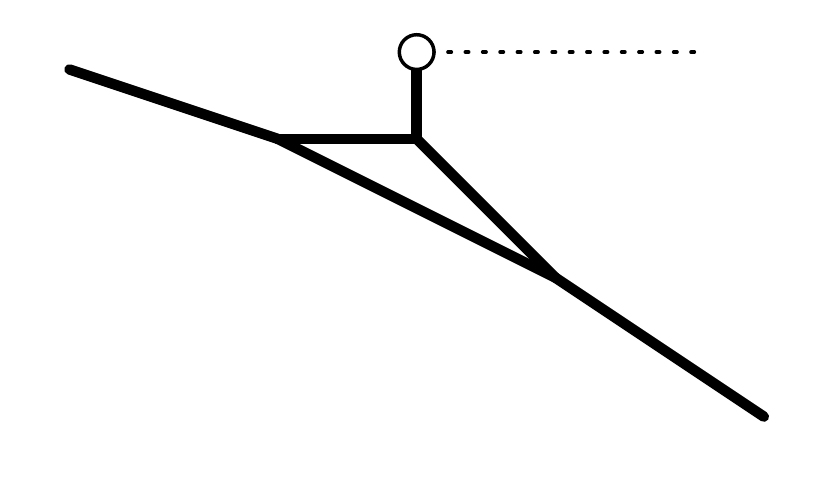}
\caption{The 5-brane diagram after moving the $(0, 1)$ 7-branes in Figure \ref{Fig:E0no1} upward.}
\label{Fig:E0no2}
\end{minipage}
\hfill
\begin{minipage}[t]{4cm}
\centering
\includegraphics[width=3cm]{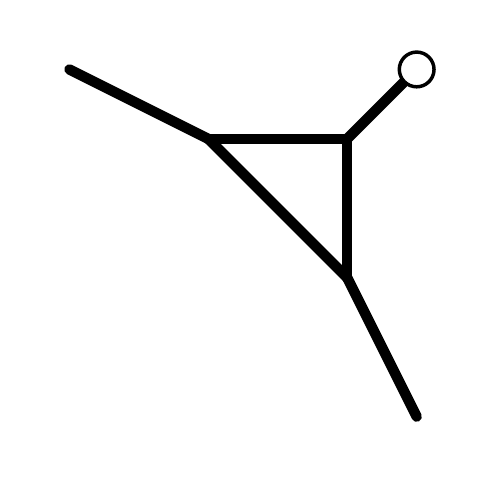}
\caption{A standard 5-brane web diagram for the $E_0$ theory.}
\label{Fig:E0no3}
\end{minipage}
\end{figure}

From the 5-brane web of the $\widetilde{E}_1$ theory in Region 3 in Figure \ref{Fig:E1tildeRegion3}, we can further take a limit of $m_0 \to \infty$ which gives rise to yet another 5d SCFT called the $E_0$ theory. 
When we take the limit $m_0 \to \infty$ for the 5-brane web in Figure \ref{Fig:E1tildeRegion3}, the middle two NS5-branes become infinitely long. Hence the degrees of freedom associated to the infinitely long NS5-branes decouple and the resulting 5d theory can be effectively described by a 5-brane web shown in Figure \ref{Fig:E0no1}. 
Here we introduced a $[0, 1]$ 7-brane at the end of the two NS5-branes. We can then consider a deformation of the theory by moving the $[0, 1]$ 7-brane upward. After the Hanany-Witten transition the 5-brane web diagram becomes the one in Figure \ref{Fig:E0no2}. 
This is equivalent to a standard web diagram of the $E_0$ theory in Figure \ref{Fig:E0no3} by an $SL(2, \mathbb{Z})$ transformation. 
Therefore, by taking the limit of $m_0 \to \infty$ from the 5-brane web of the $\widetilde{E}_1$ theory in Figure \ref{Fig:E1tildeRegion3} with the appropriate deformation and the Hanany-Witten transition, we arrive at the 5-brane web diagram of the $E_0$ theory.

\subsection{Effective coupling of $E_2$ theory}
\label{sec:effE2}
As we determined the 5-brane web diagrams in Figure \ref{Fig:E2Phase1} (or more detailed figures given in appendix \ref{sec:detailE2}), it is possible to determine the effective coupling of the $E_2$ theory for each Region. The tension of the monopole string is related to the area $a_D$ of the compact face where $A_{1, 2}$ becomes the largest, and the effective coupling is obtained by taking a derivative of $a_D$ with respect to $-u$. Then the effective coupling of the $E_2$ theory in each region can be computed and is summarized in Table \ref{Table:effE2Phase1} and Table \ref{Table:effE2Phase2}. 
\begin{table}[t]
\centering
\begin{tabular}{|c|c|c|c|c|c|}
\hline
&Region 1 & Region 2 & Region 3 & Region 4 & Region 5 \\
\hline
$\tau_{\text{eff}}$ & $2m_0 - 7u$ & $2m_0 + m_1 - 8u$ & $2m_0+m_1-8u$ & $2m_0 - 7u$ & $3m_0 - \frac{m_1}{2} - 8u$ \\
\hline
\end{tabular}
\caption{The effective coupling of the $E_2$ theory ($m_1\le 0$) for Region 1 -- Region 5.}
\label{Table:effE2Phase1}
\end{table}
\begin{table}[t]
\begin{tabular}{|c|c|c|c|c|c|}
\hline
 &Rergion 6 & Region 7 & Region 8 & Region 9 & Region 10\\
 \hline
 $\tau_{\text{eff}}$& $2m_0 - 7u$ & $2m_0 - m_1 - 8u$ & $3m_0 - \frac{3 m_1}{2} - 9u$ & $3m_0 - \frac{m_1}{2} - 8u$ & $3m_0 - \frac{m_1}{2} - 8u$\\
 \hline
\end{tabular}
\caption{The effective coupling of the $E_2$ theory ($m_1\ge 0$) for Region 6 -- Region 10.}
\label{Table:effE2Phase2}
\end{table}

In fact, it is possible to obtain a form of the effective coupling which can be valid in 
all the region in the physical Coulomb moduli, that is Region 1--10.
The explicit expression of such a form is 
\be
\tau_{\text{eff}} = \frac{1}{2}\left(5m_0 - \frac{1}{2}m_1-17u -\left|u-m_0+\frac{1}{2}m_1\right| - \left|u-m_1\right| - \left|u+m_1\right|\right). \label{eq:fulleff}
\ee 
Then one can see that in each region  
the effective coupling \eqref{eq:fulleff} precisely reproduces the effective couplings listed in Table \ref{Table:effE2Phase1} and Table \ref{Table:effE2Phase2}.

In section \ref{sec:FlowfromE2}, we showed that the $E_2$ theory becomes the $E_1$ theory after sending $m_1 \to - \infty$ with the redefinition \eqref{eq:m0E2toE1}. Indeed, the effective coupling \eqref{eq:fulleff} of the $E_2$ theory reduces to the effective coupling \eqref{eq:E1taueff} of the $E_1$ theory after the redefinition \eqref{eq:m0E2toE1}. Similarly, the $E_2$ theory flows to the $\widetilde{E}_1$ theory in the limit $m_1 \to \infty$ with the redefinition \eqref{eq:m0E2toE1tilde}. Again it is possible to see that the effective coupling \eqref{eq:fulleff} of the $E_2$ theory reduces to the effective coupling \eqref{eq:E1tildetaueff} of the $\widetilde{E}_1$ theory after the redefinition \eqref{eq:m0E2toE1tilde}.

\bigskip
\section{Conclusion and discussion}
\label{sec:concl}
In this paper, we discussed  
two different 5d $\mathcal{N}=1$ pure $Sp(1)$ theories from the perspective of a 5-brane web with an O5-plane: one has the discrete theta angle $\theta=0$ (the $E_1$ theory) and the other has $\theta=\pi$ (the $\widetilde{E}_1$ theory). From the point of view of the pure $SU(2)$ theory, these two theories clearly have different web diagrams. This means that the corresponding M-theory configurations, dictated in their Seiberg-Witten curves, are different. 
From the perspective of the pure $Sp(1)$ theory, on the other hand, their difference does not seem very clear, as a naive brane configuration with an O5-plane for two theories does not seem to show a sharp distinction in the weak coupling region. 

As the first step to distinguish these two theories in 5-brane webs, we proposed a way to compute the Seiberg-Witten curves for 5d $Sp(1)$ theory based on a web of 5-branes with an O5-plane, 
by introducing proper boundary conditions on OM5-planes originated from the O5-plane.
The Seiberg-Witten curve for the $E_1$ and $\widetilde{E}_1$ theories, we obtained, agrees with the known Seiberg-Witten curves.  Our method is also applicable for the cases of non-zero flavors; for example, as shown in appendix \ref{sec:E8}, the Seiberg-Witten curves for $Sp(1)$ theory with $N_f\le 7$ flavors were computed and show apparent $SO(2N_f)\times U(1)$ global symmetry. Moreover, their Weierstrass form exactly matches the known result computed for $SU(2)$ theory with same $N_f$ flavors revealing the global symmetry enhancement to $E_{N_f+1}$. 

Using the obtained $Sp(1)$ Seiberg-Witten curves for the $E_1$ and $\widetilde{E}_1$ theories, we then analyzed the phase structure of the curves to differentiate the two theories. In the 5d decompactification limit, we found various intriguing points, which can be summarized as follows:
\begin{itemize}
	\item The phase diagrams for the $E_1$ and $\widetilde{E}_1$ theories are clearly different which also lead to the two different effective coupling formulas as given in \eqref{eq:E1taueff} and in \eqref{eq:E1tildetaueff}.
	Their phase structure shows that two theories, in the strong coupling region, give rise to distinctive brane configurations. 
	
	\item In the weak coupling region, however, 
M5-brane configurations for the $E_1$ and $\widetilde{E}_1$ theories in the decompactification limit do not seem different.

	\item The phase diagram for 5d $Sp(1)$ theory with one flavor (the $E_2$ theory) also reveals consistent structure for the flavor decoupling to the $E_1$ and $\widetilde{E}_1$ theories (and even to the $E_0$ theory), and so does the effective coupling for the $E_2$ theory \eqref{eq:fulleff}. It naturally suggests new types of flop transitions (``generalized flop transition") in the presence of an O5-plane and each theory only allows distinctive generalized flop transitions which hence could serve as a characteristic feature that distinguishes one theory from the other. (See section \ref{sec:flop}.)
\end{itemize}

	We note that in the weak coupling region, the web configurations for the $E_1$ theory and the $\widetilde{E}_1$ theory in the decompactification limit look the same, as given in Figure \ref{Fig:E1Region1}, while the configurations for two  theories in the strong coupling region are very different. A natural question then arises whether two theories can be distinguished from the 5-brane web with an O5-plane in the weak coupling.
As the very same question arises for the brane configuration with O7$^-$-plane in the weak coupling was already address in \cite{Bergman:2015dpa}, it is instructive to discuss the O7$^-$-plane case to better understand the weak coupling brane configurations for pure $Sp(1)$ with an O5-plane.\footnote{We thank Oren Bergman for the illuminating discussion about analogy between a brane configuration with an O7$^-$-plane and a brane configurations with an O5-plane.}

Recall salient feature of a brane configuration with an O7$^-$-plane describing the 5d $\mathcal{N}=1$ pure $Sp(N)$ gauge theories with different discrete theta angle.  In a brane configuration for an $Sp(N)$ gauge theory in the ``strong coupling region'', 
an O7$^-$-plane 
is resolved and is decomposed into two $[p,q]$ 7-branes \cite{Sen:1996vd}. It is known that 
the discrete theta angle appears as two inequivalent decompositions of the O7$^-$-plane as discussed in \cite{Bergman:2015dpa}.
In the weak coupling region of the $Sp(N)$ gauge theory, on the other hand, an O7$^-$-plane 
is not resolved, which means that the difference on the discrete theta angle is not manifest from web diagram with an O7$^-$-plane.
However, this does not means that the theory itself cannot be distinguished: there is difference that is not manifest.
For example, the way that an O7$^-$-plane is resolved in the strong coupling region 
should be already encoded in the weak coupling region. In other words, the flow to the strong coupling regions, yielding a definite O7$^-$ resolution to two $[p,q]$ 7-branes, is unambiguous in the weak coupling brane configuration. One could even make a distinction between two types of O7$^-$-plane based on their inequivalent decompositions. When there is a flavor in the brane configuration, the monodromy cut of the D7-brane can generate $SL(2,\mathbb{Z})$  T-transformation which  converts one type of resolved $[p,q]$ 7-branes into the other type.

For the brane configuration with an O5-plane in the weak coupling region, the situation is partially analogous to the O7$^-$-plane case. In the weak coupling region, difference in their brane configurations for the $E_1$ theory and the $\widetilde{E}_1$ theory is not manifest, while their brane configurations in the strong region are very different. As in the O7$^-$-plane case, the flow from the strong coupling to the weak coupling encode the differences. For instance, 
as two theories are obtained by two different flavor decoupling limits of the $E_2$ theory, such different decouplings project out distinctive BPS spectra for two theories from the BPS spectrum of the $E_2$ theory. This means that the resultant BPS spectrum for each theory is different, regardless of the resulting web diagrams being similar or not in the weak coupling region. 
 In other words, if one carefully studies possible $(p,q)$ strings in brane configuration with an O5-plane, the $(p,q)$ string configurations for each case should be different. Likewise, the allowed boundary conditions for two O4-planes after T-duality are clearly different as explained in section \ref{sec:E1E1t}, and thus the states respecting such boundary conditions are of course different. One could even introduce a hidden labeling for the web diagram with an O5-plane to distinguish the $E_1$ theory from the $\widetilde{E}_1$ theory in the weak coupling region. 
We also note that adding a flavor to the $E_1$ and $\widetilde{E}_1$ web diagrams no longer distinguish the discrete theta angle because the two configurations with different theta angle can be connected through a generalized flop transition corresponding to Figure \ref{Fig:newflops} (b), like the monodromy transformation of the O7$^-$ case.

We also remark that generalized flop transitions in section \ref{sec:flop} support the transitions between brane configurations involving an O5-plane in \cite{Hayashi:2015vhy}, one of which is an S-dual of the brane configurations with an ON-plane, and the other is a natural brane configuration with an O5-plane.  

It would be interesting to generalize the analysis of the phase structure diagram for the case of more flavors which would lead to an explicit effective coupling formula, and 5d theories of other types of the orientifolds, such as $SO(N)$ gauge groups with flavors or D-type quiver theories, and also further to obtain 6d Seiberg-Witten curves.


\acknowledgments
We thank Eran Avraham, Oren Bergman, Kantaro Ohmori, Masato Taki, Kazuya Yonekura and Gabi Zafrir for useful discussions. 
SK is supported by UESTC Initial Research Grant A03017023801317. 
KL is supported in part  by the National Research Foundation of Korea Grant NRF-2017R1D1A1B06034369.
FY is supported in part by Israel Science Foundation under grant no. 352/13.
We would like to thank Physics and Geometry of F-theory 2017, 2017 Aspen Winter Conference ``Superconformal Field Theories
in $d \ge 4$'', Exceptional Groups as Symmetries of Nature `17, National Center for Theoretical Sciences
and Galileo Galilei Institute for Theoretical Physics where the part of this work is done.

\bigskip

\appendix 


\section{Seiberg-Witten curves for 5d $Sp(1)$ theory with $N_f\le 7$ flavors}
\label{sec:E8}
In section \ref{sec:E1E1t}, the method of obtaining Seiberg-Witten (SW) curve based on web diagram with an O5$^-$-plane was discussed. It is straightforward to generalize to gauge groups of higher rank as well as higher flavors. As an instructive example, we here consider 5d Seiberg-Witten curve for $Sp(1)$ theory with $N_f=7$ flavors. As the curve was obtained in \cite{Minahan:1997ch, Eguchi:2002nx,Huang:2013yta,Kim:2014nqa}, we check that its j-invariant agrees with the known result and also that a successive application of the flavor decoupling limit reproduces the curves for less flavors. 
\subsection{5d $Sp(1)$ theory with $N_f=7$ flavors}
\begin{figure}[t]
\centering
\includegraphics[width=9cm]{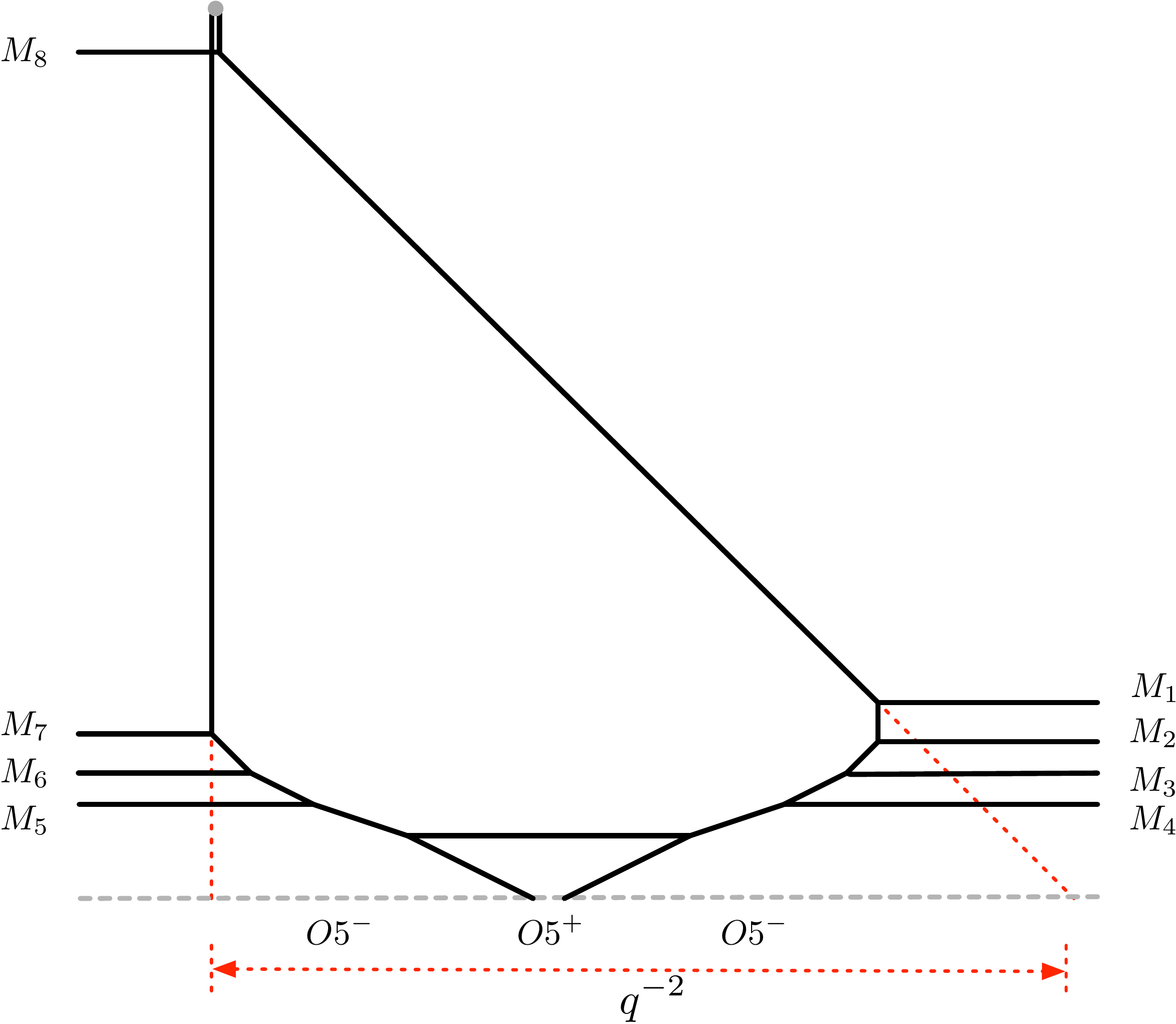}
\caption{5d $Sp(1)$ theory with $N_f=7$ flavors in a web diagram with O5-planes.}
\label{fig:O5Sp1Nf7}
\end{figure}
We now consider 5d $Sp(1)$ theory with $N_f=7$ flavors in a web diagram with O5-plane given in Figure \ref{fig:O5Sp1Nf7}.
For convenience, we introduce the following notation
\begin{align}
\chi_1 &= \sum_{i=1}^4 M_i, \qquad \chi_4 = \prod_{i=1}^4 M_i,\qquad \chi_3 = \chi_4 \sum_{i=1}^4 M_i^{-1},\cr
\widetilde\chi_1 &= \sum_{i=5}^8 M_i, \qquad \widetilde\chi_4 = \prod_{i=5}^8 M_i,\qquad \widetilde\chi_3 = \widetilde\chi_4 \sum_{i=5}^8 M_i^{-1},\cr
\chi_8 &=\chi_4\,\widetilde\chi_4,\qquad \chi^{\rm SO(16)}_1= \chi_1 + \chi_3 \,\chi_4^{-1}+\widetilde\chi_1 + \widetilde\chi_3\,\widetilde\chi_4^{-1}.
\end{align}
With the Ansatz for the Seiberg-Witten curve
\begin{align}
	&\quad t^2 \prod_{i=1}^{4} (w-M_i)(w^{-1}-M_i) 
	\cr&
	+c_1 t \Big(w^4+a w^3 + b w^2 + c w+ d + cw^{-1}+ bw^{-2}+ aw^{-3}+ w^{-4} \Big)\cr
	&+c_2 \prod_{i=5}^{8} (w-M_i)(w^{-1}-M_i)=0,
\end{align}
we impose two boundary conditions  (i) and (ii) :\\
(i) As $w\to\infty$, the leading term in $w$ is given by
\begin{align}
	(\chi_4 t^2+ c_1 t + c_2 \widetilde\chi_4 )w^4 = \chi_4(t-1)^2w^4,
\end{align}
which gives
\begin{align}
	c_1= -2 \chi_4,\qquad \qquad c_2= \chi_4\,\widetilde{\chi}_4^{-1}.
\end{align}
As discussed in \cite{Kim:2014nqa}, for the web diagram involving such jumping, the next leading term in $w$ should be proportional to $(t-1)$, which means \begin{align}
	\Big(-(\chi_3+ \chi_1\chi_4)t^2 + c_1 a t - c_2 (\widetilde\chi_3+ \widetilde\chi_1\widetilde\chi_4)\Big)w^3 \propto -(\chi_3+ \chi_1\chi_4)(t-1)(t-\alpha) w^3,
\end{align}
where $\alpha=\chi_4\widetilde{\chi}_4^{-1} (\widetilde\chi_1\widetilde\chi_4 + \widetilde\chi_3)(\chi_1\chi_4 + \chi_3)^{-1} $. 
From this, we find
\begin{align}
	a=-\frac12  \Big( \widetilde\chi_1+\widetilde\chi_4^{-1}\widetilde\chi_3 +\chi_1+ \chi_4^{-1}\chi_3 \Big)= -\frac12 \chi^{\rm SO(16)}_1.
\end{align}
The rescaling 
\begin{align}
t\to \frac{(\chi_4 \widetilde\chi_4^{-1})^\frac12}{ \prod_{i=1}^{4} (w-M_i)(w^{-1}-M_i)} t,
\end{align}
yields 
\begin{align}\label{eq:sp1nf7ans}
	&\quad t^2  
	-2\chi_8^\frac12 t \Big(w^4+a w^3 + b w^2 + c w+ d + cw^{-1}+ bw^{-2}+ aw^{-3}+ w^{-4} \Big)\cr
	&+\prod_{i=1}^{8} (w-M_i)(w^{-1}-M_i)=0.
\end{align}
(ii) We also require a double root at $w=1$ and $w=-1$ to take into account of the mirror image due to the O5-plane.
At $w=1$, \eqref{eq:sp1nf7ans} becomes
\begin{align}
	t^2-2\chi_8^\frac12 (2+ 2a+ 2b+ 2c+d)t+\prod_{i=1}^{8} (1-M_i)^2,
\end{align}
and we require it should be of a complete square form,
which means 
\begin{align}\label{eq:nf7bc1}
	\chi_8^\frac12 (2+ 2a+ 2b+ 2c+d) =  
	\prod_{i=1}^{8} (1-M_i),
\end{align}
where ($\pm$) signs are possible, but here we choose ($+$) sign\footnote{
If we denote the $(+, +)$-choice for the sign for  \eqref{eq:nf7bc1} and \eqref{eq:nf7bc2}, then there are four different choices: $(+,+)$,$(-,+)$ and $(-,-)$, $(+,-)$. The $(-, +)$-choice gives rise to the same Seiberg-Witten curve for the $(+, +)$-choice \eqref{eq:sp1nf7} but with $\chi_s\leftrightarrow\chi_c$, while the other choice $(-,-)$ (or $(+,-)$ ) leads to the Seiberg-Witten curve for the $(+, +)$-choice (or the $(-, +)$-choice) with a Wilson line that flips the instanton factor $q\to -q$, respectively, as discussed in section \ref{sec:E1E1t}. 
}

Likewise, at $w=-1$,  \eqref{eq:sp1nf7ans} becomes
\begin{align}
	t^2-2\chi_8^\frac12 (2- 2a+ 2b- 2c+d)t+\prod_{i=1}^{8} (1+M_i)^2,
\end{align}
and in order for this to be a complete square form, we have 
\begin{align}\label{eq:nf7bc2}
	\chi_8^\frac12 (2- 2a+ 2b- 2c+d) = 
	\prod_{i=1}^{8} (1+M_i).
\end{align}

It then follows from \eqref{eq:nf7bc1} and \eqref{eq:nf7bc2} that 
\begin{align}
	2\chi_8^\frac12 (2+ 2b+d)
	&= \prod_{i=1}^{8} (1-M_i)+\prod_{i=1}^{8} (1+M_i) =  2\chi_8^\frac12 \chi^{\rm SO(16)}_{\rm s},\cr
		4\chi_8^\frac12 (a+ c) 
	 &= \prod_{i=1}^{8} (1-M_i)- \prod_{i=1}^{8} (1+M_i) =- 2\chi_8^\frac12 \chi^{\rm SO(16)}_{\rm c}.
\end{align}
We then have all the parameters
\begin{align}
	a&= -\frac12 \chi_1^{\rm SO(16)}, \cr
	b&= U \quad{\rm (Coulomb~branch~modulus)},\cr
	c&=\frac12 \chi_1^{\rm SO(16)} -\frac12\chi^{\rm SO(16)}_{\rm c},\cr
	d&=-2-2U +\chi^{\rm SO(16)}_{\rm s}.
\end{align}
The SW curve for 5d $Sp(1)$ theory with $N_f=7$ flavors is given by
\begin{align}\label{eq:sp1nf7}
	& ~t^2  
	-2\chi_8^\frac12  \bigg[\big(w^4+ w^{-4}\big)-\frac12 \chi_1^{\rm SO(16)}\big( w^3+ w^{-3}\big) 
	+ U \big(w^2 + w^{-2}\big)\\
	&+ \frac12 \big(\chi_1^{\rm SO(16)} -
	\chi^{\rm SO(16)}_{\rm c}\big) \big(w+ w^{-1}\big) -2-2U +
	\chi^{\rm SO(16)}_{\rm s} \bigg]\, t
	+\prod_{i=1}^{8} (w-M_i)(w^{-1}-M_i)=0,\nonumber
\end{align}
where $M_8$ is related to the instanton factor as $M_8=q^{-2}$. 

Notice that this web configuration gives rise to a manifest $SO(16)$ symmetry, although naive global symmetry for $Sp(1)$ theory with $N_f=7$ flavors is $SO(14)\times U(1)_I$. This naive global symmetry is  enhanced to $E_8$ at the UV fixed point. As $SO(16)$ is a maximal compact subgroup of $E_8$, the brane configuration with an O5-plane already reveals partial enhancement of global symmetry
\begin{align}
SO(14)\times U(1)_I	\subset SO(16) \subset E_8. 
\end{align}
We also note that when the SW curve \eqref{eq:sp1nf7} is expressed as Weierstrass form, it coincides with the known Seiberg-Witten curve for $SU(2)$ theory with $N_f=7$ flavors \cite{Huang:2013yta,Eguchi:2002nx,Kim:2014nqa} which is written in terms of the $E_8$ characters. (We used a Mathematica package called ‘Susyno’ \cite{Fonseca:2011sy} to re-express the Seiberg-Witten curve written in terms of the $SO(16)$ characters into the curve written in terms of the $E_8$ characters.)

\subsection{Flavor decouplings}
\noindent\underline{$N_f=6$ SW curve:~}
Let us consider flavor decoupling limit. From the web diagram Figure \ref{fig:O5Sp1Nf7}, one finds that the $N_f=6$ configuration can be obtained by taking the limit where the physical masses to infinity which corresponds to the exponentiated masses $M_i= e^{-\beta m_i}$  to be 
\begin{align}
	M_1\to 0,\qquad {\rm while}~~\frac{M_8}{M_1} = q^{2}_{N_f=6}~~{\rm fixed},
\end{align}
where $q_{N_f=6}$ is the instanton factor for theory of $N_f=6$ flavors.
Given \eqref{eq:sp1nf7}, we perform this decoupling limit with the following redefinition ($q_{N_f=6}= q$) 
\begin{align}
U \to - \frac12 M_1^{-1}(1+ q^{-2}_{N_f=6})\,U,
\end{align}
which leads to the Seiberg-Witten curve for 5d $Sp(1)$ with $N_f=6$ flavors:
\begin{align}\label{eq:sp1nf6}
	& ~t^2  
	+\chi_1(q^2) \, \chi_6^\frac12 \bigg[\big( w^3+ w^{-3}\big) 
	+ U \big(w^2 + w^{-2}\big)\\
	&-  \big(1-\chi_1^{-1}(q^2) 
	\,\chi^{\rm SO(12)}_{\rm c}\big) \big(w+ w^{-1}\big) -2\big(U +
	\chi_1^{-1}(q^2)\,\chi^{\rm SO(12)}_{\rm s} \big)\bigg]\,t
	+\prod_{i=1}^{6} (w-M_i)(w^{-1}-M_i)=0,\nonumber 
\end{align}
where we have relabeled the mass parameters for the remaining six flavors to be $M_i$ ($i=1,\cdots, 6$) and 
\begin{align}
\chi_6  &= \prod_{i=1}^{6}M_i,  \qquad 
\chi_1(q^2)= q + q^{-1} =\chi_1(q)^2-2 ,
\end{align}
and $\chi_{\rm s}^{\rm SO(12)}$ and $\chi_{c}^{\rm SO(12)}$ are the characters for $SO(12)$ spinor and conjugate spinor representations, respectively
\begin{align}
\chi_{\rm s}^{\rm SO(12)} &= \frac12\,\chi_6^{-\frac12}\, \Big[ \prod_{i=1}^{6} (1+M_i) +\prod_{i=1}^{6} (1-M_i)\Big],\cr
\chi_{\rm c}^{\rm SO(12)} &=\frac12\, \chi_6^{-\frac12}\, \Big[ \prod_{i=1}^{6} (1+ M_i)-\prod_{i=1}^{6} (1-M_i) \Big].
\end{align}
Notice that the curve \eqref{eq:sp1nf6} is expressed in terms of not only $SO(12)$ but also $SU(2)_I \supset U(1)_I$ characters, hence it shows a manifest $SO(12)\times SU(2)$ symmetry which is a maximal compact subgroup of $E_7$,
\begin{align}
SO(12)\times U(1)_I	\subset SO(12)\times SU(2)_I\subset E_7.
\end{align}

As for lower flavors, one can take the same flavor decoupling limit,
\begin{align}
M_{N_f+1} \to 0 ~~\&~~q_{N_f+1} \to 0, \quad {\rm while~~} \frac{q^{2}_{N_f+1}}{M_{N_f+1}} = q^{2}_{N_f}~~{\rm fixed}.
\end{align}
The SW curve for $N_f\le 5$ flavors is then expressed in terms of with the characters 
\begin{align}
\chi_{N_f}  &= \prod_{i=1}^{N_f}M_i,\cr
\chi_{\rm s}^{SO(2N_f)} &= \frac12\,\chi_{N_f}^{-\frac12}\, \Big[\prod_{i=1}^{N_f} (1+M_i)+ \prod_{i=1}^{N_f} (1-M_i) \Big],\cr
\chi_{\rm c}^{SO(2N_f)} &=\frac12\, \chi_{N_f}^{-\frac12}\, \Big[\prod_{i=1}^{N_f} (1+M_i)- \prod_{i=1}^{N_f} (1- M_i) \Big].
\end{align}

 We list below the resulting Seiberg-Witten curves for the $Sp(1)$ theory with $N_f\le 5$ flavors which are written in terms of $SO(2N_f)\times U(1)_I$ characters. 
It is worth noting that all the Seiberg-Witten curves below agree with those given as the Weierstrass form expressed in terms of manifest $E_{N_f+1}$ characters given in 
\cite{Eguchi:2002nx,Huang:2013yta,Kim:2014nqa}. Therefore the Seiberg-Witten curves below shows that manifest $SO(2N_f)\times U(1)$ symmetry is in fact enhanced to $E_{N_f+1}$ symmetry.   \\
\noindent\underline{$N_f=5$~ SW curve:~}
\begin{align}\label{eq:sp1nf5}
	& ~t^2  
	+ q^{-1}\, \chi_5^\frac12\bigg[\big( w^3+ w^{-3}\big) 
	+ U \big(w^2 + w^{-2}\big)\\
	&-  \big(1- q\,
	\chi^{\rm SO(10)}_{\rm c}\big) \big(w+ w^{-1}\big) -2\big(U +
	 q\,\chi^{\rm SO(10)}_{\rm s} \big)\bigg]\,t
	+\prod_{i=1}^{5} (w-M_i)(w^{-1}-M_i)=0,\nonumber 
\end{align}
where $q=q_{N_f=5}$ is the instanton factor for the $Sp(1)$ theory with  $N_f=5$ flavors.

\noindent\underline{$N_f=4$~ SW curve:~}
\begin{align}\label{eq:sp1nf4}
	& ~t^2  
	+ q^{-1}\, \chi_4^\frac12\bigg[\big( w^3+ w^{-3}\big) 
	+ U \big(w^2 + w^{-2}\big)\\
	&-  \big(1-q^{} 
	\chi^{\rm SO(8)}_{\rm c}\big) \big(w+ w^{-1}\big) -2\big(U 
	+q^{}\chi^{\rm SO(8)}_{\rm s} \big)\bigg]\,t 
	+\prod_{i=1}^{4} (w-M_i)(w^{-1}-M_i)=0,\nonumber
\end{align}
where $q=q_{N_f=4}$ is the instanton factor for the $Sp(1)$ theory with $N_f=4$ flavors.

\noindent\underline{$N_f=3$~ SW curve:~}
\begin{align}\label{eq:sp1nf3}
	& ~t^2  
	+ q^{-1}\, \chi_3^\frac12\bigg[\big( w^3+ w^{-3}\big) 
	+ U \big(w^2 + w^{-2}\big)\\
	&-  \big(1-q^{} 
	\chi^{\rm SO(6)}_{\rm c}\big) \big(w+ w^{-1}\big) -2\big(U
	+ q^{}\chi^{\rm SO(6)}_{\rm s} \big)\bigg]\,t 
	+\prod_{i=1}^{3} (w-M_i)(w^{-1}-M_i)=0,\nonumber
\end{align}
where $q=q_{N_f=3}$ is the instanton factor for the $Sp(1)$ theory with $N_f=3$ flavors.

\noindent\underline{$N_f=2$~ SW curve:~}
\begin{align}\label{eq:sp1nf2}
	& ~t^2  
	+ q^{-1}\, \chi_2^\frac12\bigg[\big( w^3+ w^{-3}\big) 
	+ U \big(w^2 + w^{-2}\big)\\
	&-  \big(1- q^{} 
	\chi^{\rm SO(4)}_{\rm c}\big) \big(w+ w^{-1}\big) -2\big(U 
	+ q^{}\chi^{\rm SO(4)}_{\rm s} \big)\bigg]\,t 
	+\prod_{i=1}^{2} (w-M_i)(w^{-1}-M_i)=0, \nonumber
\end{align}
where $q=q_{N_f=2}$ is the instanton factor for the $Sp(1)$ theory with $N_f=2$ flavors.

\noindent\underline{$N_f=1$~ SW curve:~}
\begin{align}\label{eq:sp1nf1}
	& ~t^2  
	+ q^{-1}\, M_1^\frac12\bigg[\big( w^3+ w^{-3}\big) 
	+ U \big(w^2 + w^{-2}\big)\\
	&-  \big(1- q^{} 
	\chi^{\rm SO(2)}_{\rm c}\big) \big(w+ w^{-1}\big) -2\big(U+ 
	 q^{}\chi^{\rm SO(2)}_{\rm s} \big)\bigg]\,t 
	+ (w-M_1)(w^{-1}-M_1)=0,\nonumber
\end{align}
where $q=q_{N_f=1}$ is the instanton factor for the $Sp(1)$ theory with $N_f=1$ flavors and
\begin{align}
\chi_{\rm c}^{\rm SO(2)} = M_1^{\frac12},\qquad
\chi_{\rm s}^{\rm SO(2)} = M_1^{-\frac12}.
\end{align}

\noindent\underline{$N_f=0$, $E_1$ SW curve:~} From \eqref{eq:sp1nf1}, we take the limit
\begin{align}
M_{1} \to \infty ~~\&~~q_{N_f=1} \to 0, \quad {\rm while~~} \frac{q^{-2}_{N_f=1}}{M_{1}} = q^{-2}_{{E}_1}~~{\rm fixed}.	
\end{align} 
With the rescaling $t\rightarrow m_1 t$, one finds 
\begin{align}
		t^2 + q^{-1}_{E_1}\Big[ (w^3+w^{-3} )+ U (w^2+w^{-2}) -(1-q_{E_1})(w+w^{-1})- 2U\Big] t + 1=0,
\end{align}
which agrees with the $E_1$ curve \eqref{eq:E1curve}, as expected.

\noindent\underline{$N_f=0$, $\widetilde{E}_1$ SW curve:} From \eqref{eq:sp1nf1}, we take the limit
\begin{align}
M_{1} \to 0 ~~\&~~q_{N_f=1} \to 0, \quad {\rm while~~} 
 \frac{q^{2}_{N_f=1}}{M_{1}} = q^{2}_{\widetilde{E}_1}~~{\rm fixed}.
\end{align} 
This then yields
\begin{align}\label{eq:E1tildeapp}
	t^2 + q^{-1}_{\widetilde{E}_1} \Big[(w^3+w^{-3})  + U (w^2+ w^{-2})-(w+w^{-1}) -2(U+q_{\widetilde{E}_1})\Big] t + 1=0,
\end{align}
which is the $\widetilde{E}_1$ curve  \eqref{eq:naivesp1curve}.

\noindent\underline{${E}_0$ SW curve:} For ${E}_0$ theory, one needs to take a special limit on the $\widetilde{E}_1$ curve \eqref{eq:E1tildeapp}. As discussed in section \ref{sec:FlowfromE2}, it is the limit that takes the $\widetilde{E}_1$ brane web given in Figure \ref{Fig:E1tildeRegion3} away from the O5-plane, while keeping the area associated with the Coulomb modulus. It corresponds to taking $q$ and $U$ on \eqref{eq:E1tildeapp} very large while $U^3q^{-1}$ fixed 
\begin{align}
	q_{\widetilde{E}_1} \rightarrow L^3,\qquad
	U_{\widetilde{E}_1} \rightarrow L \,U_{E_0},\qquad {\rm where }\quad L\rightarrow \infty.
\end{align}
with the rescaling $w+w^{-1}\rightarrow L \,\big( w+w^{-1}\big)$. 
This yields 
\begin{align}
	t^2 +  \Big[(w^3+w^{-3})  + U (w^2+ w^{-2}) +3(w+ w^{-1}) -2(1-U)\Big] t + 1=0,
\end{align}
or in terms of $x= w+w^{-1}$, the $E_0$ Seiberg-Witten curve is written as
\begin{align}
	t^2	+\big( x^3 + U\, x^2-2\big)\,t +1=0.
\end{align}


\section{5-brane web diagrams and phase structure of $E_2$}\label{sec:E2Web}

In this appendix, we first see the consitency of the phase diagram of the $E_2$ theory obtained in section \ref{sec:E2} with the phase diagram in \cite{Morrison:1996xf} which has been obtained by using 5-brane web diagrams of the $E_2$ theory without an O5-plane. Appendix \ref{sec:detailE2} summarizes the detailed structure of the 5-brane web diagrams of the $E_2$ theory with an O5-plane obtained by using the Seiberg-Witten curves in section \ref{sec:E2Phase}.

\subsection{Phase diagram for the 5-brane web without an O5-plane of the $E_2$ theory}
In section \ref{sec:E2}, we have discussed the phase diagram of the $E_2$ theory through the 5-brane web diagram with an O5-plane. The $E_2$ theory can be also described by a 5-brane web without an O5-plane, which we can regard as a 5-brane web for an $SU(2)$ gauge theory with one flavor rather than the $Sp(1)$ gauge theory with one flavor. Therefore it is instructive to see the relation of the phase diagram obtained from the 5-brane web with an O5-plane and the phase diagram obtained from the 5-brane web diagram without an O5-plane. 

\begin{figure}
\centering
\subfigure[Phase diagram]{\includegraphics[width=7cm]{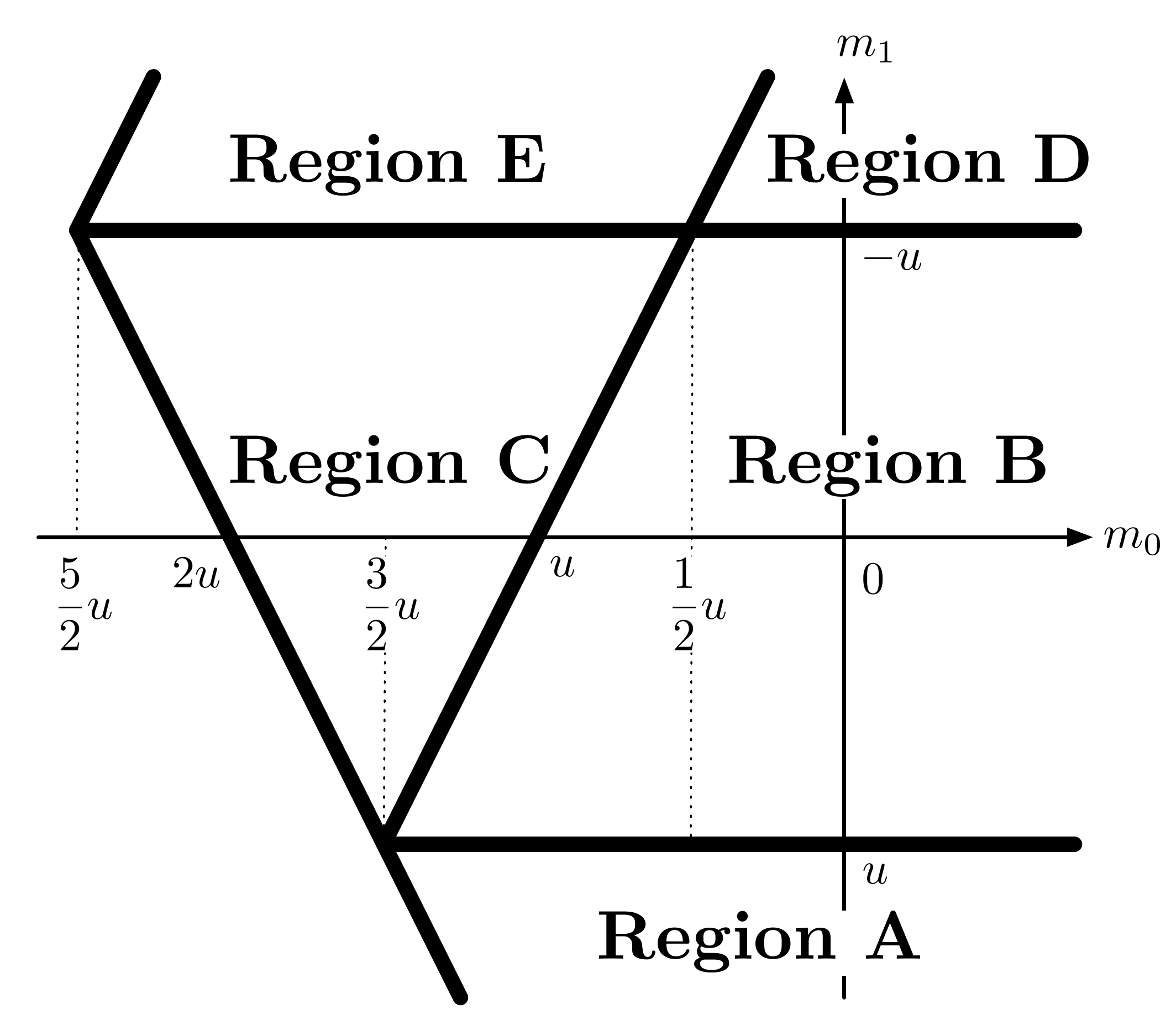}}\\
\subfigure[Region A]{\includegraphics[width=2cm]{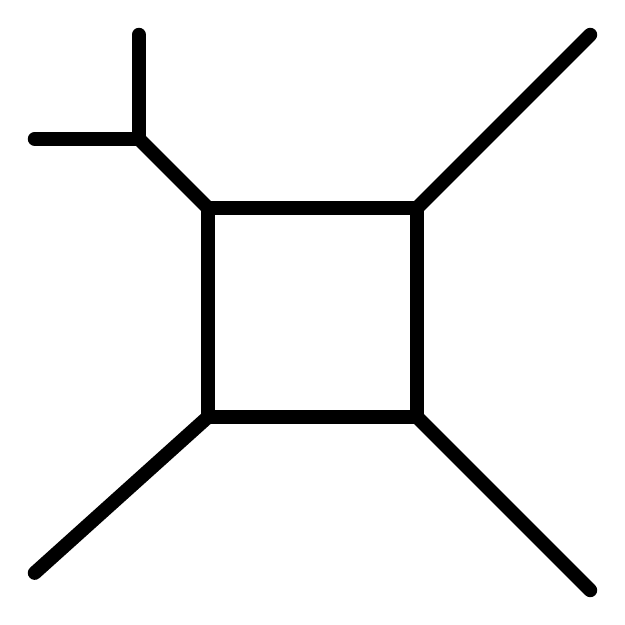}}
\subfigure[Region B]{\includegraphics[width=2cm]{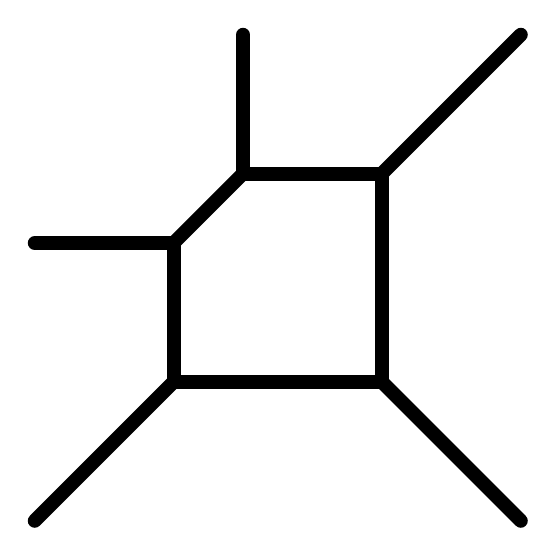}}
\subfigure[Region C]{\includegraphics[width=2cm]{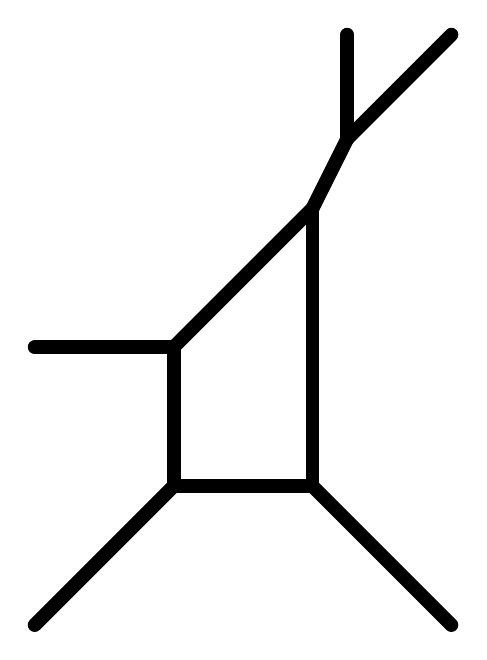}}
\subfigure[Region D]{\includegraphics[width=2cm]{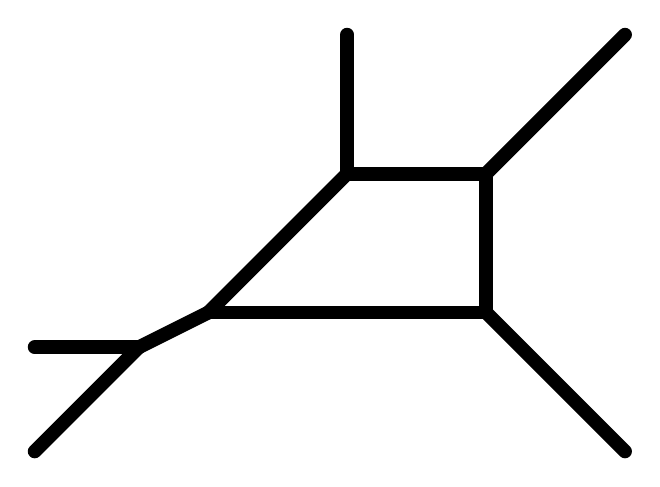}}
\subfigure[Region E]{\includegraphics[width=2cm]{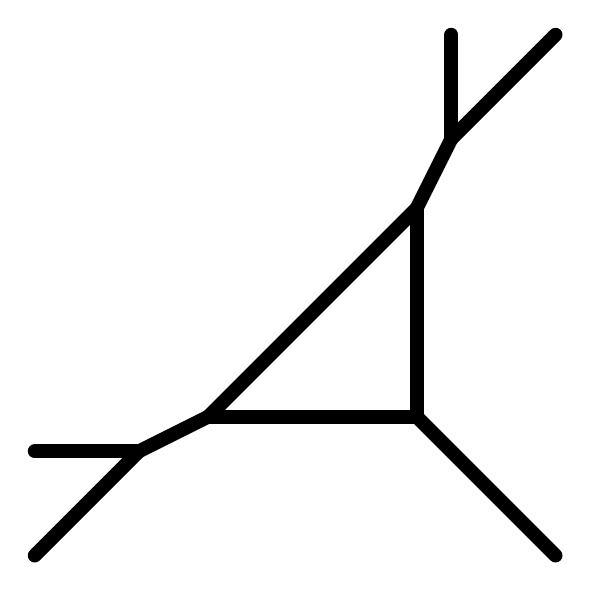}}
\caption{The phase diagram and the corresponding 5-brane web diagrams of the $E_2$ theory when we use a 5-brane web without an O5-plane.}
\label{Fig:SU2Phase}
\end{figure}

We have seen in section \ref{sec:effE2} that some regions for the $E_2$ theory are characterized by the same effective coupling. When we write down the corresponding 5-brane web diagram without an O5-plane, those regions with the same effective coupling should give rise to the same 5-brane web diagram. Whenever the effective coupling changes, the transition is accompanied by a flop transition in terms of the 5-brane web without an O5-plane. Therefore, we rename the regions so that the regions with the same effective coupling becomes the same region. Therefore. we denote the Region 2, 3 by Region A, the Region 1, 4, 6 by Region B, Region 5, 9, 10 by Region C, the Region 7 by Region D and the Region 8 by Region E. Then the phase diagram for the 5-brane web for the $SU(2)$ gauge theory with one flavor is given in Figure \ref{Fig:SU2Phase} (a).
Note that some resgions are combined together compared to the phase diagram in Figure \ref{Fig:E2Phase3}. 

Furthermore, we know that the Region 2 and the Region 3 flows to the $E_1$ theory when $|m_1| \to \infty$ and the Region 7 and 8 become the two phases of the $\widetilde{E}_1$ theory when $|m_1| \to \infty$. Therefore, in terms of the 5-brane web without an O5-plane, the 5-brane diagram in those regions should be consistent with the decoupling behavior. These features turns out to uniquely determine the 5-brane web diagram without an O5-plane in each region of the Region A, B, C, D, E. The results are summarized in Figure \ref{Fig:SU2Phase} (b) -- (f) \cite{Bergman:2013ala, Aharony:1997ju}.

In fact the phase diagram in Figure \ref{Fig:SU2Phase} (a) is completely consistent with the phase diagram obtained in \cite{Morrison:1996xf} where the same phase diagram is obtained both from the field theory and from geometry of the corrsponding Calabi--Yau threefold which is given by a local Calabi--Yau threefold with a compact surface of blowing up two points in $\mathbb{P}^2$. 

\begin{figure}
\centering
\includegraphics[width=9cm]{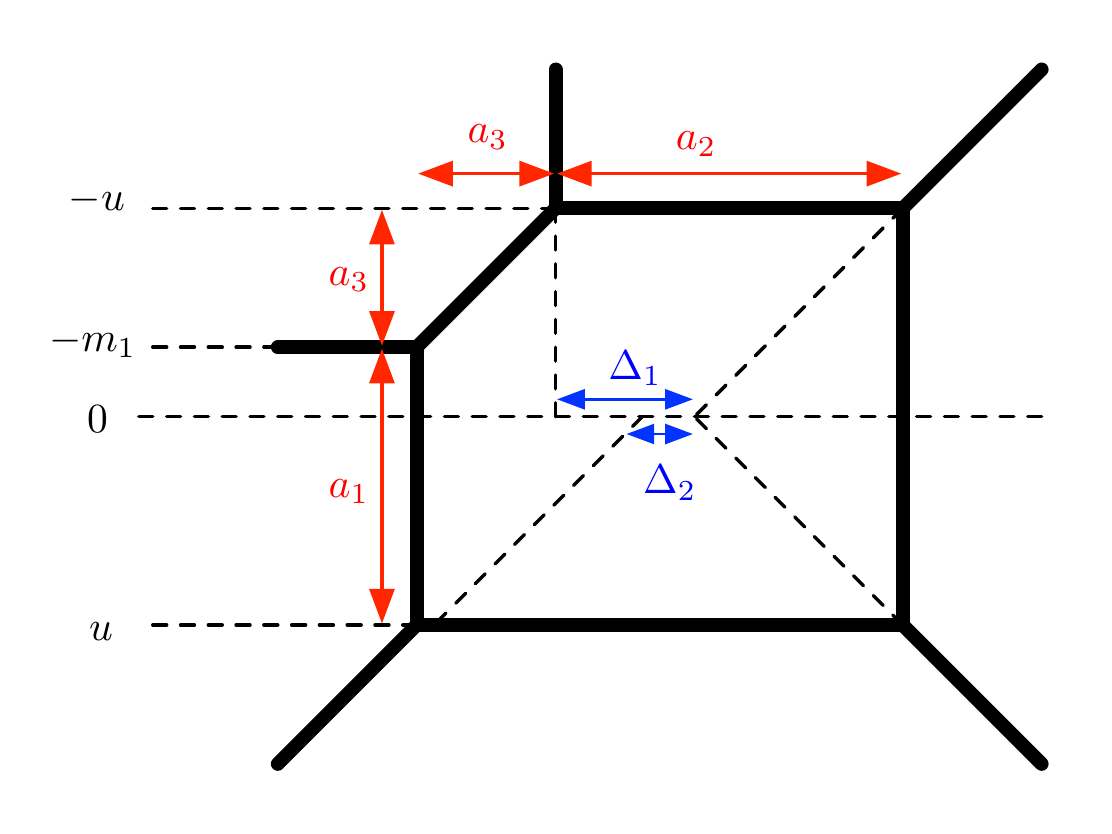}
\caption{A parameterization of a 5-brane web diagram of the $E_2$ theory.}
\label{Fig:SU2parameters}
\end{figure}

Due to the relation between a toric variety and a 5-brane web diagram in \cite{Leung:1997tw}, we can understand the phase structure of the $SU(2)$ gauge theory with one flavor from the 5-brane web diagram without an O5-plane realizing the $SU(2)$ gauge theory with one flavor. For example, we can start from the 5-brane web diagram in Figure \ref{Fig:SU2Phase} (c). The we parameterize the web as in Figure \ref{Fig:SU2parameters}. 
While $a_1, a_2, a_3$ are lengths of the corresponding 5-branes (or the volumes of the two-cycles in the dual Calabi--Yau threefold), from the 5-brane web diagram, we can read off the gauge theory parameters. The parameter $u$ and $m_1$ in Figure \ref{Fig:SU2parameters} are the Coulomb branch modulus and the mass parameter for the one flavor respectively. While the inverse of the gauge coupling squared represented by $m_0$ is given by $m_0 = \frac{1}{2}(\Delta_1 + \Delta_2)$. Then the lengths of the 5-branes can be written by the gauge theory parameters as
\be
a_1 = - u - m_1, \quad a_2 = \frac{1}{2}(2m_0 - 2u - m_1), \quad a_3 = -u + m_1. \label{eq:geometrygauge}
\ee

Since the lengths of the 5-branes should be positive, the phase describing the diagram in Figure \ref{Fig:SU2parameters} should satisfy
\be
a_1 > 0 , \quad a_2 > 0, \quad a_3 > 0. \label{eq:SU2w1flvrsPhase1}
\ee
The flop transition with repsect to a two-cycle with the size $a_3$ yields the diagram in Figure \ref{Fig:SU2Phase} (b). Then the parameter region after the flop transition becomes
\be
a_1 + a_2 > 0, \quad a_2 + a_3 > 0, \quad -a_3 > 0.\label{eq:SU2w1flvrsPhase2}
\ee
On the other hand, it is also possible to consider other two flops from the 5-brane web in Figure \ref{Fig:SU2parameters}. We can perform a flop transition with respect to a two-cycle with the size $a_2$, yeilding the 5-brane web in Figure \ref{Fig:SU2Phase} (d) or a flop transition with respect to a two-cycle with the size $a_1$, yielding the 5-brane web in Figure \ref{Fig:SU2Phase} (e) . Then the parameters region in each case becomes
\be
a_1 > 0, \quad -a_2 > 0, \quad a_2 + a_3 > 0 \label{eq:SU2w1flvrsPhase3}
\ee
for the 5-brane web in Figure \ref{Fig:SU2Phase} (d) and 
\be
-a_1 > 0, \quad a_2 > 0, \quad a_1 + a_3 > 0\label{eq:SU2w1flvrsPhase4}
\ee
for the 5-brane web in Figure \ref{Fig:SU2Phase} (e). From the 5-brane web in Figure \ref{Fig:SU2Phase} (d), we can further perform a flop transition with respect to a two-cycle with the size $a_1$ and it gives the 5-brane web in Figure \ref{Fig:SU2Phase} (f). The parameter region in this case is 
\be
-a_1 > 0, \quad -a_2 > 0, \quad a_1 + a_2 + a_3 > 0.\label{eq:SU2w1flvrsPhase5}
\ee
When we rewrite the regions in all the phases \eqref{eq:SU2w1flvrsPhase1} -- \eqref{eq:SU2w1flvrsPhase5} by using the gauge theory parameters in \eqref{eq:geometrygauge}, the five regions are described by 
\begin{align}\label{eq:SU2Phases}
\text{Region } A &: \quad m_1 < 0, \quad m_1 < u < 0, \quad m_0 > -\frac{1}{2}m_1 + 2u,\cr
\text{Region } B &: \quad u < 0, \quad u < m_1 < -u, \quad m_0 > \frac{1}{2}m_1 + u,\cr
\text{Region } C &: \quad u < 0, \quad u < m_1 < -u, \quad -\frac{1}{2}m_1 + 2u < m_0 < \frac{1}{2}m_1 + u,\cr
\text{Region } D &: \quad m_1 > 0, \quad -m_1 < u < 0, \quad m_0 > \frac{1}{2}m_1 + u,\cr
\text{Region } E &: \quad m_1 > 0, \quad -m_1 < u < 0, \quad \frac{1}{2}m_1 + 3u < m_0 < \frac{1}{2}m_1 + u.
\end{align}
Then the phase diagram by using \eqref{eq:SU2Phases} exactly reproduces the phase diagram in Figure \ref{Fig:SU2Phase}. Therefore, our analysis of the $E_2$ theory from the 5-brane with an O5-plane is completely consistent with the analysis of the phase structure of the $E_2$ theory in \cite{Morrison:1996xf}.

By using the parameterization \eqref{eq:geometrygauge} as well as the relation of the flop transitions, it is also possible to obtain the effective coupling for the 5-brane web diagram in Figure \ref{Fig:SU2Phase} (b) -- (f). The result is summarized in Table \ref{Table:SU2eff}. 
\begin{table}[t]
\centering
\begin{tabular}{|c|c|c|c|c|c|}
\hline
&Region $A$ & Region $B$ & Region $C$ & Region $D$ & Region $E$ \\
\hline
$\tau_{\text{eff}}$ & $2m_0 + m_1 - 8u$ & $2m_0 - 7u$ & $3m_0 - \frac{m_1}{2} - 8u$ & $2m_0 - m_1 - 8u$ & $3m_0 - \frac{3 m_1}{2} - 9u$ \\
\hline
\end{tabular}
\caption{The effective coupling for the $E_2$ theory realized by the 5-brane web diagrams in Figure \ref{Fig:SU2Phase} (b) -- (f).}
\label{Table:SU2eff}
\end{table}
Again, the result completely reproduces the effective coupling calculated from the 5-brane web diagrams with an O5-plane in Table \ref{Table:effE2Phase1} and \ref{Table:effE2Phase2}.

\subsection{Detailed structure of the 5-brane web with an O5-plane of the $E_2$ theory}
\label{sec:detailE2}
\begin{figure}
\centering
\begin{minipage}{0.48\hsize}
\includegraphics[width=7.4cm]{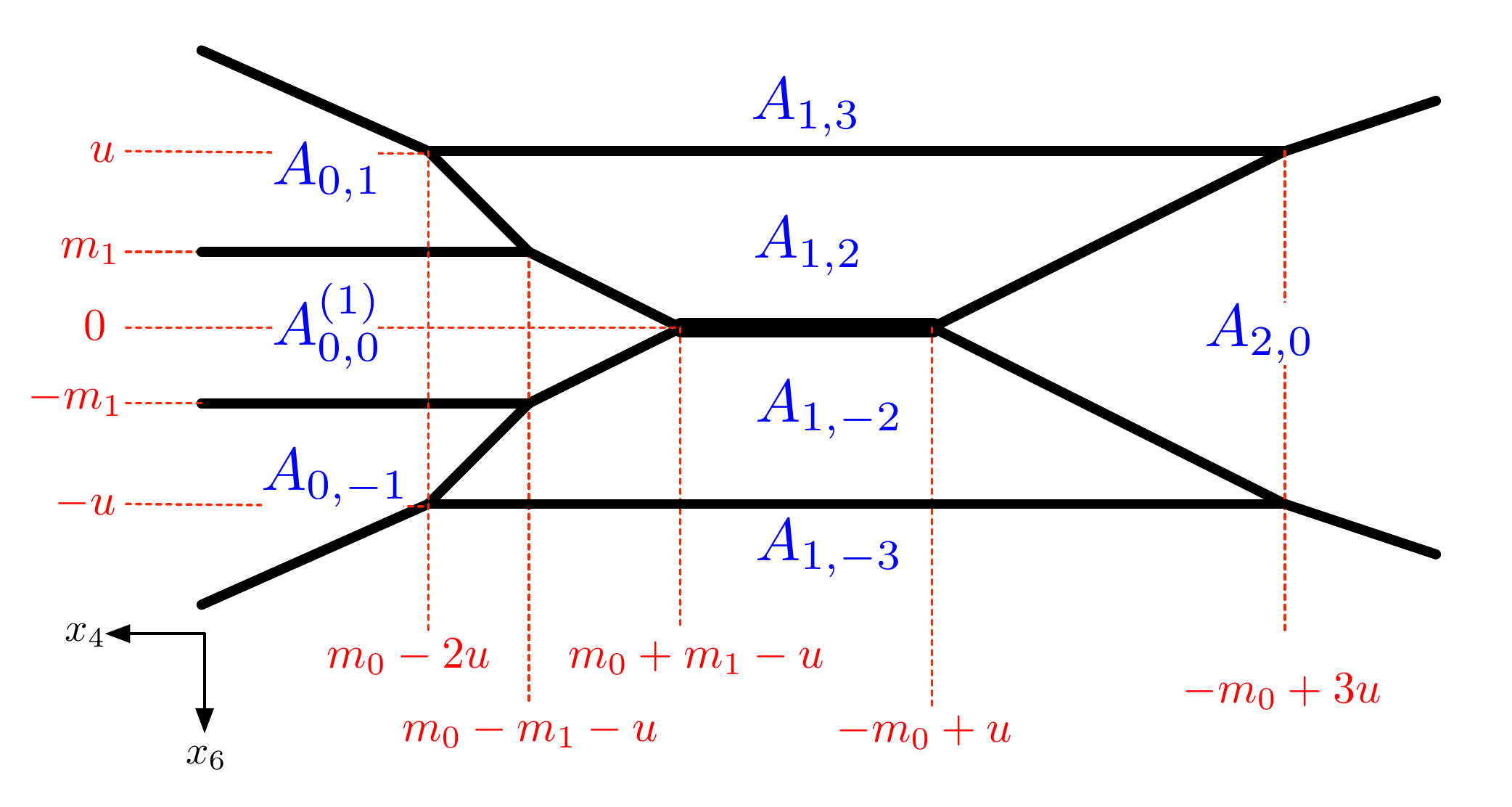}
\vspace{-1cm}
\caption{The regions for the largest $A_{k, l}^{(m)}$ in Region 1 of the $E_2$ theory. }
\label{Fig:E2Region1}
\vspace{1cm}
\end{minipage}
\hspace{0.02\hsize}
\begin{minipage}{0.48\hsize}
\includegraphics[width=7.4cm]{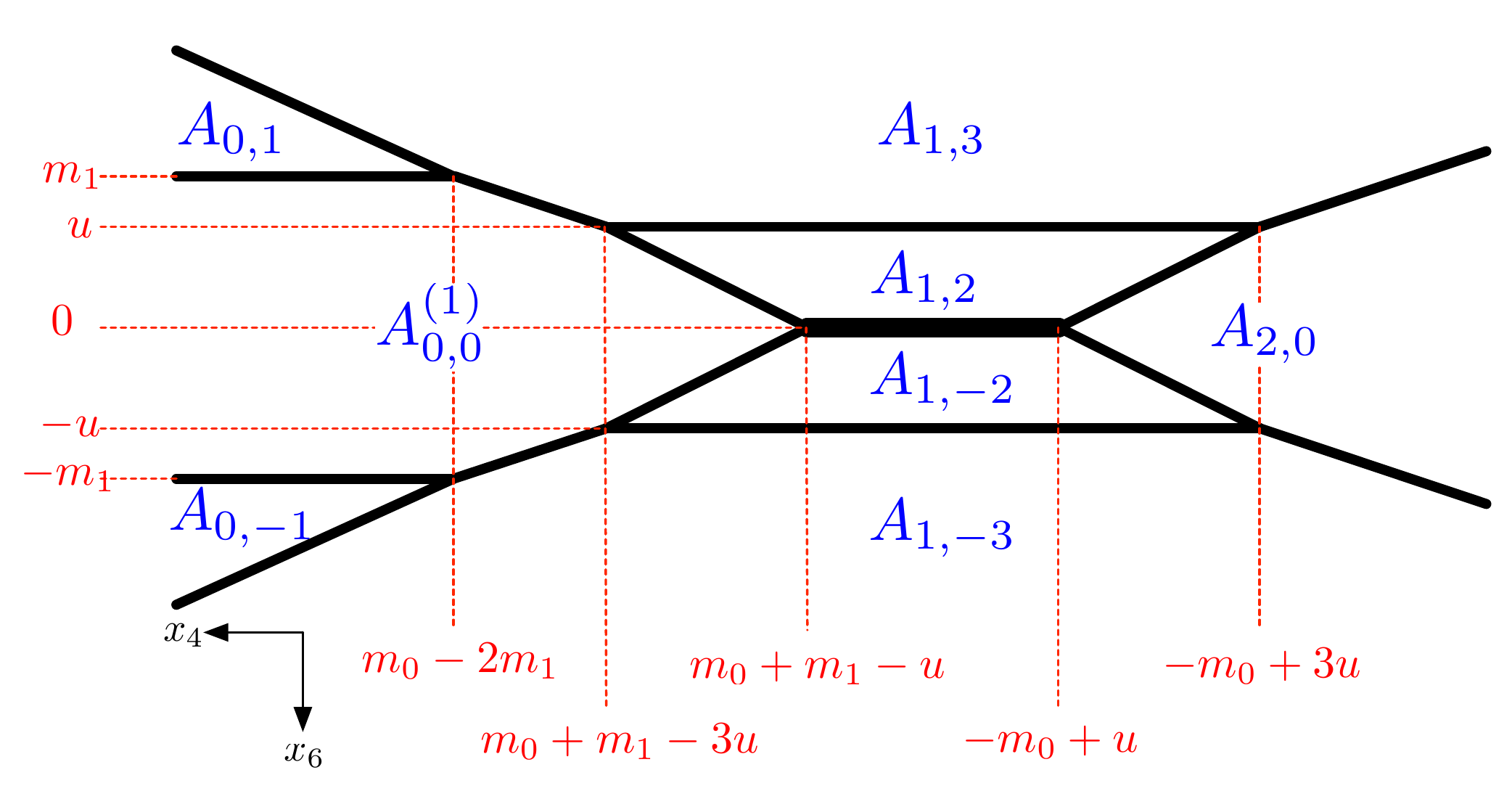}
\vspace{-1cm}
\caption{The regions for the largest $A_{k, l}^{(m)}$ in Region 2 of the $E_2$ theory. }
\label{Fig:E2Region2}
\vspace{1cm}
\end{minipage}
\begin{minipage}{0.48\hsize}
\includegraphics[width=7.4cm]{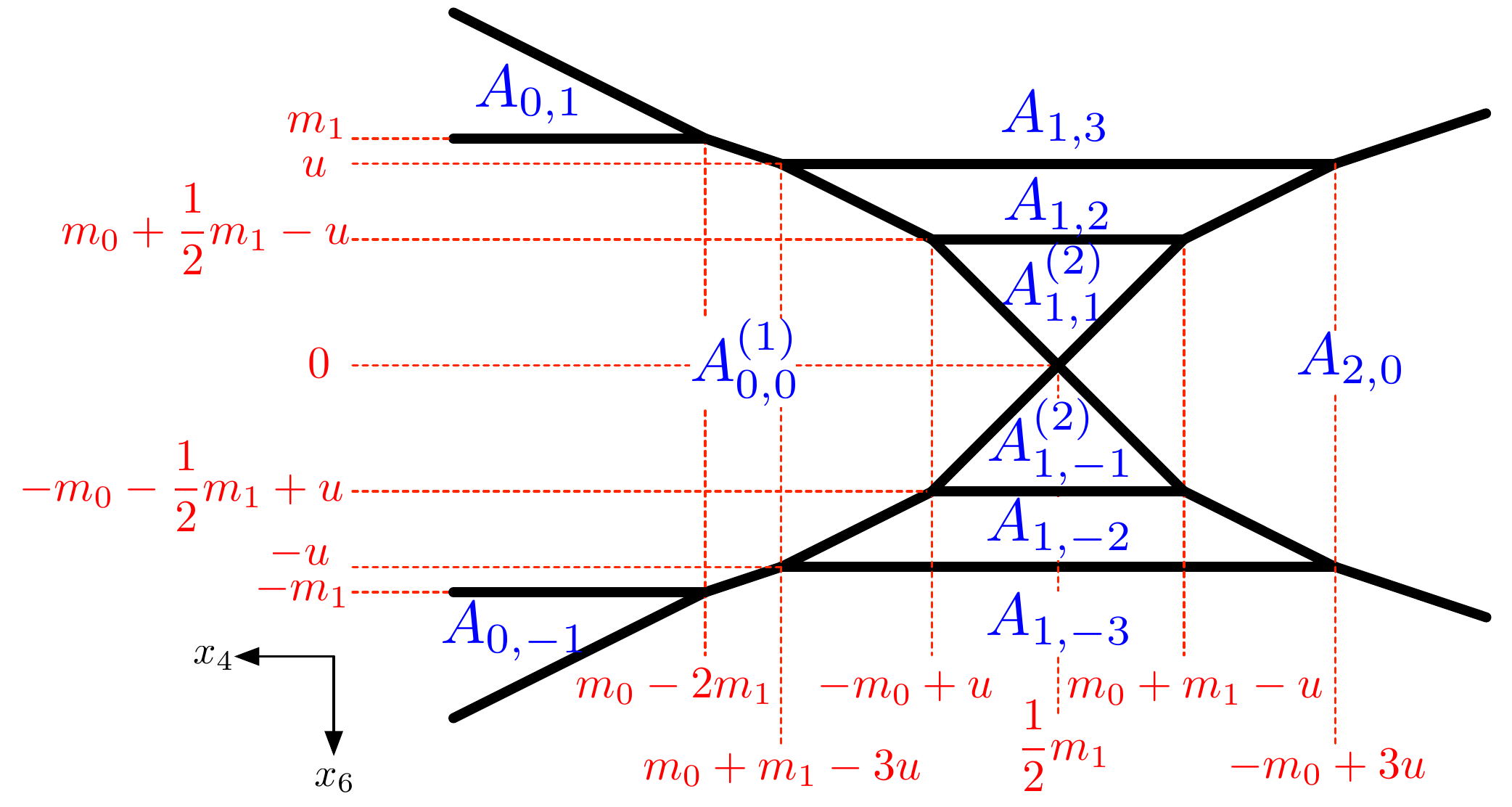}
\vspace{-1cm}
\caption{The regions for the largest $A_{k, l}^{(m)}$ in Region 3 of the $E_2$ theory. }
\label{Fig:E2Region3}
\vspace{1cm}
\end{minipage}
\hspace{0.02\hsize}
\begin{minipage}{0.48\hsize}
\includegraphics[width=7.4cm]{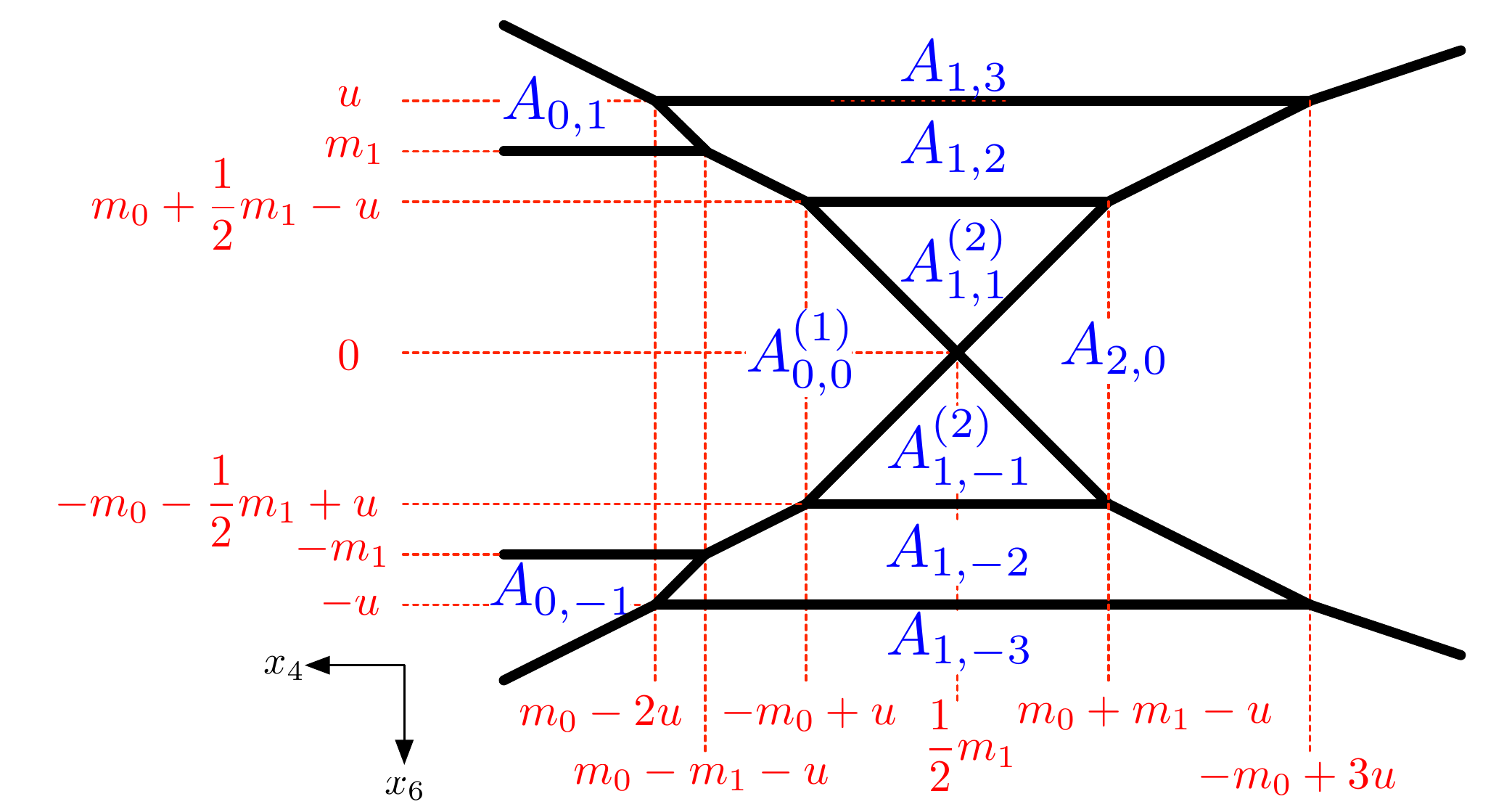}
\vspace{-1cm}
\caption{The regions for the largest $A_{k, l}^{(m)}$ in Region 4 of the $E_2$ theory. }
\label{Fig:E2Region4}
\vspace{1cm}
\end{minipage}
\begin{minipage}{0.48\hsize}
\includegraphics[width=7.4cm]{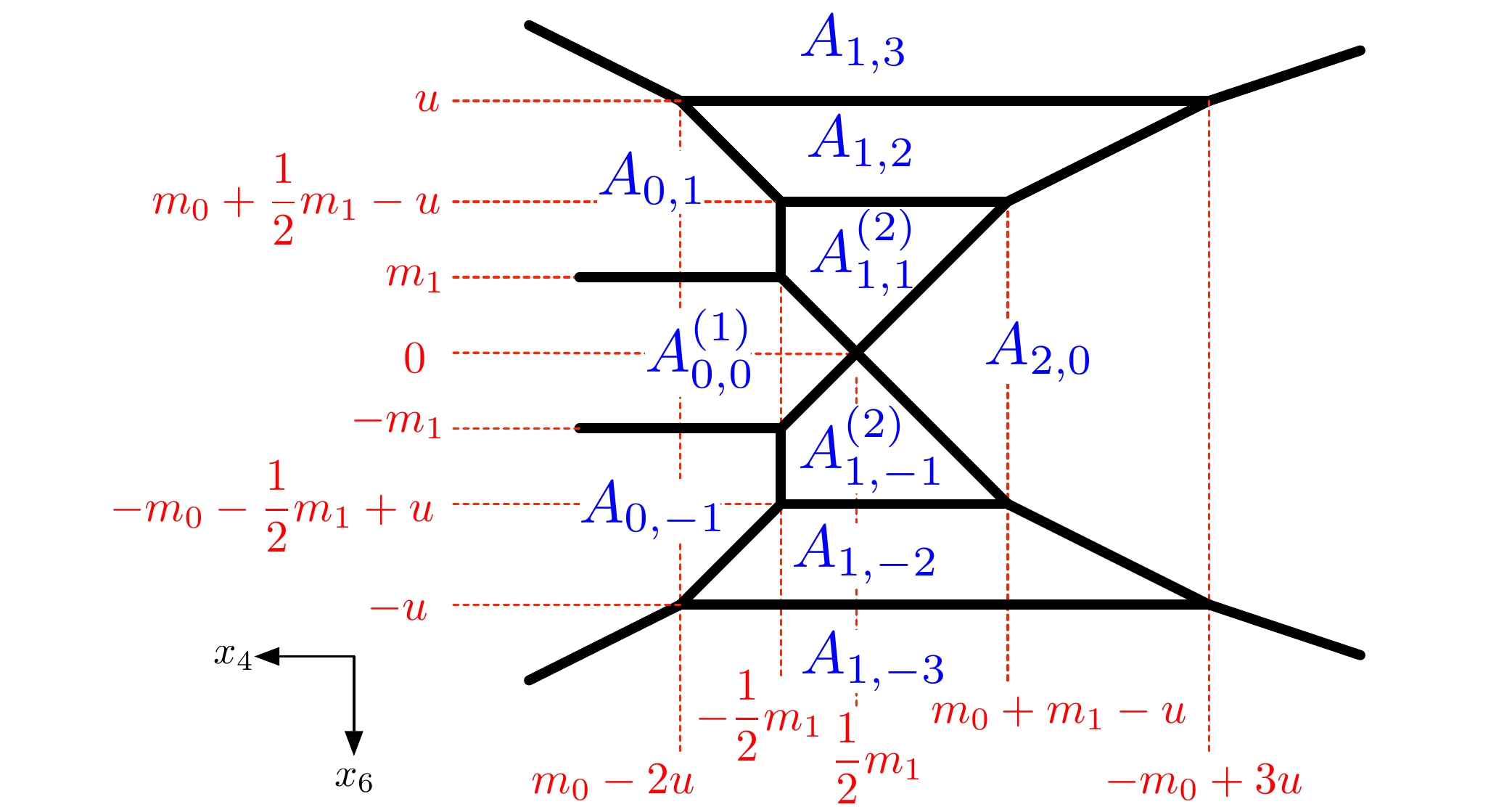}
\vspace{-1cm}
\caption{The regions for the largest $A_{k, l}^{(m)}$ in Region 5 of the $E_2$ theory. }
\label{Fig:E2Region5}
\vspace{1cm}
\end{minipage}
\hspace{0.02\hsize}
\begin{minipage}{0.48\hsize}
\includegraphics[width=7.4cm]{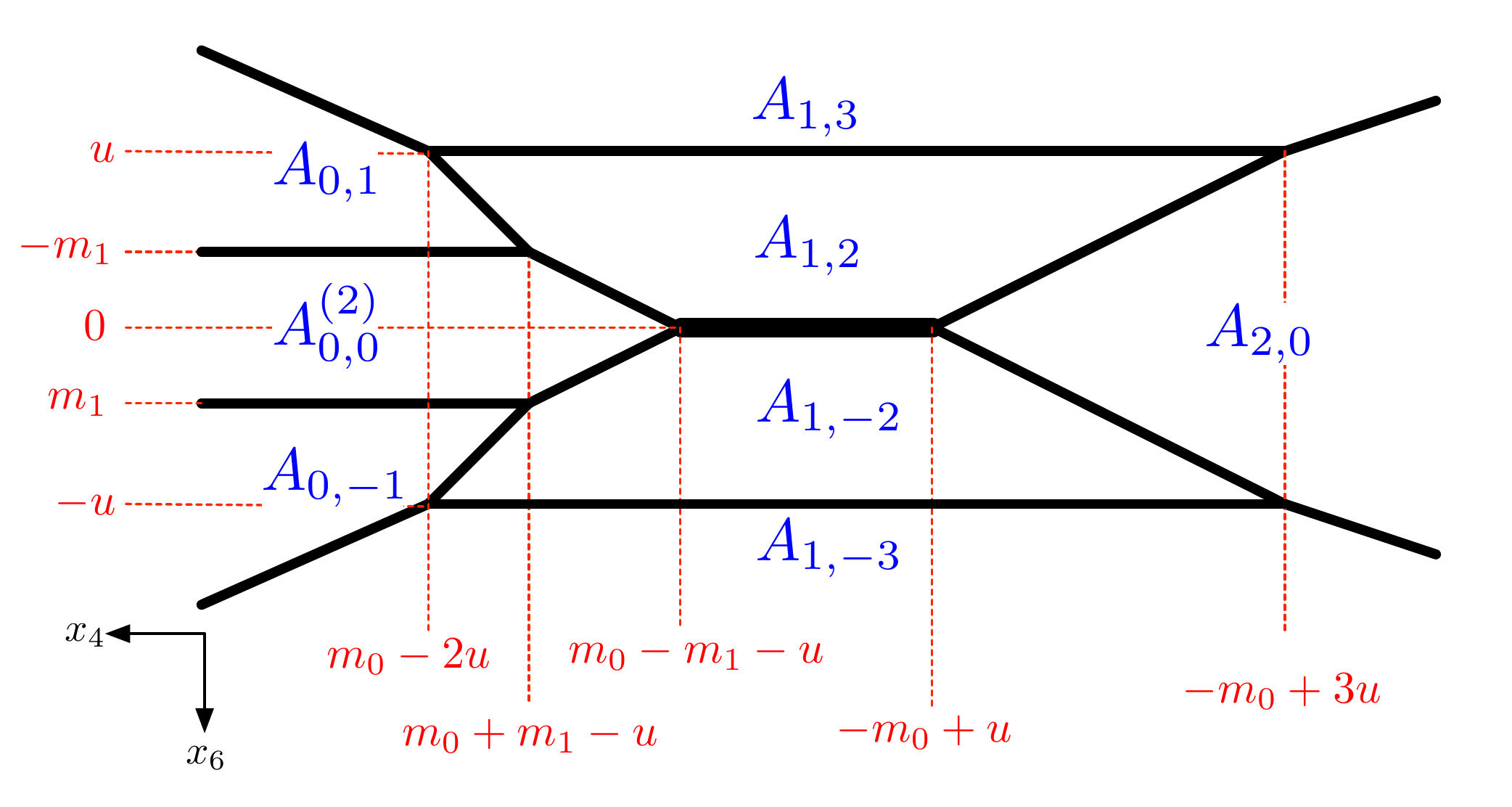}
\vspace{-1cm}
\caption{The regions for the largest $A_{k, l}^{(m)}$ in Region 6 of the $E_2$ theory. }
\label{Fig:E2Region6}
\vspace{1cm}
\end{minipage}
\begin{minipage}{0.48\hsize}
\includegraphics[width=7.4cm]{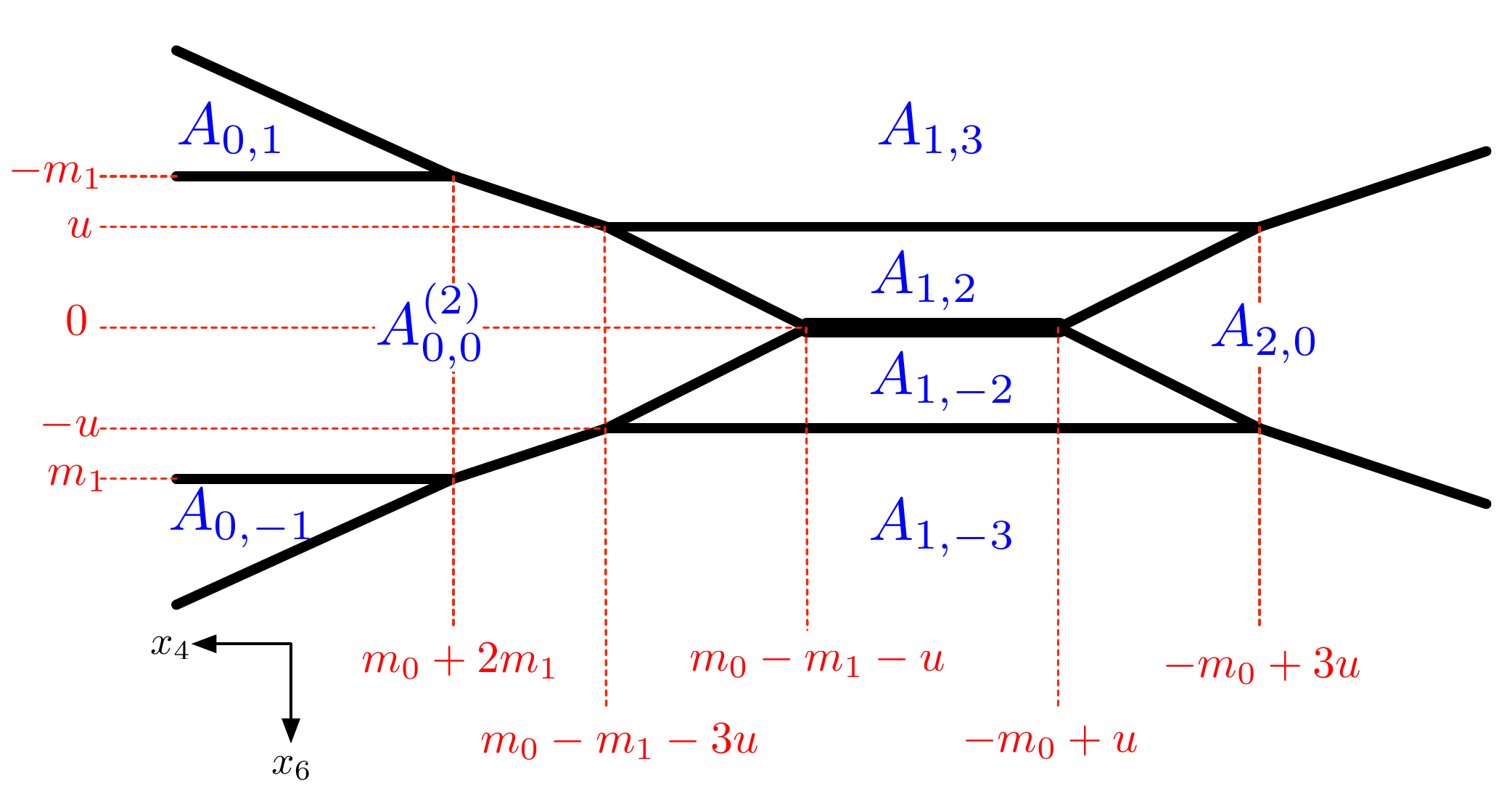}
\vspace{-1cm}
\caption{The regions for the largest $A_{k, l}^{(m)}$ in Region 7 of the $E_2$ theory. }
\label{Fig:E2Region7}
\end{minipage}
\hspace{0.02\hsize}
\begin{minipage}{0.48\hsize}
\includegraphics[width=7.4cm]{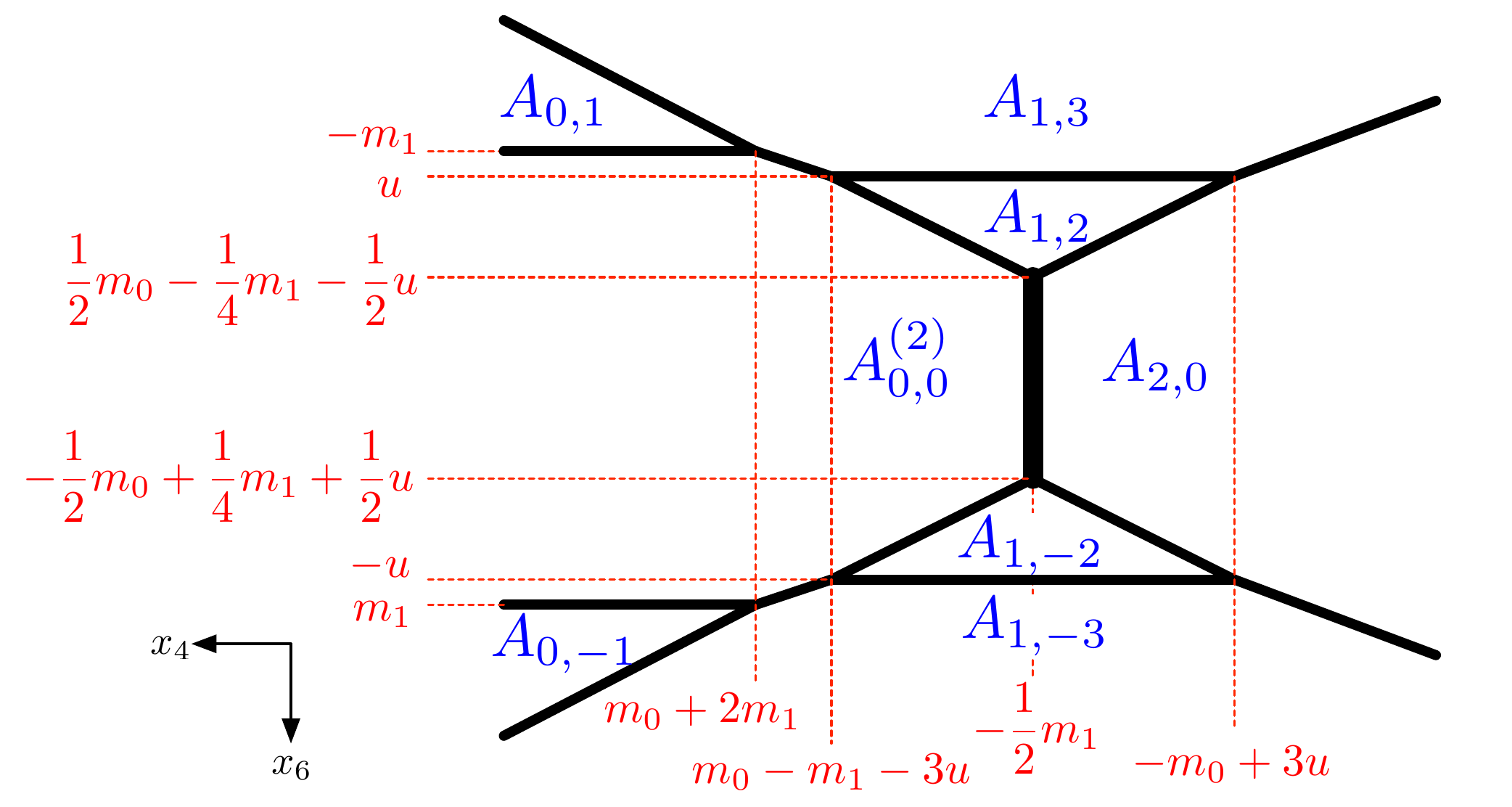}
\vspace{-1cm}
\caption{The regions for the largest $A_{k, l}^{(m)}$ in Region 8 of the $E_2$ theory. }
\label{Fig:E2Region8}
\end{minipage}
\end{figure}

\begin{figure}
\centering
\begin{minipage}{0.48\hsize}
\includegraphics[width=7.4cm]{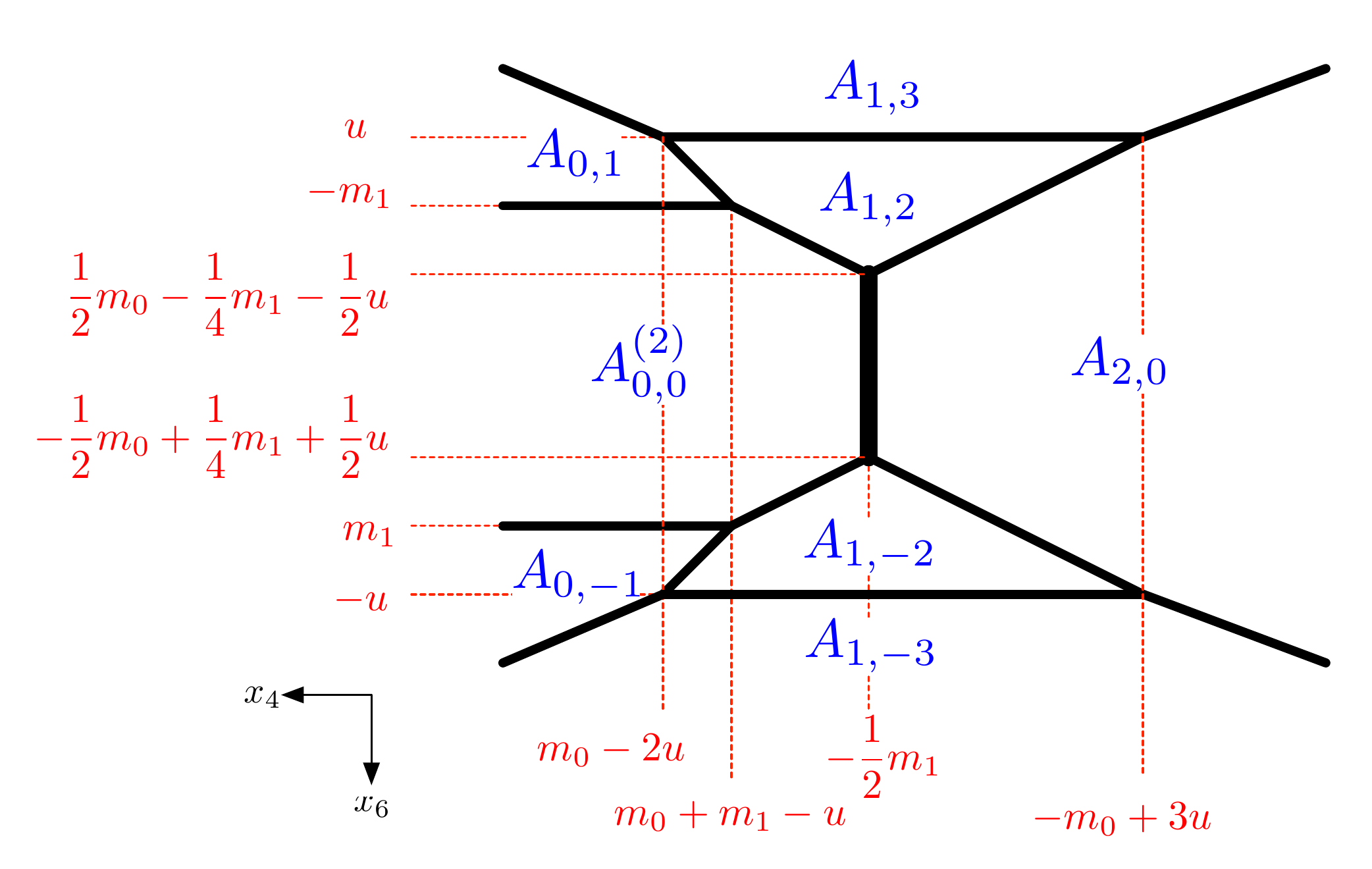}
\vspace{-1cm}
\caption{The regions for the largest $A_{k, l}^{(m)}$ in Region 9 of the $E_2$ theory. }
\label{Fig:E2Region9}
\vspace{1cm}
\end{minipage}
\hspace{0.02\hsize}
\begin{minipage}{0.48\hsize}
\includegraphics[width=7.4cm]{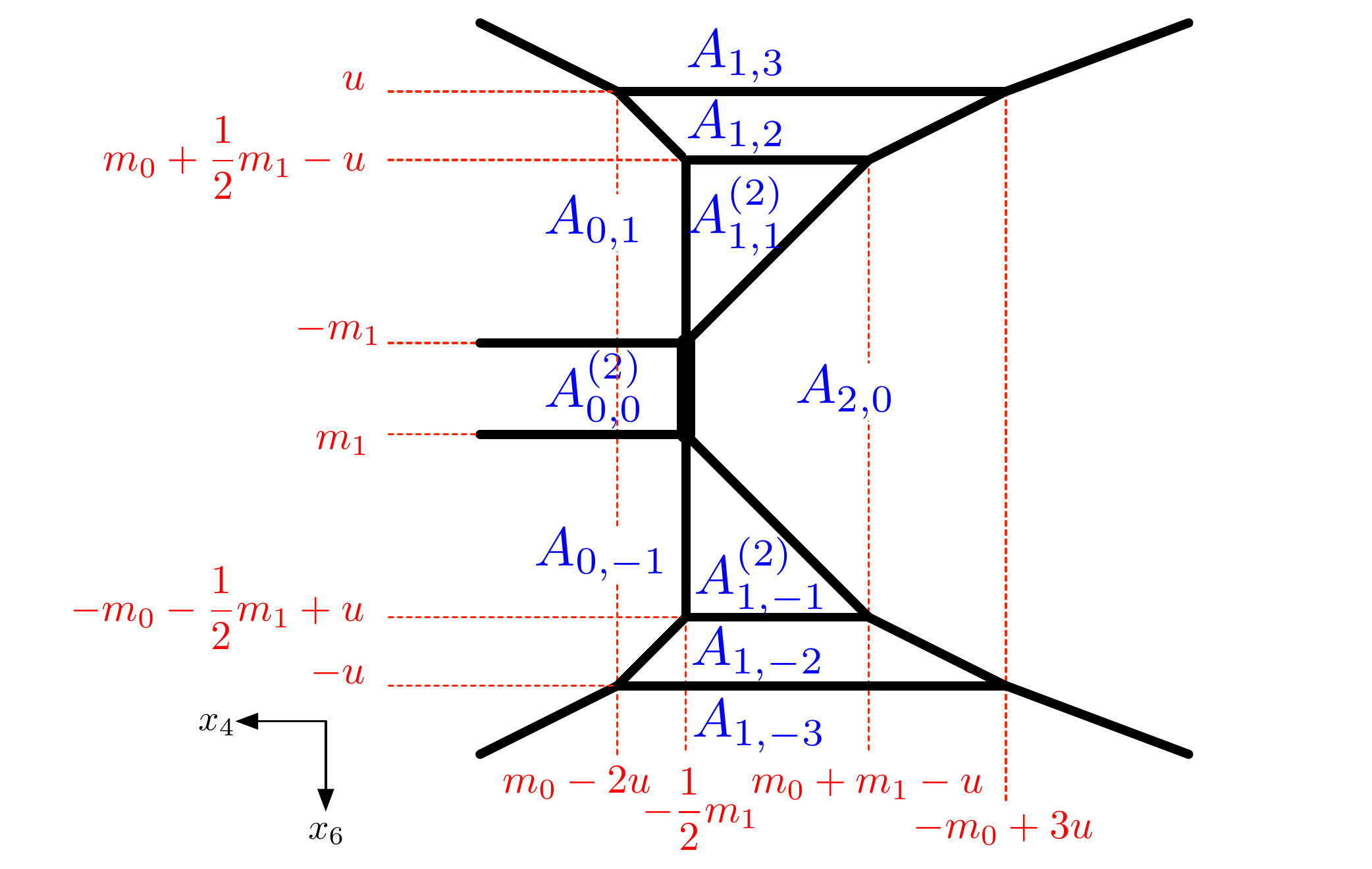}
\vspace{-1cm}
\caption{The regions for the largest $A_{k, l}^{(m)}$ in Region 10 of the $E_2$ theory. }
\label{Fig:E2Region10}
\vspace{1cm}
\end{minipage}
\begin{minipage}{0.48\hsize}
\includegraphics[width=7.4cm]{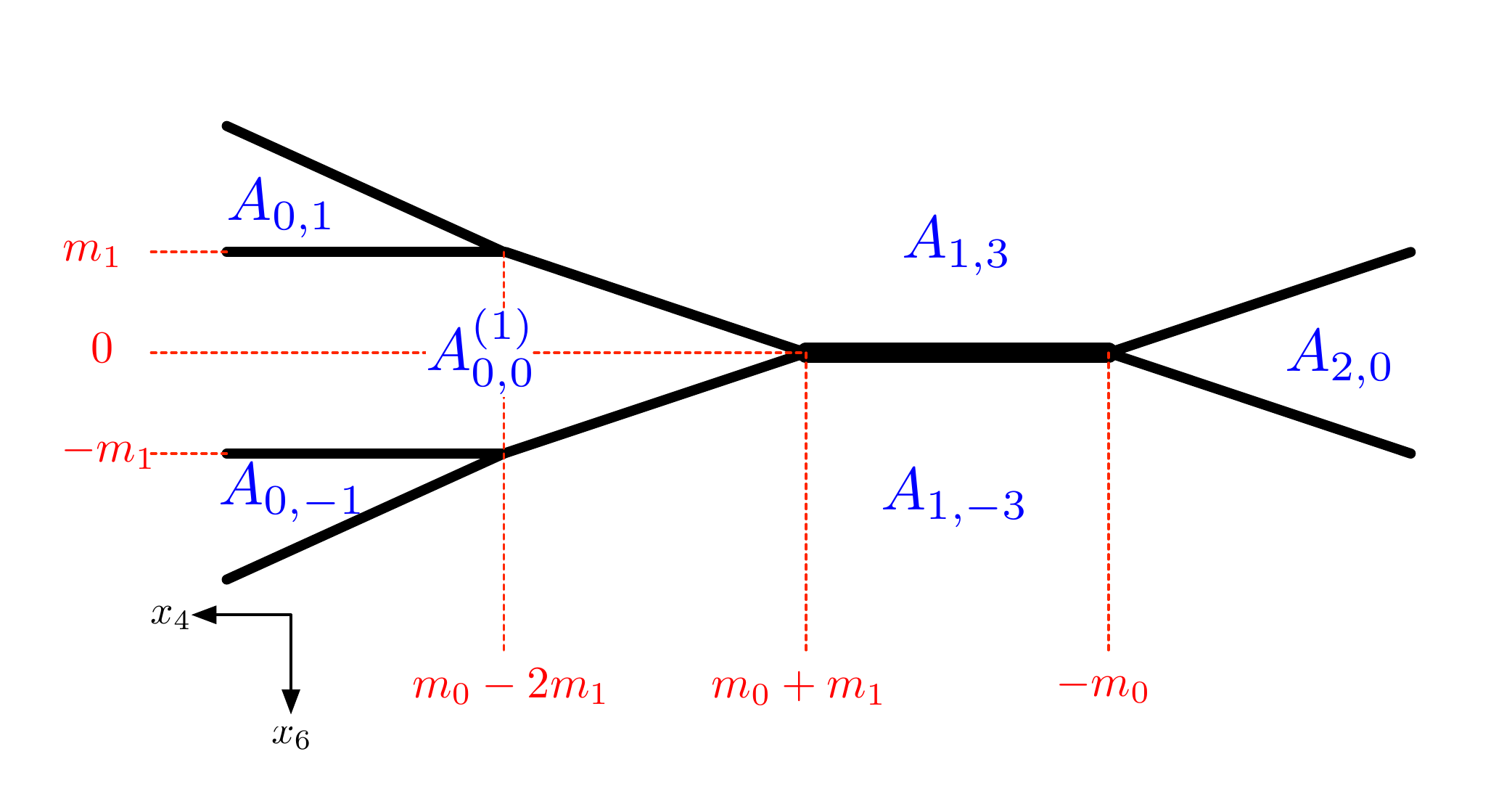}
\vspace{-1cm}
\caption{The regions for the largest $A_{k, l}^{(m)}$ in Region 11 of the $E_2$ theory. }
\label{Fig:E2Region11}
\vspace{1cm}
\end{minipage}
\hspace{0.02\hsize}
\begin{minipage}{0.48\hsize}
\includegraphics[width=7.4cm]{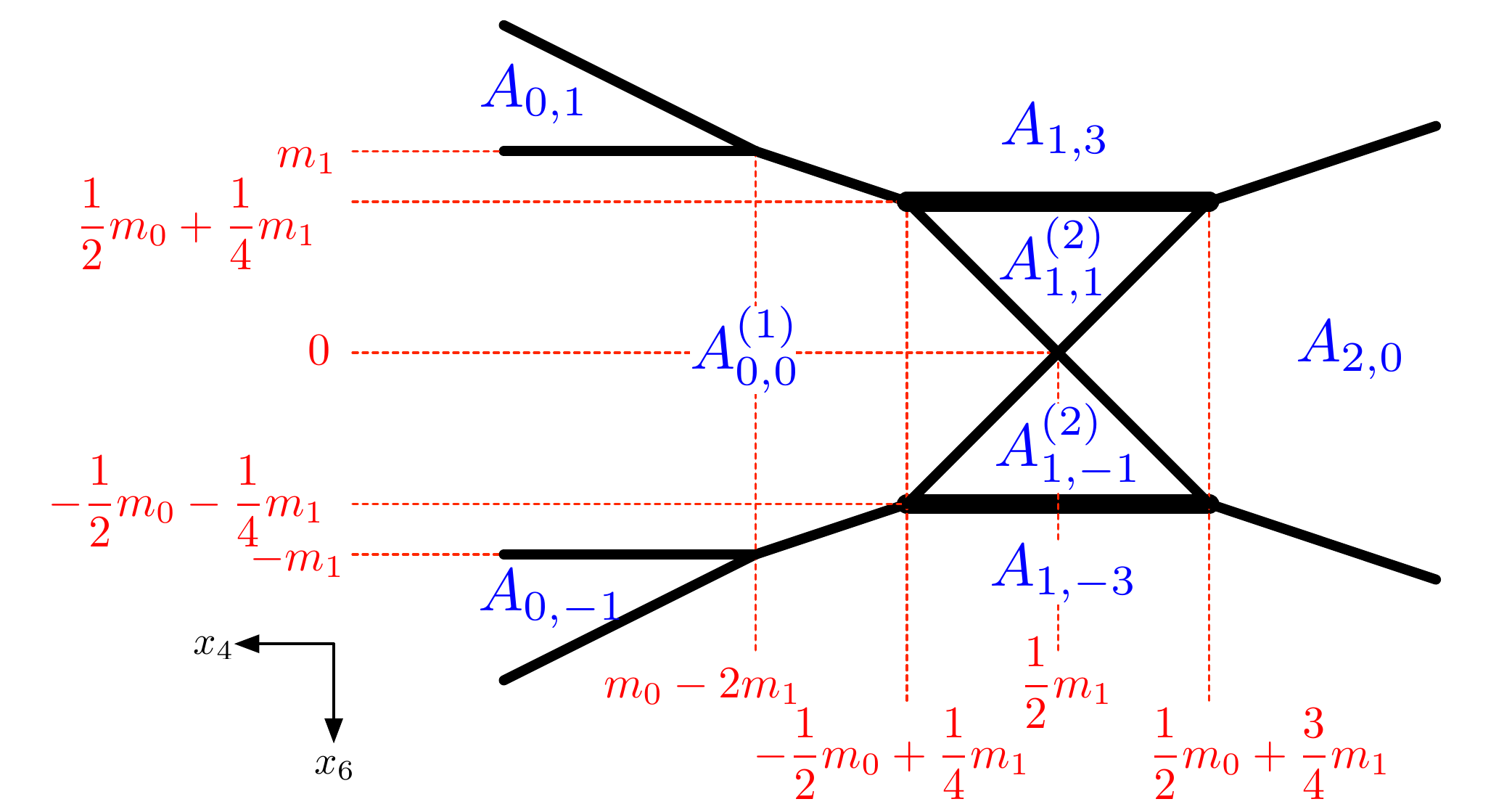}
\vspace{-1cm}
\caption{The regions for the largest $A_{k, l}^{(m)}$ in Region 12 of the $E_2$ theory. }
\label{Fig:E2Region12}
\vspace{1cm}
\end{minipage}
\begin{minipage}{0.48\hsize}
\includegraphics[width=7.4cm]{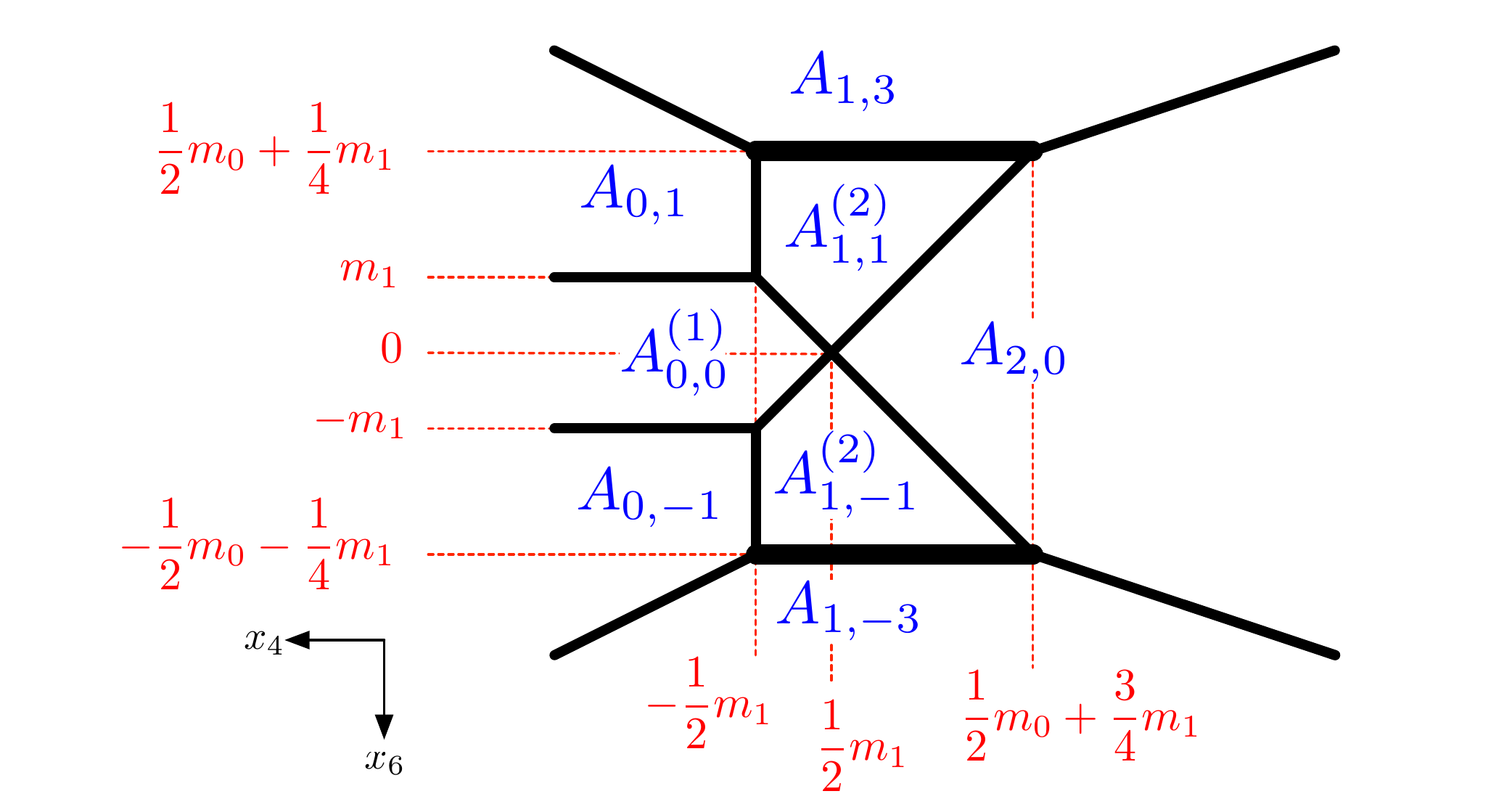}
\vspace{-1cm}
\caption{The regions for the largest $A_{k, l}^{(m)}$ in Region 13 of the $E_2$ theory. }
\label{Fig:E2Region13}
\vspace{1cm}
\end{minipage}
\hspace{0.02\hsize}
\begin{minipage}{0.48\hsize}
\includegraphics[width=7.4cm]{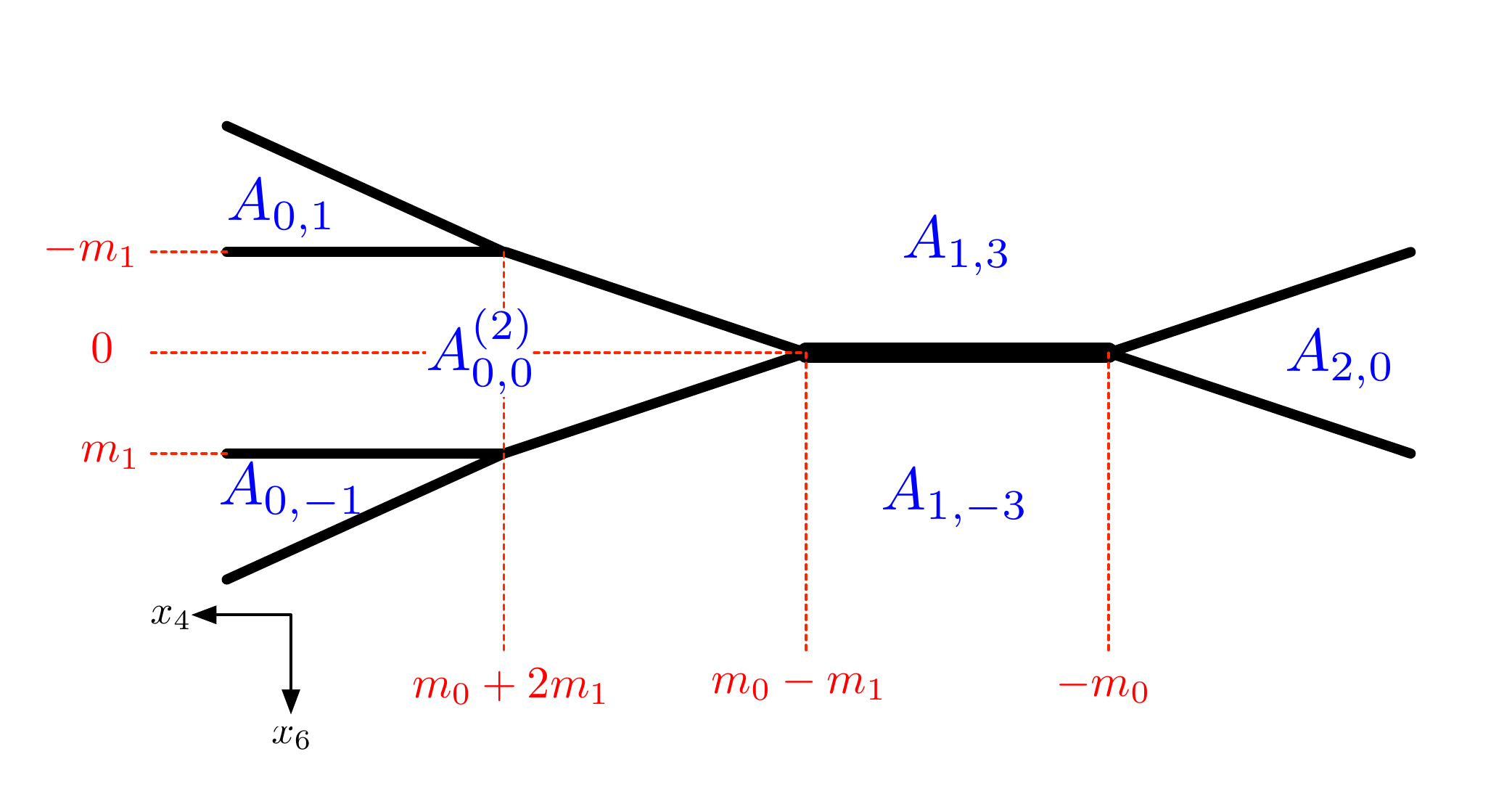}
\vspace{-1cm}
\caption{The regions for the largest $A_{k, l}^{(m)}$ in Region 14 of the $E_2$ theory. }
\label{Fig:E2Region14}
\vspace{1cm}
\end{minipage}
\begin{minipage}{0.48\hsize}
\includegraphics[width=7.4cm]{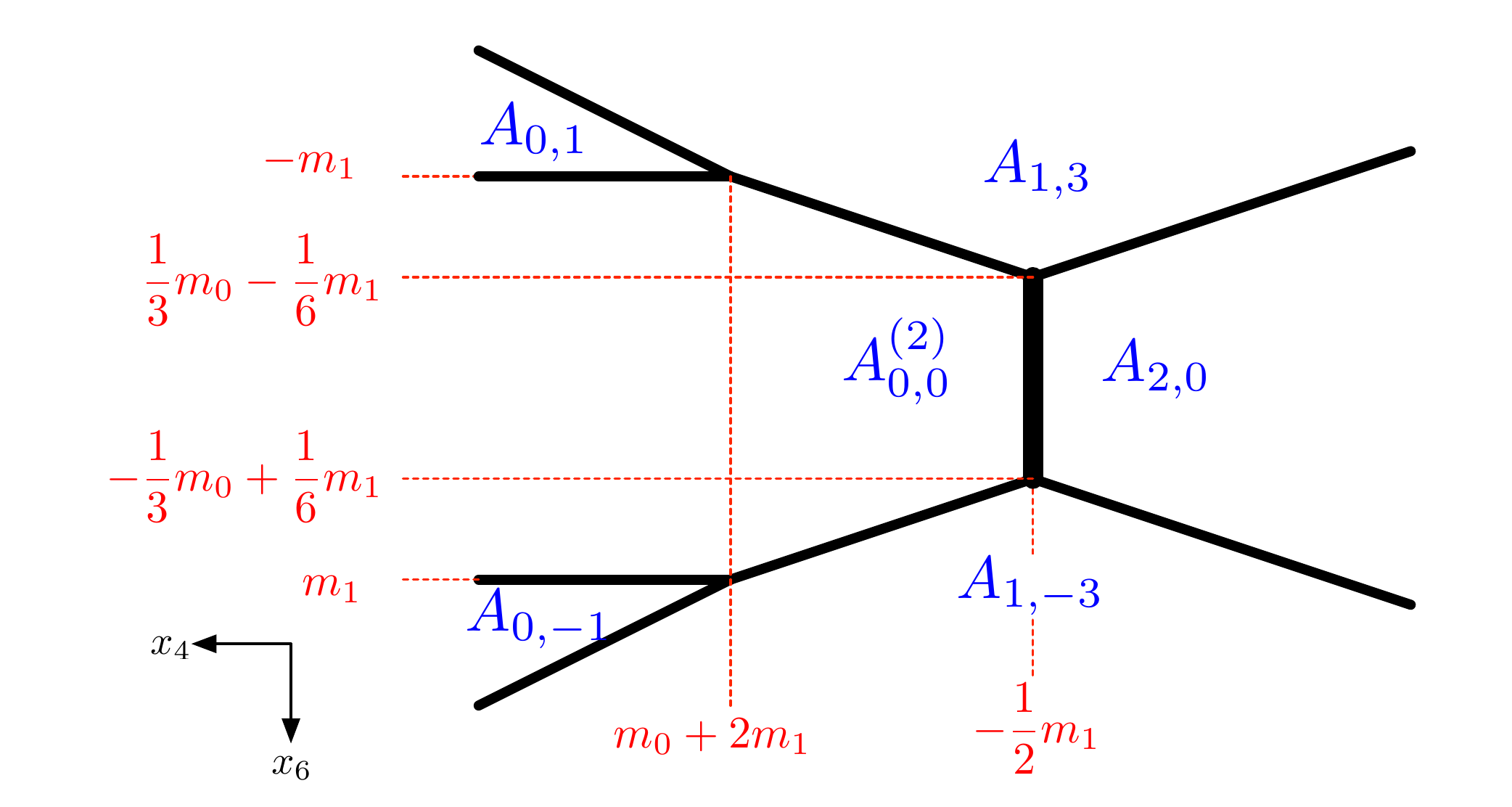}
\vspace{-0.9cm}
\caption{The regions for the largest $A_{k, l}^{(m)}$ in Region 15 of the $E_2$ theory. }
\label{Fig:E2Region15}
\end{minipage}
\hspace{0.02\hsize}
\begin{minipage}{0.48\hsize}
\includegraphics[width=7.4cm]{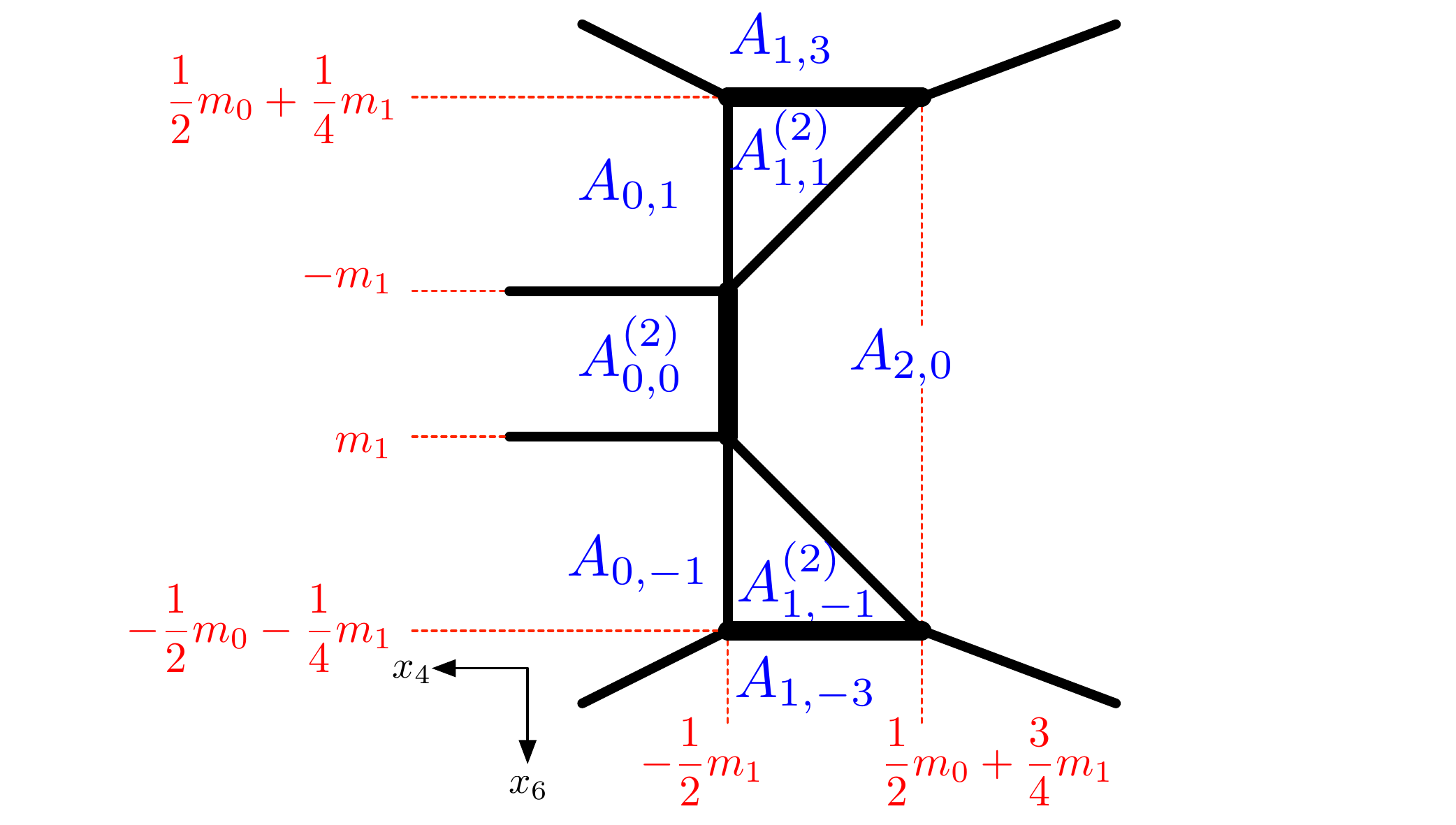}
\vspace{-1.1cm}
\caption{The regions for the largest $A_{k, l}^{(m)}$ in Region 16 of the $E_2$ theory. }
\label{Fig:E2Region16}
\end{minipage}
\end{figure}

In section \ref{sec:E2Phase}, we described the qualitative structure of the 5-brane diagram with an O5-plane. We here summarize more detailed structure of the 5-brane web diagrams by indicating the largest $A_{k, l}^{(m)}$'s in the $(x_4, x_6)$-space for Region 1 -- 16 in \eqref{eq:E2region1to16} as well as the locations of their boundaries. The results are depicted in Figure \ref{Fig:E2Region1} -- Figure \ref{Fig:E2Region16}.


\bigskip

\providecommand{\href}[2]{#2}\begingroup\raggedright\endgroup

\end{document}